\documentclass[prb,twocolumn,preprintnumbers,amsmath,amssymb]{revtex4}

\usepackage{graphicx}

\begin{document}
\title{\bf  Mutual Composite Fermion and composite Boson approaches to balanced and imbalanced bilayer quantum Hall system:
        an electronic analogy of the Helium 4 system }
\author{\bf  Jinwu Ye  }
\affiliation{ Department of Physics, The Pennsylvania State
University, University Park, PA, 16802 }
\date{\today}

\begin{abstract}
     We use both Mutual Composite Fermion (MCF) and Composite Boson (CB) approach
     to study balanced and im-balanced Bi-Layer Quantum Hall systems (BLQH) and make critical
     comparisons between the two approaches. We find the CB approach is superior to the MCF
     approach in studying ground states with different kinds of broken symmetries.
     In the phase representation of the CB theory,
     we first study the Excitonic superfluid state (ESF). The theory  puts spin and charge degree freedoms in the same footing,
     explicitly bring out the
     spin-charge connection and classify all the possible excitations in a systematic way.
     Then in the dual density representation of the CB theory, we study
     possible intermediate phases as the distance increases.
     We propose there are two critical distances $ d_{c1} < d_{c2} $ and three phases as the distance increases.
     When $ 0 < d < d_{c1} $, the system is in the ESF state which breaks the internal $ U(1) $ symmetry,
     when $ d_{c1} < d < d_{c2} $, the system is in an Pseudo-spin density wave ( PSDW ) state which breaks
     the translational symmetry,
     there is a first order transition at $ d_{c1} $ driven by the collapsing of magneto-roton minimum
     at a finite wavevector in
     the pseudo-spin channel. When $ d_{c2} < d < \infty $, the system becomes two weakly coupled
     $ \nu =1/2 $ Composite Fermion Fermi Liquid ( FL) state.
     There is also a first order transition at $ d= d_{c2} $.
     We construct a quantum Ginzburg Landau action to describe the transition from ESF to PSDW
     which break the two completely different symmetries. By using the QGL action, we explicitly
     show that the PSDW takes a square lattice and analyze in detail the properties of
     the PSDW at zero and finite temperature.
     We also suggest that the correlated hopping of vacancies in the active and passive layers
     in the PSDW state leads
     to very large and temperature dependent drag consistent with the experimental data.
     Then we study the effects of imbalance on both ESF and PSDW.
     In the ESF side, the system supports continuously changing
     fractional charges as the imbalance changes. In the PSDW side, there are two quantum
     phase transitions from the commensurate excitonic solid to an
     in-commensurate excitonic solid and then to the excitonic superfluid state.
     We also comment on the effects of disorders and compare our results with the previous
     work. The very rich and interesting phases and phase transitions in the pseudo-spin channel
     in the BLQH is quite similar to those in $ ^{4} He $ system with the distance
     playing the role of the pressure. A BLQH system in a periodic potential is also discussed.
     The Quantum Hall state to Wigner crystal transition in single layer Quantum Hall system is
     studied.

\end{abstract}
\maketitle

\section{Introduction}
{\sl (1) Experimental Observations }

   Extensive attention has been lavished on Fractional
   Quantum Hall Effects (FQHE) in multicomponent systems
   since the pioneering work by Halperin \cite{bert}. These components could be the spins of electrons
   when the Zeeman coupling is very small or layer indices in multi-layered system.
   In particular, spin-polarized Bilayer Quantum Hall systems at total filling factor
  $ \nu_{T} =1 $ have been under enormous experimental and theoretical investigations over the last decade
  \cite{rev}.  When the interlayer separation $ d $ is sufficiently large, the bilayer
  system decouples into two separate compressible $ \nu=1/2 $ layers.
  Earlier experiments exhibited a strong suppression of the tunneling current at low biases
  \cite{supp}. However, when $ d $ is smaller than a critical distance $ d_{c} $, even in the absence
  of interlayer tunneling, the system undergoes a quantum phase transition into
  a novel spontaneous interlayer coherent incompressible phase \cite{rev}.
  At low temperature, with extremely small interlayer tunneling amplitude,
  Spielman {\sl et al} discovered
  a very pronounced narrow zero bias peak in this interlayer coherent
  incompressible state \cite{gold}. M. Kellogg {\sl et al } also observed
  quantized Hall drag resistance at $ h/e^{2} $ \cite{hall}.
  In recent counterflow experiments, it was found that
  both linear longitudinal and Hall resistances take activated
  forms and vanish only in the low temperature limit \cite{counterflow}.

{\sl (2) Theoretical achievements }

    Starting from Halperin's $ 111 $ wavefunction Eqn.\ref{111} which describes a bi-layer system
   with $ N_{1} $ ( $ N_{2} $ ) electrons in the top (bottom) layer ( the total number of
   electrons is $ N=N_{1}+N_{2} $ ), using various methods, several authors discovered a
   Neutral Gapless Mode (NGM) with linear dispersion relation $ \omega \sim v k $ and
   that there is a finite temperature Kosterlitz-Thouless (KT)
   phase transition associated with this NGM \cite{fer,wen,jap}.
   By treating the two layer indices as two pseudo-spin indices,
   Girvin, Macdonald and collaborators mapped the bilayer system
   into a Easy Plane Quantum Ferromagnet (EPQFM) \cite{yang,moon,rev} ( which is equivalent to
   the Excitonic Superfluid ).
   They established the mapping by projecting the Hamiltonian of
   the BLQH onto the Lowest Landau Level (LLL) and then using
   subsequent Hartree-Fock (HF) approximation and gradient expansion ( called LLL+HF in the following ).
   In the picture of EPQFM \cite{rev},
   the canonical ensemble with definite $ S^{z}= N_{1}-N_{2}=0 $ is replaced
   by Grand canonical ensemble with fluctuating $ S^{z} $. The relative fluctuation of $ S^{z} $ is at
   the order of $ 1/\sqrt{N} \rightarrow 0 $ as $ N \rightarrow \infty $.
   By drawing the analogy with superconductivity where canonical ensemble with definite number of Cooper pairs
   is shown to be equivalent to BCS wavefunction which is a grand canonical ensemble
   with indefinite number of Cooper pairs,
   the authors in \cite{yang,moon,rev} argued
   this trial wavefunction is a good approximation to the exact ground state. The low energy
   excitations above the ground state is given by an effective $ 2+1 $ dimensional $ XY $ model.
   There are 4 flavors of topological defects called " merons " which carry fractional charges $ \pm 1/2 $ and
   also have $ \pm $ vorticities. They have logarithmic divergent self energies and are bound into pairs
   at low temperature. The lowest energy excitations carry charge $ \pm e $ which
   are a meron pair with opposite vorticity, but
   the same charge. There is a finite temperature phase transition at $ T_{KT} $ where
   bound states of the 4 flavors of merons are broken into free merons.
   The large longitudinal resistivity $ (\sim 1 k\Omega ) $ observed in \cite{gold} at very low temperature
   indicated that these meron pairs  are highly mobil.
   In the presence of small tunneling,
   they \cite{yang,rev} found that when the applied in-plane magnetic field
   is larger than a critical field $ B^{*}_{||} $,
   there is a phase transition from a commensurate state to an incommensurate state ( C-IC)
   with broken translational symmetries. When $ B > B^{*}_{||} $, there is a finite temperature KT transition
   which restores the translation symmetry by means of dislocations
   in the domain wall structure in the incommensurate phase.
   Starting from the EPQFM approach,
   several groups investigated $ I-V $ curves in the presence of small tunneling \cite{balents}.
   In addition to the work mentioned above, there are also many other
   works done on BLQH. For example, several authors applied different versions of
   composite fermion Chern-Simon theory to study BLQH systems in
   \cite{benice,kim,jain1}.

{\sl (3) Discrepancies between theory and the experiments. }

    Despite the intensive theoretical research in \cite{fer,wen,jap,yang,moon,rev},
    there are still many serious discrepancies between theory and the experiments.
    According to the present theories, in the Excitonic Superfluid ( ESF ) state, there should be
    the interlayer tunneling Josephson effect, finite temperature KT transition and vanishing linear longitudinal and Hall resistances
    in the counterflow channel below the KT transition temperature.
    Unfortunately, all these characteristics have never been observed in the experiments.
    Although it appears certain that
    the dramatic conductance peak observed at $ d < d_{c} $ is due
    to the collective NGM in the ESF state, no theory can explain
    the magnitude of zero bias peak conductance which, though
    enormously enhanced, does not exceed $ 10^{-2} \frac{e^{2}}{h}
    $, its width and its dependence on proximity to the ESF state
    boundary ( see, however, the most recent work \cite{transfer} ).
    All the previous calculations predicted that the in-plane field will split the zero bias peak into two side
    peaks, but the experiments showed that although there are two tiny shoulders appearing,
    the central peak stay. The excess dissipation observed in the counterflow
    experiments \cite{counterflow} appears to vanish only as $
    T \rightarrow 0 $ limit. The origin of the excess dissipation
    remains an important unresolved question. But it was argued in \cite{counterflow2} that
    the origin may come from disorders induced mobil vortices.

{\sl (4) Open problems in SLQH and BLQH }

   One of the remaining outstanding problems in single layer quantum Hall ( SLQH ) system is to understand the quantum
   phase transitions from QH to QH or QH to insulating state as
   tuning the magnetic field at a fixed electron density or vice versa.
   Similarly, one of the remaining outstanding problems in BLQH is
   to understand
   novel phases and quantum phase transitions as changing the distance
   between the two layers. When the distance is smaller than $ d_{c1} $, the system is in the
   ESF state, while it is
   sufficiently large, the system becomes two weakly coupled $ \nu =1/2 $ Composite Fermion Fermi Liquid (FL) state.
   There could be a direct first order transition between the two states as indicated in some numerical calculations
   \cite{phasetran}.
   However, the experimental observations
   that both zero voltage tunneling peak \cite{gold} and the Hall drag resistivity \cite{hall}
   develop very gradually when $ d \sim d_{c1} $
   suggest the transition at $ d= d_{c1} $ is a 2nd order phase transition.
   Although there are very little dissipations
   in both the ESF and FL, the experiment \cite{drag} discovered strong enhancement of drag and dissipations
   in a large intermediate distance range. These experimental observations
   suggest that there must be intermediate phases between the two phases.
   Unfortunately, so far, the nature of the intermediate phase,
   especially the quantum phase transition between the ESF and the
   intermediate phase was not systematically investigated. The
   outstanding problems in  both SLQH and BLQH will be addressed in this
   paper.

{\sl (5) Advances in Helium 4 system}

     A superfluid is a fluid that can flow through the tiniest channels or cracks without viscosity.
     The phenomenon of superfluidity was first observed in the two isotopes of
     Helium: $^{4}He $ and $^{3}He$ which become superfluids below the transition
     temperatures $ T_{c} =2.18 K $ and $ T_{c} = 2.4 mK $ respectively \cite{he4,he3}.
     Recently, a PSU group lead by Chan found some signature of a
     possible supersolid state in $^{4}He $ at low temperature $ T < 200 \ mK $ and
     high pressure $ p > p_{c} \sim 25 \ bar $. The authors in
     \cite{andreev,ches,ander} argued that there may be vacancies even at $ T=0 $
     whose condensation leads to the supersolid $^{4}He $.
     In \cite{qgl}, the author constructed a Ginsburg Landau (GL ) theory to map out the $^{4}He$ phase diagram,
     analyze carefully the conditions for the  existence of the supersolid (SS) and
     study all the phases and phase transitions in a unified framework.
     I introduced a single parameter $ g $ which is the coupling between the normal solid
     (NS)component  and superfluid (SF) component in the GL theory. If
     $ g=g_{v} < 0 $, I explicitly showed that there are two scenarios
     (1) If $| g_{v} |$ is sufficiently small, then the
     normal solid is a commensurate solid ( C-NS ). The C-NS
     still does not have a particle-hole symmetry, it is a vacancy-like NS where the excitation energy
     of a vacancy is lower than that of an interstitial, therefore named NS-v.
     Then for the first time, we construct a quantum Ginsburg-Landau
     theory (QGL) to study the SF to NS-v transition and explicitly derive the Feymann relation from the QGL  (2)
     If $ |g_{v}| $ is sufficiently large, then a  vacancy induced SS  ( SS-v ) exists at sufficient low temperature.
     The critical temperature $ T_{SS-v} $ becomes an effective measure of the coupling strength $ g_{v} $.
     The analogy between ESF in the pseudospin channel in the BLQH and the superfluid in
     $ ^{4}He $ was explored by many previous work
     \cite{fer,wen,jap,yang,moon,rev}. In this paper, we push this analogy to deeper and broader
     contexts. We find that the very rich and
     interesting phases and phase transitions in the pseudo-spin channel in the BLQH is
     quite similar to those in $ ^{4} He $ system with the distance
     playing the role of the pressure.

{\sl (6) The analogy between BLQH and Helium 4 to be explored in
this paper }

   In this paper, we use both Mutual Composite Fermion ( MCF) \cite{jain,frad,hlr}  and
   Composite Boson ( CB) approaches \cite{off,cb} to study balanced and im-balanced BLQH systems.
   We identify many problems with MCF approach. Then we develop a simple and
   effective CB theory which naturally avoids all the problems suffered in the MCF approach.
   The CB theory not only can describe many crucial properties of the ESF elegantly, it
   can also be applied to study the possible novel intermediate phase with broken translational symmetry
   and the quantum phase
   transitions between these different ground states. Therefore the CB approach
   is superior to the MCF approach in describing the BLQH where ground states with different kinds of broken symmetries
   may happen as the distance changes.
     By using this CB theory in its phase representation, we first study the excitonic
     superfluid ( ESF ) state. The theory naturally puts spin and charge degree freedoms in the same footing, explicitly
     bring out the spin-charge connection and classify all the possible
     excitations in a systematic way.
     Then by using the CB theory in its dual density
     representation and inspired by the insights achieved from the QGL theory  developed in \cite{qgl}
     to describe the superfluid to solid and supersolid transition in $^{4} He $,
     we explore  many similarities and also some difference between  the
     BLQH and the $^{4} He $ system.
     Our understanding on phases and phase transitions in $ ^{4}He $
     system can shed considerably lights on the phase and phase
     transitions in BLQH with the pressure playing the role of the distance.
     For balanced BLQH, there are two critical distances $ d_{c1} < d_{c2} $.
     When $ 0 < d < d_{c1} $, the system is in the ESF state,
     when $ d_{c1} < d < d_{c2} $, the system is in the Pseudo-spin density wave
     (PSDW) state, there is a first order transition at $ d_{c1} $
     driven by magneto-roton minimum collapsing at a finite wavevector in the pseudo-spin channel.
     When $ d_{c2} < d < \infty $, the PSDW melts into two weakly coupled
     $ \nu =1/2 $ Composite Fermion Fermi Liquid (CFFL) state.
     There is a  also first order transition at $ d= d_{c2} $.
     However, disorders could smear the two first order transitions into two second order transitions.
     The transition from the ESF to the PSDW is unusual because the
     two states break two completely different symmetries: the global internal $
     U(1) $ symmetry and  the translational symmetry respectively.
     We construct an effective theory in the dual density representation to describe
     this novel quantum phase transition. We find the transition is very similar to the
     superfluid to normal solid transition in $ ^{4} He $ system
     with the distance playing the role of the pressure.

{\sl (7) Experimental setup}

   Consider a bi-layer system with $ N_{1} $ ( $ N_{2} $ ) electrons in left ( right ) layer and
  with interlayer distance $ d $ in the presence of magnetic field $ \vec{B} = \nabla \times \vec{A} $ ( Fig.1):
\begin{eqnarray}
   H & = & H_{0} + H_{int}    \nonumber  \\
   H_{0} & = &  \int d^{2} x c^{\dagger}_{\alpha}(\vec{x})
   \frac{ (-i \hbar \vec{\nabla} + \frac{e}{c} \vec{A}(\vec{x}) )^{2} }
      {2 m }  c_{\alpha}(\vec{x})                \nonumber   \\
    H_{int} & = &  \frac{1}{2} \int d^{2} x d^{2} x^{\prime} \delta \rho_{\alpha} (\vec{x} )
               V_{\alpha \beta} (\vec{x}-\vec{x}^{\prime} )  \delta \rho_{\beta } ( \vec{x^{\prime}} )
\label{first}
\end{eqnarray}
  where electrons have {\em bare} mass $ m $ and carry charge $ - e $, $ c_{\alpha}, \alpha=1,2 $ are
  electron  operators in top and bottom layers, $ \delta \rho_{\alpha}(\vec{x}) = c^{\dagger}_{\alpha} (\vec{x})
  c_{\alpha} (\vec{x} ) -n_{\alpha}, \alpha=1,2 $ are normal ordered electron densities on each layer.
  The intralayer interactions
  are $ V_{11}=V_{22}= e^{2}/\epsilon r $, while interlayer interaction is $ V_{12}=V_{21}= e^{2}/ \epsilon
  \sqrt{ r^{2}+ d^{2} } $ where $ \epsilon $ is the dielectric constant. For imbalanced bi-layers,
  $ n_{1} \neq n_{2} $, but the background positive charges are still the same in the two layers,
  the chemical potential term is already included in $ H_{int} $ in Eqn.\ref{first}.
  There are two limits: (1) Weak tunneling limit $ \Delta_{SAS} \ll
  \frac{ e^{2} d }{\epsilon} $. (2) Strong tunneling limit $ \Delta_{SAS}
  \gg
  \frac{ e^{2} d }{\epsilon} $ where the bilayer system becomes
  essentially the same as a single layer. In this paper, we only
  consider the non-trivial weak tunneling limit at total filling
  factor $ \nu_{T}= \nu_{1} + \nu_{2} =1 $. For simplicity, we set $ \Delta_{SAS}=0 $ in
  the Eqn.\ref{first}.

\vspace{0.25cm}

\begin{figure}
\includegraphics[width=8cm]{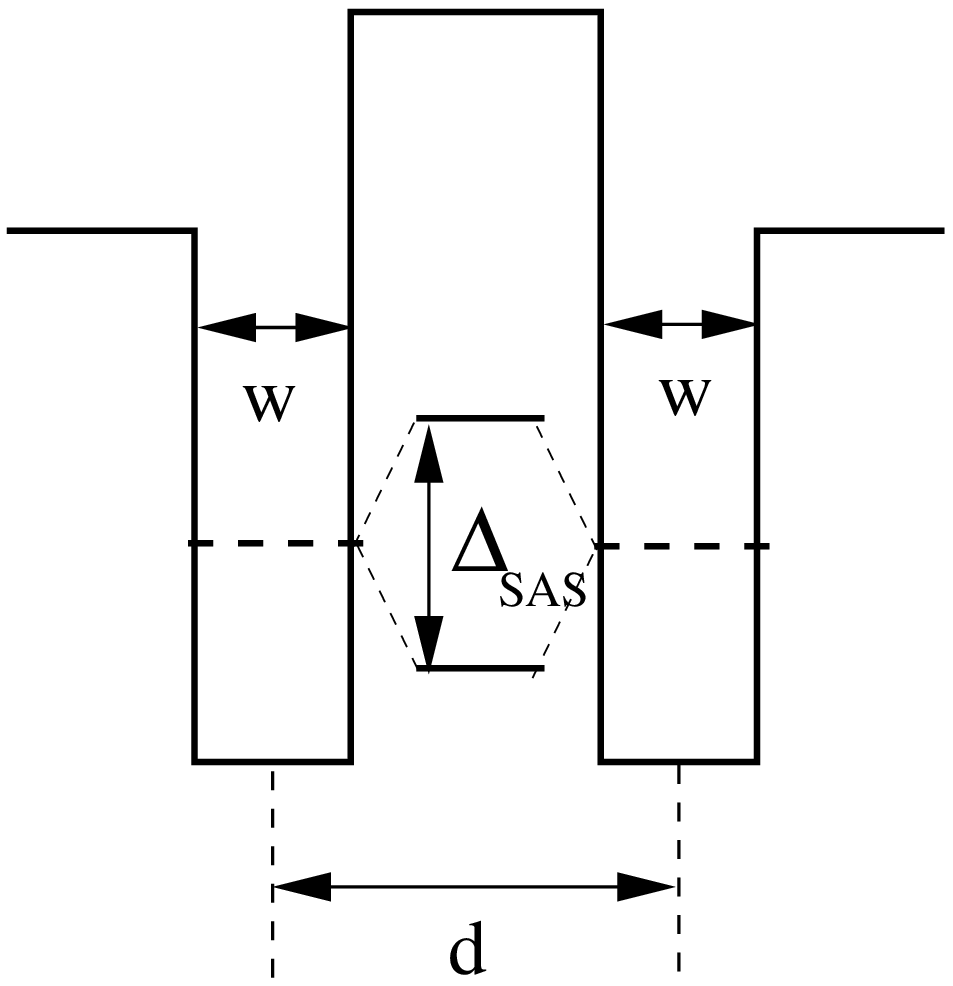}
\caption{ A  fabricated bilayer quantum Hall system. The quantum
well width $ w $, the well center to well center distance $ d $ and
the magnetic length $ l $ ( which is the average distance between
electrons in each well) are all of the same order $ \sim 100 \AA $.
$ \Delta_{SAS} $ is the tunneling between the two wells. }
\label{fig1}
\end{figure}

\vspace{0.25cm}

{\sl (8) The organization of the paper }

   The rest of the paper is organized as follows. In section II,
   we use a MCF theory to study the BLQH.
   We achieve some limited success, but also run into many troublesome problems.
   In section III, we use a CB theory which puts spin and charge sector
   on the same footing to study the BLQH. We demonstrate why this CB theory
   naturally avoid all the problems suffered
   in the MCF approach. We also compare the CB approach with the EPQFM approach
   and point out advantages and limitations of both approaches. In
   section IV which is the key section of this paper,
   we push the CB theory further to analyze carefully the instability in the
   pseudo-spin channel which lead to the PSDW state and study its properties at both zero and finite
   temperatures. We also point out the important physical
   consequences of  possible quantum fluctuation generated vacancies
   in the PSDW state. We also compare our results with previous ones on the intermediate phase.
   We study the effects of imbalance on both ESF and PSDW in the corresponding sections III and IV.
   In the final section, we summarized the main results of the paper and concluded that
   CB theory is superior to MCF approach in BLQH systems with ground states of different broken symmetries.
   In Appendix A, we evaluate the meron fractional charges from its
   trial wavefunction in imbalanced case.
   In Appendix B, we investigate one of the outstanding problems
   in SLQH: the Quantum Hall state to Wigner crystal state transition in SLQH.
   In appendix C, we study phases and phase transitions in a BLQH in
   a periodic potential and two other closely related interesting models.

\section{ Mutual Composite fermion approach: Limited success and failure}

    In parallel to advances in bi-layer QH systems, much progress has been made on novel physics involving
   quasi-particles and vortices in high temperature superconductors.
   Anderson employed a single-valued singular
   gauge transformation to study the quasi-particle energy spectrum in the vortex lattice state \cite{and}.
   By employing the Anderson transformation, the author studied the quasi-particle transport
   in random vortex array in the mixed state \cite{random}. The author also extended the Anderson singular gauge
   transformation for static vortices to a mutual singular gauge transformation for quantum fluctuation
   generated dynamic vortices \cite{quantum}. By using this dynamic
   gauge transformation, the author investigated the zero temperature quantum phase transition
   from $ d $-wave superconductor
   to underdoped side by assuming the transition is driven by the condensations of quantum
   fluctuation generated vortices \cite{quantum}.
   In this section, by employing essentially the same singular gauge transformation used
   to study the interactions between quasi-particles and vortices in high temperature
   superconductors, we revisit the bi-layer QH systems.

{\sl (1) Singular Gauge Transformation:}

  Performing a single-valued singular gauge transformation (SGT) \cite{quantum,ex}:
\begin{equation}
  U= e^{i \tilde{\phi} \int d^{2} x \int d^{2} x^{\prime}
  \rho_{1} (\vec{x} ) arg(\vec{x}-\vec{x}^{\prime} ) \rho_{2} ( \vec{x}^{\prime} ) }, ~~~~~~~\tilde{\phi}=1
\label{sing}
\end{equation}
   we can transform the above Hamiltonian into:
\begin{equation}
   H_{0} = \int d^{2} x \psi^{\dagger}_{\alpha}(\vec{x}) \frac{ (-i \hbar \vec{\nabla} + \frac{e}{c} \vec{A}(\vec{x})
             - \hbar \vec{ a }_{\alpha}(\vec{x}) )^{2} }
      {2 m }  \psi_{\alpha}(\vec{x})
\label{second}
\end{equation}
  where the transformed fermion is given by:
\begin{eqnarray}
  \psi_{1}(\vec{x}) & =  & U c_{1}(\vec{x}) U^{-1} = c_{1}(\vec{x})
  e ^{ i \int d^{2} x^{\prime} arg(\vec{x}-\vec{x}^{\prime} ) \rho_{2} ( \vec{x}^{\prime} ) }
                                 \nonumber  \\
  \psi_{2}(\vec{x}) & = & U c_{2}(\vec{x}) U^{-1} = c_{2}(\vec{x})
  e ^{ i \int d^{2} x^{\prime} arg(\vec{x}^{\prime} -\vec{x} ) \rho_{1} ( \vec{x}^{\prime} ) }
\label{inter}
\end{eqnarray}
  and the two mutual Chern-Simon (CS) gauge fields $ a_{\alpha} $ satisfies:
  $ \nabla \cdot \vec{a}_{\alpha}=0, \nabla \times \vec{ a }_{\alpha} = 2 \pi \rho_{\bar{\alpha}}(\vec{x}) $ ( See Fig.2a).
  Obviously, the interaction term is unaffected by the singular-gauge
  transformation.
  Note that it is $ arg(\vec{x}-\vec{x}^{\prime} ) $ appearing in $ \psi_{1}(\vec{x}) $, while
  $ arg(\vec{x}^{\prime} -\vec{x} ) $ appearing in $ \psi_{2}(\vec{x}) $ in Eqn.\ref{inter}.
  This subtle difference is crucial to prove all the commutation relations are kept intact by
  the single-valued singular gauge transformation Eqn.\ref{sing}. Note also that
  $ arg(\vec{x}^{\prime} -\vec{x} ) $ works equally well in Eqn.\ref{sing}.

    It is easy to check that in single layer system where $ \rho_{1}= \rho_{2} =\rho $,
    Eqn.\ref{sing} reduces to the conventional singular gauge transformation
    employed in \cite{hlr}:
\begin{equation}
  \psi_{a}(\vec{x}) = e ^{ i 2 \int d^{2} x^{\prime} arg(\vec{x}-\vec{x}^{\prime} )
  \rho_{a} ( \vec{x}^{\prime} ) } c_{a}( \vec{x})
\label{dec}
\end{equation}
   where $ \rho_{a} ( \vec{x} ) = c^{\dagger}_{a}( \vec{x} ) c_{a}( \vec{x} ) $
   is the electron density in layer $ a =1,2 $.
    It puts {\em two} flux quanta in the opposite direction
    to the external magnetic field at the position of each electron ( Fig. 1 b). On the average, a CF feels
    a reduced effective field which is the external magnetic field minus the attached flux quanta.

    For $ \nu_{T} =1 $ bi-layer system,
    Eqn.\ref{inter} puts {\em one } flux quantum in one layer in the opposite direction
    to the external magnetic field at the position directly above or below
    each electron in the other layer ( Fig.2 a).

\vspace{0.25cm}

\begin{figure}
\includegraphics[width=8cm]{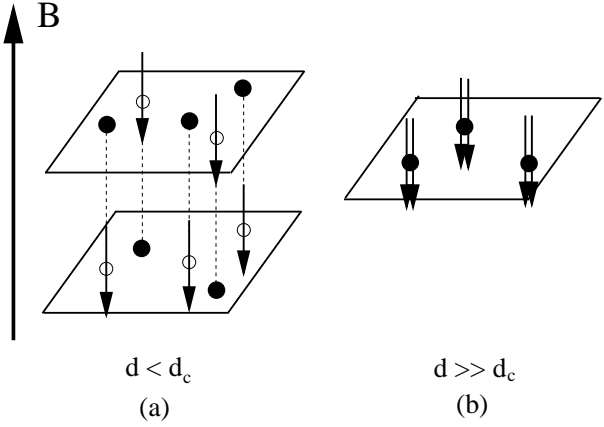}
\caption{ Contrast the flux attachment in Eqn.\ref{sing} (a) with
    that  in Eqn.\ref{dec} (b). In (a), there is one flux quantum in layer 1 when there is an electron directly
          downstairs in layer 2 and there is one flux quantum in layer 2 when there is an electron directly
          upstairs in layer 1. One can compare Fig. 1a with Fig. 2.
          For simplicity, we only show three electrons in the top layer and two electrons in the bottom layer.  }
 \label{fig2}
\end{figure}

\vspace{0.25cm}

    On the average, a Mutual Composite Fermion (MCF)
    in each layer feels a reduced effective field which is the external magnetic field
    minus the inserted flux quanta in this layer.
    In single layer system, it is essential to attach even  number of flux quanta to keep Fermi
    statistics intact.  The two attached flux quanta
    are moving together with the associated electron.
    However, in bi-layer system, inserting one flux quantum does keep Fermi statistics intact. The inserted
    one flux quantum in one layer is moving together with its associated electron
    in the other layer.  If choosing $ \tilde{\phi} $=1/2,
    the transformation is not single-valued, the statistics
    is changed from fermion to boson, this choice will be pursued in the next section
    on Composite boson approach.

{\sl (2) Mean field theory:}

 In the following, we put $ \hbar=c=e= \epsilon = 1 $.
  At total filling factor $ \nu_{T}=1 $, $ \nabla \times \vec{A}
   = 2 \pi n $ where $ n=n_{1}+n_{2} $ is the total average electron density. By absorbing the average
   values of C-S gauge fields $ \nabla \times < \vec{a}_{\alpha} > = 2 \pi
   n_{\bar{\alpha}} $ into the external gauge potential $ \vec{A}^{*}_{\alpha} =\vec{A}- < \vec{a}_{\alpha} > $,
   we have:
\begin{equation}
   H_{0} = \int d^{2} x \psi^{\dagger}_{\alpha}(\vec{x}) \frac{ (-i \vec{\nabla} + \vec{A}^{*}_{\alpha}(\vec{x})
             - \delta \vec{ a }_{\alpha}(\vec{x}) )^{2} }
      {2 m }  \psi_{\alpha}(\vec{x})
\label{original}
\end{equation}
  where $ \nabla \times \vec{A}^{*}_{\alpha} = 2 \pi n_{\alpha} $ and $  \nabla \times  \delta \vec{ a }_{\alpha} =
  2 \pi \delta \rho_{\bar{\alpha}}(\vec{x}) $ are the deviations from the corresponding
  average density ( In the following,
  we will simply use $ a_{\alpha} $ to stand for these deviations).

   When $ d < d_{c} $, the strong inter-layer interactions renormalize the bare mass into
   two effective masses $ m^{*}_{\alpha} $ \cite{shankar}.
   MCF in each layer feel effective magnetic
   field $ B^{*}_{\alpha} =\nabla \times \vec{A}^{*}_{\alpha} = 2 \pi n_{\alpha} $, therefore fill exactly
   one MCF Landau level. The energy gaps are simply the cyclotron gaps of the MCF
   Landau levels $ \omega^{*}_{c \alpha} = \frac{ B^{*}_{\alpha} }{ m^{*}_{\alpha} } $.

{\sl (3) Fractional charges:}

 Let's look at the charge of quasi-particles created
   by MCF field operators $ \psi^{\dagger}_{\alpha} ( \vec{x} ) $.
   Intuitively, inserting $ \psi^{\dagger}_{1} ( \vec{x} ) $ on layer 1 not only inserts an electron in layer 1
    at position $ \vec{x} $, but also pushes $ \nu_{2} = \frac{N_{2}}{N} $ electrons in layer 2 into its boundary,
    therefore induces a local charge deficit at $\vec{x} $ in layer 2
   which carries charge $ \nu_{2} $, the total
   charge at $ \vec{x} $ is $ e^{*}_{1}= -1+ \nu_{2} =- \nu_{1} $. If $ \vec{x} $ and $ \vec{y} $ are two points
   far apart, then the product of operators $ \psi^{\dagger}_{1} ( \vec{x} )
   \psi_{1}( \vec{y} ) $ in layer 1 will create a pair of fractional charges $ \pm e^{*}_{1} $ at positions
   $ \vec{x} $ and $ \vec{y} $ with the charge gap $ \hbar \omega^{*}_{c1} $.
   Similarly, inserting $ \psi^{\dagger}_{2} ( \vec{x} ) $ in layer 2 will create a charge
   $ e^{*}_{2}= -1+ \nu_{1} =- \nu_{2} $ at $ \vec{x} $ ( note that $ e^{*}_{1} + e^{*}_{2} = -1 $ ).
   The product of operators $ \psi^{\dagger}_{2} ( \vec{x} )
   \psi_{2}( \vec{y} ) $ in layer 2 will create a pair of fractional charges $ \pm e^{*}_{2} $ at
   $ \vec{x} $ and $ \vec{y} $ with the charge gap $ \hbar \omega^{*}_{c2} $.
   Only when the two layers are
   identical $ N_{1}= N_{2} $, $ e^{*}_{1}= e^{*}_{2} = -1/2 $ which carries fractional charge of even denominators.
   In general, the imbalanced bilayer system supports continuously changing total fractional charges
   in the thermodynamic limit.

    The above arguments give the correct {\em total} fractional charges $ \pm e^{*}_{1}, \pm e^{*}_{2} $.
    However, it can not determine the {\em relative} charge  distributions between the two layers.
    At mean field level, the energies of all the possible relative charge differences are degenerate.
    The lowest energy configuration can only be determined by fluctuations.
    The above arguments are at most intuitive. The much more rigorous and elegant topological arguments
    can only be given in the composite boson theory of the next section.

{\sl (4) Fluctuations:}

   When considering fluctuations around the MCF mean field theory,
   it is convenient to go to Lagrangian \cite{shankar}:
\begin{eqnarray}
   {\cal L} & = & \psi^{\dagger}_{\alpha} ( \partial_{\tau} -i a^{\alpha}_{0} ) \psi_{\alpha} +
           \psi^{\dagger}_{\alpha} \frac{ (-i \vec{\nabla} + \vec{A}^{\alpha}
             - \vec{ a }^{\alpha})^{2} } {2 m^{*}_{\alpha} }  \psi_{\alpha}   \nonumber  \\
         & + & i a^{\alpha}_{0} n_{\alpha} +
      \frac{i q}{ 2 \pi } (\sigma_{1})_{\alpha \beta} a^{\alpha}_{0} {a}^{\beta}_{t}
             \nonumber   \\
         & + &   \frac{ q }{ 4 \pi } ( a^{1}_{t} a^{1}_{t} +  a^{2}_{t} a^{2}_{t}
               + 2 e^{-q d} a^{1}_{t} a^{2}_{t} )
\end{eqnarray}
   where the constraints have been used to rewrite the Coulomb interactions and
   $ a^{\alpha}_{t} $ is the transverse spatial component of gauge field in Coulomb gauge
   $ \nabla \cdot \vec{a}_{\alpha} =0 $ \cite{hlr,ye1,ye2}
  ( In Lorenz invariant gauge the second to the last term becomes the mutual C-S term \cite{quantum}
  $  \frac{i}{ 4 \pi} (\sigma_{1})_{\alpha \beta} \epsilon_{\mu \nu \lambda}  a^{\alpha}_{\mu}
     \partial_{\nu} a^{\beta}_{\lambda} $ where $ \alpha, \beta $ refer to
    layer indices, while $\mu, \nu, \lambda $ refer to space-time indices ).
  The Hamiltonian has local $ U(1)_{1} \times U(1)_{2} $ gauge symmetry
  which corresponds to the invariance under
  $ \psi_{\alpha}(\vec{x}) \rightarrow e^{i \theta_{\alpha}( \vec{x} ) }
  \psi_{\alpha}(\vec{x}),
  a^{\alpha}_{\mu} \rightarrow a^{\alpha}_{\mu} + \partial_{\mu} \theta_{\alpha} $.

  Integrating out MCF $ \psi_{1}, \psi_{2} $ to one-loop and carefully expanding the interlayer Coulomb
  interaction to the necessary order in the long-wavelength limit leads to:
\begin{eqnarray}
  {\cal L} & = & \frac{ i q}{ 2 \pi}  a^{+}_{0} a^{+}_{t} +  \frac{q}{ 4 \pi } (  a^{+}_{t} )^{2}
                         \nonumber  \\
          & + &   \frac{ \epsilon_{+} }{4} q^{2} ( a^{+}_{0} )^{2}
         + \frac{1}{4}  ( \epsilon_{+} \omega^{2} + ( \chi_{+} -\frac{ d }{ 2 \pi} ) q^{2} ) ( a^{+}_{t} )^{2}
                        \nonumber  \\
         & +  & \frac{ \epsilon_{+} }{4}  q^{2} ( a^{-}_{0} )^{2}
         + \frac{1}{4} ( \epsilon_{+} \omega^{2} +
               ( \chi_{+} + \frac{ d}{ 2 \pi} ) q^{2} ) ( a^{-}_{t} )^{2}
                            \nonumber   \\
         &  +  &   \frac{ \epsilon_{-} }{2} q^{2}  a^{+}_{0} a^{-}_{0}
         + \frac{1}{2} ( \epsilon_{-} \omega^{2} +  \chi_{-} q^{2} ) a^{+}_{t} a^{-}_{t} + \cdots
\label{ngm}
\end{eqnarray}
  where $ \cdots $ are higher gradient terms and
   $ a^{\alpha}_{t} $ is the transverse component of gauge field in Coulomb gauge
   $ \nabla \cdot \vec{a}_{\alpha} =0 $\cite{hlr,ye1,ye2}, $ a^{\pm}_{\mu}=a^{1}_{\mu} \pm a^{2}_{\mu} $ and
  $ \epsilon_{\pm}= \frac{1}{2} (\epsilon_{1} \pm \epsilon_{2})
    ,\chi_{\pm}= \frac{1}{2} (\chi_{1} \pm \chi_{2} ) $.
  $ a^{+} $ ( $ a^{-} $ ) stands for the total ( relative ) density
  fluctuation.  $ a^{-}_{\mu} $ is the NGM identified previously \cite{fer,wen,jap}.
  The dielectric constants $ \epsilon_{\alpha} = \frac{ m^{*}_{\alpha} }{ 2 \pi B_{\alpha} } $ and the
  susceptibilities $ \chi_{\alpha}= \frac{1}{ 2 \pi m^{*}_{\alpha} } $ were calculated in single
  layer system in \cite{frad}.

   The first two terms are C-S term and Coulomb
    interaction term for $ + $ gauge field which take
    exactly the same forms as in a single layer system \cite{hlr}.
    The third and the fourth terms are
    non-relativistic Maxwell terms for $ + $ and $ - $ modes respectively.
    The last two terms couple $ + $ mode to $ - $ mode. Integrating out
    $ + $ modes leads to $ \epsilon^{2}_{-} q^{3} ( a^{-}_{0} )^{2} +  \epsilon_{-}
    ( \epsilon_{-} \omega^{2} + \chi_{-} q^{2} ) ( i q a^{-}_{0} a^{-}_{t} ) $ which
    are subleading to the Maxwell term of $ a^{-}_{0}, a^{-}_{t} $.
    In fact, these terms break Time reversal and Parity, in principle,
    a C-S term $   i q a^{-}_{0} a^{-}_{t} $ will be generated under RG sense.
   However, the coefficient of this generated C-S term could be so small
   that it can be neglected except at experimentally unattainable low temperatures.




{\sl (5) Neutral Gapless modes:}

   For simplicity, we only consider the balanced case and will comment
   on im-balanced case later. Note that $ \nabla \times \vec{a}^{-} = 2 \pi \delta \rho $ where $ \delta \rho=
    \delta \rho_{2} (\vec{x}) - \delta \rho_{1} (\vec{x} )  $
    is the relative density fluctuation of the two layers.
    Introducing a variable $ \phi $ which
    is conjugate to $ \delta \rho( \vec{x} ) $, namely $ [ \phi(\vec{x}), \delta \rho (\vec{x}^{\prime}) ]
    =i \delta( \vec{x}- \vec{x}^{\prime} ) $,
    we can write a spin-wave Hamiltonian density:
\begin{equation}
  {\cal H} = \frac{1}{2} \chi^{-1}_{s} ( \delta \rho )^{2} + \frac{1}{2} \rho_{s} ( \nabla \phi )^{2}
\label{h}
\end{equation}

   If $ \delta \rho $ is treated as a continuous variable,
   then $ \phi $ is a free field varying from
   $ -\infty $ to $ \infty $. By integrating out $ \delta \rho $,
   we get the $ \phi $ representation:
\begin{equation}
  {\cal L}_{\phi} = \frac{1}{2} \chi_{s} ( \partial_{\tau}{\phi} )^{2} + \frac{1}{2} \rho_{s} ( \nabla \phi )^{2}
\label{p}
\end{equation}

   Integrating out $ \phi $,
   we get an effective action density in the $ \delta \rho $ representation which is dual
   to the above $ \phi $ representation:
\begin{equation}
  {\cal L}_{\rho}
  = \frac{1}{2} \chi^{-1}_{s} ( \delta \rho )^{2} + \frac{1}{2} \rho^{-1}_{s} ( \frac{ \omega }{ q } )^{2}
  ( \delta \rho )^{2}
\label{r}
\end{equation}

   Plugging the constraint  $ \nabla \times \vec{a}^{-} = 2 \pi \delta \rho $
   into Eqn.\ref{r}, we get:
\begin{equation}
  {\cal L}_{a} =  \frac{1}{2} ( \rho^{-1}_{s}  \omega^{2}  +  \chi^{-1}_{s} q^{2} )  ( a^{-}_{t} )^{2}
\label{gauge}
\end{equation}

    This is consistent with the well-known fact that
    a {\em pure} $ 2 +1 $ dimensional $ U(1) $ gauge field
    is dual to a $ 2+ 1 $ dimensional Gaussian model which does not have any
    topological excitations \cite{dual}.  Comparing Eqn.\ref{gauge} with Eqn.\ref{ngm}
    ( for simplicity, we take $ N_{1}= N_{2} $ ),
   we get $ \rho_{s} =  \hbar \omega^{*}_{c}/ \pi,
   \chi_{s}= [ 2 \pi^{2} ( \chi+ \frac{ d }{ 2 \pi} )]^{-1} $. So the spin stiffness scales as the cyclotron gap and
   the finite charge gap of MCF implies finite spin stiffness. In order to compare with experimental data
   in \cite{gold,hall}, we have to put back $ \hbar, c , e, \epsilon $ and find
   the spin-wave velocity:
\begin{equation}
  v^{2}= \rho_{s}/\chi_{s} = \frac{ ( \omega^{*}_{c} )^{2} }{ \pi n } +
   ( \frac{ \alpha c}{\epsilon} ) ( \frac{d}{l} ) \frac{ \omega^{*}_{c} }{ \sqrt{ 2 \pi n} }
\label{v}
\end{equation}
   where $ n $ is the total density, $ l $ is the magnetic length
   and $ \alpha \sim 1/137 $ is the fine structure constant.
   Note that the correct expansion of interlayer Coulomb interaction is crucial to get the second term.

  By measuring the transport properties at finite temperature, the authors in \cite{hall} found
  the activation gap $ E_{A} \sim 0.4 K $.
   By setting $ E_{A}= \hbar \omega^{*}_{c} $ and plugging the experimental parameters
  $ n=5.2 \times 10^{10} cm^{-2}, d/l=1.61, \epsilon=12.6 $ into Eqn.\ref{v}, we find that
  the first term is $ 1.65 \times 10^{10}  ( cm/s )^{2} $, the second term is $ 2.54 \times 10^{12}  ( cm/s )^{2} $
  which is  two orders of magnitude bigger than the first term. Finally, we find
  $ v \sim 1.59 \times 10^{6} cm/s $ which is dominated by the second term. This value is
  in good agreement with $ v \sim 1.4 \times 10^{6} cm/s $ found in \cite{gold}.

{\sl (6) Topological excitations:}

   As discussed in the previous paragraph, at mean field theory,
   there are four kinds of gapped excitations with total fractional charges
   $ \pm e^{*}_{1} = \pm \nu_{1} $ and $ \pm e^{*}_{2} = \pm \nu_{2} $ which, for example,
   can be excited by finite temperature close to $ \hbar \omega^{*}_{c} $.
   But their relative charge
   distributions between the two layers are undetermined. In fact, all the possible
   excitations can be characterized
   by their $ ( a^{+}, a^{-} ) $ charges $ ( q_{+}, q_{-} ) $. For example,
   the four kinds of excitations are denoted by
   $ ( \pm \nu_{1}, q_{-} ), ( \pm \nu_{2}, q_{-} ) $.
   For $ \nu_{1}=\nu_{2} $, they reduce to two sets: $ ( \pm \frac{1}{2}, q_{-} ) $.
   If $ \delta \rho $ in Eqn.\ref{h} is treated as a discrete variable,
   then $ 0 < \phi < 2 \pi $ is an angle variable.
   $ q_{-} $ must be integers $ 0, \pm 1, \cdots $.
   Exchanging $ a^{+} $ leads to $ 1/r $ interaction between the four sets of beasts.
   While exchanging $ a^{-} $ leads to logarithmic interactions which lead to a bound state
   between two beasts with opposite $ q_{-} \neq 0 $.
   The energy of this bound state with length $ L $
   is $ E_{b}= \Delta_{+} + \Delta_{-} + q^{2}_{+} \frac{ e^{2} }{ L } +
    q^{2}_{-} \hbar \omega^{*}_{c} \ln L/l $
    where $ \Delta_{+} $ and $ \Delta_{-} $ are
   the core energies of QH and QP respectively.

     An important question to ask is what is the gluing conditions ( or selection
   rules ) of $ ( q_{+}, q_{-} ) $ for the realizable
   excitations ? Namely, what is spin ( $ q_{-} $ ) and charge ( $ q_{+} $ ) connection ?
   Two specific questions are: (1)
   Is there a charge neutral vortex excitation with $ (0, \pm 1 ) $ ?
   Being charge neutral, this kind of excitation is a bosonic excitation.
   (2) Is there a charge $ \pm 1/2 $ and spin neutral excitation $ ( \pm 1/2, 0 ) $ ?
   If they do exist, then the QP and QH pair with $ q_{-} = 0 $
   have the lowest energy $ \hbar \omega^{*}_{c} = \Delta_{+} + \Delta_{-} $.
   They decouple from $ a^{-} $ gauge field, therefore
   interact with each other only with $ 1/r $ interaction
   and are asymptotically free even at $ T=0 $ just like those in single layer system.
   Their charges are evenly distributed in the two layers
   ( namely carry fractional charges $ \pm 1/4 $ in each separate layer ).
   They are deconfined $ \pm 1/2 $ excitations which are completely different excitations
   from merons which are confined logarithmically.

    Unfortunately, the spin-charge connection  is far from obvious in this  MCF approach.
    So little can be said about these two possibilities.
    In the composite boson approach to be presented in the next section, both interesting
    possibilities are ruled out.


{\sl (7)  Extension to other filling factors:}

   Let's briefly discuss $ \nu_{T}=1/2 $ bilayer system. It is well known that this state is described by
  Halperin's 331 state \cite{bert}. In this state, the singular gauge transformation $
  U= e^{i\int d^{2} x \int d^{2} x^{\prime}
   U_{\alpha \beta} \rho_{\alpha} (\vec{x} ) arg(\vec{x}-\vec{x}^{\prime} )
  \rho_{\beta} ( \vec{x}^{\prime} ) }  $ where the matrix $
   U= \frac{1}{2} \left ( \begin{array}{cc}
       2 &  1   \\
    1 &  2   \\
   \end{array}   \right ) $ attached
   two intralayer flux quanta and one interlayer flux quantum to electrons to form Entangled Composite
   Fermions ( ECF ). At layer 1 ( layer 2 ), the filling factor of ECF is $ \nu^{*}_{1}= \frac{N_{1}}{N_{2} } $
    (  $ \nu^{*}_{2}= \frac{N_{2}}{N_{1} } $ ). Only when the two layers are identical $ N_{1}=N_{2} $,
    we get $ \nu^{*}_{1} =\nu^{*}_{2}=1 $ QH states on both layers, therefore $ \nu_{T}=1/2 $ system lacks
    interlayer coherence.

{\sl (8) Comparison with another version of CF approach:}

   It is instructive to compare our MCF picture developed in this section
   with the earlier pictures proposed in \cite{kim}.
   When $ N_{1} = N_{2} $, the authors in \cite{kim} attached $ \tilde{\phi} = 2 $ flux quanta of
   layer 2 to electrons in layer 1 or vice versa
   to form interlayer composite fermions so that at mean field theory, CF in each layer form a compressible
   Fermi liquid. They conjectured that $ a^{-} $ gauge field fluctuation mediates an attractive interaction
   between CF in different layers which leads to a ( likely p-wave) pairing instability.
   This pairing between CF in different layers opens an energy gap.
   But no systematic theory is developed along this
   picture. In the MCF picture studied in this section, there is a {\em charge} gap which is equal to
   the cyclotron gap even at mean field theory which is robust against any gauge field fluctuation.

{\sl (9) Interlayer tunneling:}

    The interlayer tunneling term is:
\begin{equation}
  H_{t} = t c^{\dagger}_{1}( \vec{x} ) c_{2} ( \vec{x} ) + h. c.
\label{tun}
\end{equation}
     Substituting Eqn.\ref{inter} into above equation leads to:
\begin{eqnarray}
  H_{t} & = & t \psi^{\dagger}_{1}( \vec{x} )
  e ^{ i \int d^{2} x^{\prime} [ arg(\vec{x}-\vec{x}^{\prime} ) \rho_{2} ( \vec{x}^{\prime} )
           - arg(\vec{x}^{\prime} - \vec{x} ) \rho_{1} ( \vec{x}^{\prime} ) ] } \psi_{2}( \vec{x} )
                \nonumber  \\
         & + &  h. c.
\label{tun2}
\end{eqnarray}
     This is a very awkward equation to deal with.

  The authors in \cite{wen} pointed out that the
  tunneling process of an electron from one layer to the other corresponds
  to an instanton in the $ 2+1 $ dimensional compact QED.
  They applied the results of Polyakov on instanton-
  anti-instanton plasma on $ 2+1 $ dimensional compact QED and found the effective action:
\begin{equation}
   {\cal L}= \frac{1}{g^{2}} ( \partial_{\mu} \chi )^{2} + \frac{ t }{ 2 \pi l^{2}}
          \cos \chi
\label{ins}
\end{equation}
   where $ g $ is the $ U(1) $ gauge coupling constant given in  Eqn.\ref{p}. In the original work of
   Polyakov, $ \chi $ is a non-compact field $ -\infty < \chi < \infty $. The compactness of $ \chi $ was forced
   in an ad hoc way in \cite{wen}. Note that the compactness of QED has nothing to do with the compactness
   of $ \chi $.

{\sl (10) Summary of limited success of MCF theory:}

   We use a Mutual Composite Fermion (MCF)
   picture to explain the interlayer coherent incompressible
   phase at $ d < d_{c} $. This MCF is a generalization
   of Composite Fermion (CF) in single layer QH systems to bilayer QH systems \cite{jain,hlr,frad,shankar}.
   In the mean field picture, MCF in each layer fill exactly $ \nu^{*} =1 $
   MCF Landau level. There are four kinds of gapped quasi-particles (QP) and quasi-holes (QH) with
   total fractional charges  $ \pm e^{*}_{1}= \pm \nu_{1}, \pm e^{*}_{2}= \pm \nu_{2} $
   ( For $ \nu_{1}=\nu_{2}=1/2 $, they reduce to two sets ).
   When considering the fluctuations above the $ \nu^{*}= 1 $ MCF QH states, we identify
   the NGM with linear dispersion relation $ \omega \sim v k $
   and determine $ v $ in terms of experimental measurable quantities.
   When interpreting the activation gap $ E_{A} $ found in \cite{hall} in terms
   of the cyclotron gap of the $ \nu^{*} =1 $ MCF Landau level, we calculate $ v $ and find its value is in
   good agreement with the value determined in \cite{gold}.
    Tentatively, we intend to classify all the possible excitations
    in terms of $ a^{+} $ and $ a^{-} $ charges $ ( q_{+}, q_{-} ) $.

{\sl (11)  Serious problems with MCF approach:}

   Despite the success of the MCF approach mentioned above, there are many serious
   drawbacks of this approach. We list some of them in the following.

 (a) The determination of the fractional charges is at most intuitive. A convincing determination can
   only be firmly established from the CB approach to be discussed in the next section.
   Furthermore, the MCF still carries charge 1, while the QP or QH carry charge $ \pm 1/2 $.
   An extension of Murthy-Shankar formalism \cite{shankar} in SLQH to BLQH may be needed to express
   physics in terms of these $ \pm 1/2 $ QP and QH.

 (b) It is easy to see that the spin wave dispersion in Eqn.\ref{v} remains
   linear $ \omega \sim v k $ even in the
   $ d \rightarrow 0 $ limit. This contradicts with the well established fact that in the
   $ d \rightarrow 0 $ limit, the linear dispersion relation
   will be replaced by quadratic Ferromagnetic spin-wave
   dispersion relation $ \omega \sim  k^{2} $ due
   to the enlarged $ SU(2) $ symmetry at $ d \rightarrow 0 $.
   This is because the flux attachment singular gauge transformation Eqn.\ref{sing}
   breaks $ SU(2) $ symmetry at very beginning even in the $ d \rightarrow 0 $ limit.

 (c) The broken symmetry in the ground state is not obvious
    without resorting to the $ (111) $ wavefunction. The physics of
    exciton pairing can not be captured.
    The origin of the gapless mode is not clear.

  (d)  The compactness of the angle $ \phi $ in Eqn.\ref{p} was put in by hand
    in an ad hoc way.

  (e) The spin-charge connection in $ ( q_{+}, q_{-} ) $ can not be determined.

  (f) The interlayer tunneling term can not be  derived in a straight-forward way. See section III-E.

  (g)   In the imbalanced case, there are two MCF cyclotron gaps $ \hbar \omega^{*}_{c \alpha} $
    at mean field theory. However, there is only one charge gap in the system.
    It is not known how to reconcile this discrepancy within MCF approach.

  (h)  It is not known how to push the MCF theory further to study the very interesting and
  novel Pseudo-spin density wave state to be discussed in section IV.

    In the following, we will show that the alternative CB approach not only can achieve
   all the results in this section, but also can get rid of all
   these drawbacks naturally. Most importantly, it can be used to
   address the novel state: Pseudo-spin density wave state in intermediate distances.

\section{Composite Boson approach: Excitonic superfluid state }

     Composite boson approach originated from  Girvin and Macdonald's off-diagonal
   long range order \cite{off}, formulated in terms of Chern-Simon Ginsburg- Landau theory \cite{cb}.
   It has been successfully applied to Laughlin's series $ \nu= \frac{1}{ 2s+1 } $ \cite{cb}.
   It has also been applied to BLQH \cite{jap,moon}.
   Unfortunately, it may not be applied to study Jain's series at
   $ \nu= \frac{ p}{ 2sp \pm 1 } $ with $ p \neq 1 $ and $ \nu =1/2 $ Fermi Liquid system.
   In this section, we applied CSGL theory to study both balanced and im-balanced bi-layer QH system.
   Instead of integrating out the charge degree of freedoms which was done in all the previous
   CB approach \cite{jap,yang}, we keep charge and
   spin degree of freedoms on the same footing and explicitly stress the spin and charge
   connection. We study how the imbalance affects various physical quantities such as
   spin wave velocity, the meron pair distance and energy, the critical in-plane field
   for commensurate-incommensurate transition, etc.

     We can rewrite $ H_{int} $ in Eqn.\ref{first} as:
\begin{eqnarray}
   H_{int} & = & \frac{1}{2} \delta \rho_{1} V  \delta \rho_{1} +  \frac{1}{2} \delta \rho_{2} V  \delta \rho_{2}
   + \delta \rho_{1} \tilde{ V } \delta \rho_{2}
                    \nonumber  \\
   & =  & \frac{1}{2} \delta \rho_{+} V_{+} \delta \rho_{+}
   + \frac{1}{2} \delta \rho_{-} V_{-} \delta \rho_{-}
\end{eqnarray}
     where $ V=V_{11}= V_{22}=\frac{2 \pi e^{2}}{\epsilon q },
     \tilde{V}= V_{12}= \frac{2 \pi e^{2}}{\epsilon q } e^{-qd},
     V_{\pm}= \frac{ V \pm \tilde{V} }{2} $
     and $ \delta \rho_{\pm}= \delta \rho_{1} \pm \delta \rho_{2} $.

     Performing a singular gauge transformation \cite{quantum,ex}:
\begin{equation}
  \phi_{a}(\vec{x}) = e ^{ i \int d^{2} x^{\prime} arg(\vec{x}-\vec{x}^{\prime} )
  \rho ( \vec{x}^{\prime} ) } c_{a}( \vec{x})
\label{singb}
\end{equation}
     where $ \rho ( \vec{x} ) = c^{\dagger}_{1}( \vec{x} ) c_{1}( \vec{x} ) +
     c^{\dagger}_{2}( \vec{x} ) c_{2}( \vec{x} )  $ is the total density of the bi-layer system.
     Note that this transformation treats both $ c_{1} $ and $ c_{2} $ on the same footing ( See Fig.3).
     This is reasonable only when the distance between the two layers is sufficiently small.

\vspace{0.25cm}

\begin{figure}
\includegraphics[width=8cm]{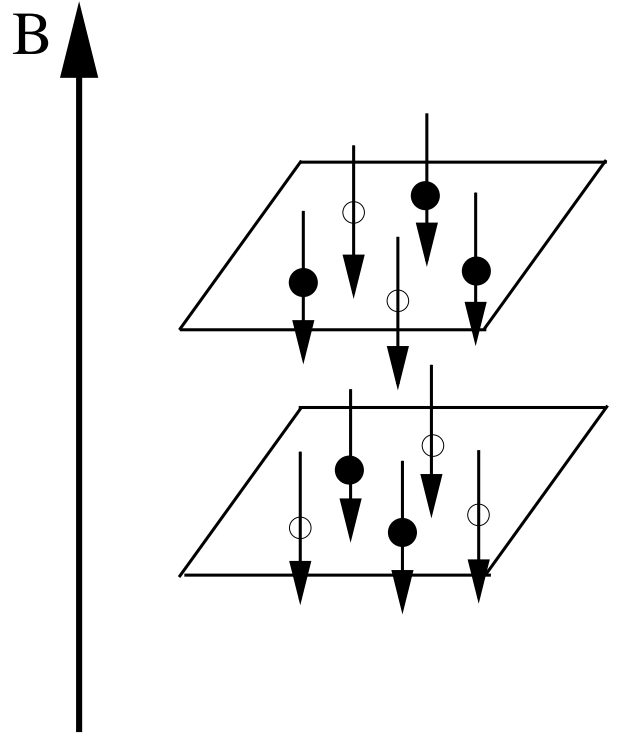}
\caption{ The flux attachment given by Eqn.\ref{singb}. There is one
flux quantum penetrating
    both layers wherever there is one electron. It does not matter this electron is in layer 1 or 2.
    One can compare this Fig. with Fig. 1a.
    For simplicity, we only show three electrons in the top layer and two electrons in the bottom layer.   }
\label{fig3}
\end{figure}

\vspace{0.25cm}

     It can be shown that $ \phi_{a}(\vec{x}) $ satisfies all the boson commutation relations.
     It can be shown that this SGT is the same as that in Eqn.\ref{sing} if we choose $ \tilde{\phi}
     =1/2 $ and replace both $ \rho_{1} $ and $ \rho_{2} $ by $ \rho= \rho_{1}+ \rho_{2} $.
     Because the SGT with $ \tilde{\phi} =1/2 $ in Eqn.\ref{sing} is not single-valued, the statistics
     is changed from fermion to boson.

     We can transform the Hamiltonian Eqn.\ref{first} into the Lagrangian in Coulomb gauge:
\begin{eqnarray}
   {\cal L} & = & \phi^{\dagger}_{a}( \partial_{\tau}- i a_{0} ) \phi_{a}
    + \phi^{\dagger}_{a}(\vec{x}) \frac{ (-i \hbar \vec{\nabla} + \frac{e}{c} \vec{A}(\vec{x})
             - \hbar \vec{ a }(\vec{x}) )^{2} }{2 m }  \phi_{a}(\vec{x})
                      \nonumber  \\
     & + & \frac{1}{2} \int d^{2} x^{\prime} \delta \rho (\vec{x} )
               V_{+} (\vec{x}-\vec{x}^{\prime} )  \delta \rho ( \vec{x^{\prime}} )
                           \nonumber  \\
    & + & \frac{1}{2} \int d^{2} x^{\prime} \delta \rho_{-} (\vec{x} )
               V_{-} (\vec{x}-\vec{x}^{\prime} )  \delta \rho_{-} ( \vec{x^{\prime}} )
    -\frac{ i }{ 2 \pi} a_{0} ( \nabla \times \vec{a} )
\label{boson}
\end{eqnarray}

   In Coulomb gauge, integrating out $ a_{0} $ leads to the constraint: $ \nabla \times \vec{a}
   = 2 \pi  \phi^{\dagger}_{a} \phi_{a}  $. Note that if setting $
   V_{-} =0 $, then the above equation is identical to a single
   layer with spin in the absence of Zeeman term, so the Lagrangian
   has a $ SU(2) $ pseudo-spin symmetry. The $ V_{-} $ term breaks
   the $ SU(2) $ symmetry into $ U(1) $ symmetry. In the BLQH $
   V_{-} > 0 $, so the system is in the Easy-plane limit.

    By taking full advantage of the easy-plane anisotropy shown in Eqn.\ref{boson},
    we can write the two bosons in terms of magnitude and phase
\begin{equation}
  \phi_{a}= \sqrt{ \bar{\rho}_{a} + \delta \rho_{a} } e^{i \theta_{a} }
\label{decom}
\end{equation}

   The boson commutation relations imply that
   $ [ \delta \rho_{a} ( \vec{x} ), \theta_{b}( \vec{x} ) ] =
     i \hbar \delta_{ab} \delta( \vec{x}-\vec{x}^{\prime} ) $.

  We can see Eqn.\ref{boson} has a local $ U(1) $ gauge symmetry $ \theta_{a} \rightarrow \theta_{a} + \chi,
  a_{\mu} \rightarrow a_{\mu} + \partial_{\mu} \chi $ and also a global $ U(1) $ symmetry
  $ \theta_{1} \rightarrow \theta_{1} + \chi, \theta_{2} \rightarrow \theta_{2} - \chi $. We denote
  the symmetry by $ U(1)_{L} \times U(1)_{G} $.

    Absorbing the external gauge potential $ \vec{A} $ into $ \vec{a} $ and
    substituting Eqn.\ref{decom} into Eqn.\ref{boson}, we get:
\begin{eqnarray}
   {\cal L} & = & \frac{1}{2} \partial_{\tau} \delta \rho^{+} +
                \frac{1}{2} \bar{\rho}_{+} i \partial_{\tau} \theta_{+}
              + \frac{i}{2} \delta \rho^{+} ( \partial_{\tau} \theta^{+}- 2 a_{0} )
                         \nonumber   \\
     & +  & \frac{ \bar{\rho}_{a} + \delta \rho_{a} }{2m}
         ( \nabla \theta_{a} - \vec{a} )^{2} +
    + \frac{1}{2} \delta \rho^{+} V_{+} (\vec{q} )  \delta \rho^{+} -\frac{ i }{ 2 \pi} a_{0} ( \nabla \times \vec{a} )
                           \nonumber  \\
    & + &  \frac{i}{2} ( \delta \rho^{-} + \bar{\rho}_{-} ) \partial_{\tau} \theta^{-}
    + \frac{1}{2} \delta \rho^{-} V_{-} (\vec{q} )  \delta \rho^{-}
\label{boson2}
\end{eqnarray}
    where $ \delta \rho^{\pm} = \delta \rho_{1} \pm  \delta
    \rho_{2}, \bar{\rho}_{\pm}= \bar{\rho}_{1} \pm \bar{\rho}_{2},
    \theta_{\pm} = \theta_{1} \pm \theta_{2} $.
    They satisfy the commutation relations
    $ [ \delta \rho_{\alpha} ( \vec{x} ), \theta_{\beta}( \vec{x}^{\prime} ) ] =
     2 i \hbar \delta_{\alpha \beta} \delta( \vec{x}-\vec{x}^{\prime} ) $ where $ \alpha, \beta= \pm $.

    It is easy to see that the symmetry is $ U(1)_{+} \times U(1)_{-} $ where $ U(1)_{+} $ is a local
    gauge symmetry, while $ U(1)_{-} $ is a global symmetry.
    We also note that it is $  \delta \rho^{-} + \bar{\rho}_{-}= ( \delta \rho_{1} - \delta \rho_{2} )
     +  (\bar{\rho}_{1}- \bar{\rho}_{2} ) = \rho_{1}- \rho_{2} $ which is conjugate to the phase $ \theta_{-} $,
   namely, $ [ \delta \rho^{-}( \vec{x} ) + \bar{\rho}_{-}, \theta_{-}( \vec{x}^{\prime} ) ]= 2 i \hbar
   \delta( \vec{x}-\vec{x}^{\prime} ) $.
     In the following, for the simplicity of notation,
     we redefine $ \delta \rho^{-} =  \rho_{1}- \rho_{2} $, then the last two terms in
      Eqn.\ref{boson2} becomes:
\begin{equation}
     \frac{i}{2} \delta \rho^{-} \partial_{\tau} \theta^{-}
    + \frac{1}{2} \delta \rho^{-} V_{-} (\vec{q} )  \delta \rho^{-} - h_{z} \delta \rho^{-}
\end{equation}
      where $ h_{z}= V_{-} \bar{\rho}_{-} = V_{-} ( \bar{\rho}_{1}-  \bar{\rho}_{2} ) $ plays a role like a Zeeman field.

     By expressing the spatial gradient term in Eqn.\ref{boson2} in terms of
     $ ( \delta \rho_{+} , \theta_{+} ) $ and $ ( \delta \rho_{-} , \theta_{-} ) $,
     we find
\begin{eqnarray}
  {\cal L}  &  =  & i \delta \rho^{+} ( \frac{1}{2} \partial_{\tau} \theta^{+}-  a_{0} ) +
          \frac{ \bar{\rho} }{2m} [ \frac{1}{2} \nabla \theta_{+} + \frac{1}{2} (\nu_{1}-\nu_{2} ) \nabla \theta_{-}
          - \vec{a} ]^{2}
                      \nonumber  \\
    & +  & \frac{1}{2} \delta \rho^{+} V_{+} (\vec{q} )  \delta \rho^{+}
          - \frac{ i }{ 2 \pi} a_{0} ( \nabla \times \vec{a} )
                       \nonumber   \\
    & + & \frac{i}{2} \delta \rho^{-}  \partial_{\tau} \theta^{-} +
          \frac{ \bar{\rho} f }{2m}  ( \frac{1}{2} \nabla \theta_{-} )^{2}
          + \frac{1}{2} \delta \rho^{-} V_{-} (\vec{q} )  \delta \rho^{-}
          - h_{z} \delta \rho^{-}
\label{main}
\end{eqnarray}
    where $ f= 4 \nu_{1} \nu_{2} $ which is equal to 1 at the balanced case.

     Plus the extra terms due to the magnitude fluctuations in the spatial gradient terms:
\begin{eqnarray}
     \frac{ \delta \rho_{a} }{2m} ( \nabla \theta_{a} - \vec{a} )^{2}
    & = & \frac{ \delta \rho^{+} }{2m} [  ( \frac{1}{2} \nabla \theta_{+} - \vec{a} )^{2}
      +  ( \frac{1}{2} \nabla \theta_{-} ) ]^{2}
             \nonumber  \\
    & + & \frac{ \delta \rho^{-} }{2m}  ( \frac{1}{2} \nabla \theta_{+} - \vec{a} ) \cdot
     \nabla \theta_{-}
\label{extra}
\end{eqnarray}

  Note that in Eqn.\ref{extra} both $ \delta \rho^{+} $ and $ \delta \rho^{-} $
  couple to terms with two spatial gradients, therefore can be dropped relative to
  the terms in Eqn.\ref{main} which contains just one temporal gradient.

   In the following, we will discuss balanced and imbalanced cases separately:

\subsection{ Balanced case $ \nu_{1}= \nu_{2}= 1/2 $ }

  Putting  $ \nu_{1}= \nu_{2}= 1/2 $ and $ h_{z}= 0 $ into Eqn.\ref{main}, we get the
  Lagrangian in the balanced case:
\begin{eqnarray}
  {\cal L}  &  =  & i \delta \rho^{+} ( \frac{1}{2} \partial_{\tau} \theta^{+}-  a_{0} ) +
          \frac{ \bar{\rho} }{2m} ( \frac{1}{2} \nabla \theta_{+} - \vec{a} )^{2}
                     \nonumber  \\
     & + & \frac{1}{2} \delta \rho^{+} V_{+} (\vec{q} )  \delta \rho^{+}
          - \frac{ i }{ 2 \pi} a_{0} ( \nabla \times \vec{a} )   \nonumber   \\
    & + & \frac{i}{2} \delta \rho^{-}  \partial_{\tau} \theta^{-} +
          \frac{ \bar{\rho} }{2m} ( \frac{1}{2} \nabla \theta_{-} )^{2}
          + \frac{1}{2} \delta \rho^{-} V_{-} (\vec{q} )  \delta \rho^{-}
\label{balance}
\end{eqnarray}

  In the balanced case, the symmetry is enlarged to $ U(1)_{L} \times U(1)_{G} \times Z_{2} $
  where the global $ Z_{2} $ symmetry is the exchange symmetry between layer 1 and layer 2.
  At temperatures much lower than the vortex excitation energy, we can neglect vortex configurations
  in Eqn.\ref{balance} and only consider the low energy spin-wave excitation. The charge sector
  ( $ \theta^{+} $ mode ) and spin sector ( $ \theta^{-} $ mode ) are essentially decoupled.

 {\sl (1) Off-diagonal algebraic order in the charge sector:}

    The charge sector is essentially the same as the CSGL action in BLQH.
    Using the constraint $ a_{t} = \frac{ 2 \pi \delta \rho^{+} }{ q } $, neglecting
   vortex  excitations in the ground state and integrating out $ \delta \rho^{+} $ leads to
    the effective action of $ \theta_{+} $:
\begin{equation}
    {\cal L}_{c} = \frac{1}{8} \theta_{+} (- \vec{q}, - \omega )
       [ \frac{ \omega^{2} + \omega^{2}_{ \vec{q} } }{ V_{+}( q) + \frac{ 4 \pi^{2} \bar{\rho} }{m}
     \frac{1}{ q^{2} } } ] \theta_{+} ( \vec{q},  \omega )
\label{cy}
\end{equation}
    where $ \omega^{2}_{ \vec{q} } = \omega^{2}_{c} + \frac{ \bar{\rho} }{m} q^{2} V_{+}( q) $
     and $ \omega_{c} = \frac{ 2 \pi \bar{\rho} }{ m } $ is the cyclotron frequency.

    From  Eqn.\ref{cy}, we can find the equal time correlator of $ \theta_{+} $:
\begin{eqnarray}
     < \theta_{+}( - \vec{q} ) \theta_{+} ( \vec{q} ) > & = &
      \int^{\infty}_{-\infty}
      \frac{ d \omega }{ 2 \pi}
      < \theta_{+} (- \vec{q}, - \omega ) \theta_{+} ( \vec{q}, \omega ) >
                                 \nonumber    \\
     & =  & 2 \times \frac{2 \pi}{ q^{2} }  + O( \frac{1}{ q } )
\label{eq}
\end{eqnarray}
     which leads to the algebraic order:
\begin{equation}
     < e^{ i ( \theta_{+}(\vec{x}) - \theta_{+}( \vec{y} ) ) } > = \frac{1}{ | x-y |^{2} }
\label{alg}
\end{equation}

   Note that if we define $ \tilde{\theta}_{+} = \frac{ \theta_{1}+ \theta_{2} }{ 2} = \theta_{+}/2 $,
   then $ < e^{ i ( \tilde{ \theta}_{+}(\vec{x}) - \tilde{\theta}_{+}( \vec{y} ) ) } > =
   \frac{1}{ | x-y |^{1/2} } $ which takes exactly the same form as that in $ \nu=1 $ SLQH.
   However, when considering vortex excitations to be discussed in the following
   $  e^{ i  \tilde{ \theta}_{+}(\vec{x}) }   $  may not be single valued, therefore the fundamental
   angle variable is $ \theta_{+} $ instead of $ \tilde{ \theta}_{+} $.

{\sl (2) Spin-wave excitation: }

  While the spin sector
  has a neutral gapless mode. Integrating out $ \delta \rho^{-} $ leads to
\begin{equation}
    {\cal L}_{s} = \frac{1}{ 2 V_{-}( \vec{q} ) } ( \frac{1}{2} \partial_{\tau} \theta^{-} )^{2} +
     \frac{ \bar{\rho} }{2m} ( \frac{1}{2} \nabla \theta_{-} )^{2}
\label{xy}
\end{equation}
    where the dispersion relation of spin wave can be extracted:
\begin{equation}
     \omega^{2} = [ \frac{ \bar{\rho} }{m}  V_{-}( \vec{q} ) ] q^{2} = v^{2}(q)  q^{2}
\label{disper}
\end{equation}
     In the long wavelength limit:
\begin{equation}
  V_{-} ( \vec{q} ) = \frac{ \pi e^{2} }{\epsilon}( d - \frac{1}{2} d^{2} q + \frac{1}{6} d^{3} q^{2} + \cdots ),
         ~~~ qd \ll 1
\label{expan}
\end{equation}
   The spin wave velocity is:
\begin{equation}
   v^{2}_{0}=  v^{2}(q=0) = \frac{ \bar{\rho} }{m} \frac{ \pi e^{2} }{\epsilon}  d  = \frac{ e^{2} }{ m \epsilon}
       \sqrt{ \frac{ \pi \bar{\rho} }{2} } \frac{ d}{l}
\label{vb}
\end{equation}

  Eqn.\ref{vb} shows that the spin wave velocity should increase as $ \sqrt{ d } $ when $ d < d_{c} $.
  At $ d=0 $, $ v=0 $. This is expected, because at $ d=0 $ the $ U(1)_{G} $ symmetry is enlarged to
  $ SU(2)_{G} $, the spin wave of isotropic ferromagnet $ \omega \sim k^{2} $.
  By plugging the experimental parameters $ m \sim 0.07 m_{e} $ which is the band mass of
  $ Ga As $, $ \bar{ \rho} =5.2 \times 10^{10} cm^{-2}, d/l=1.61, \epsilon=12.6 $ into Eqn.\ref{vb},
  we find that $ v \sim 1.14 \times 10^{7} cm/s $. This value is
  about 8 times larger than the experimental value.
  Although Quantum fluctuations will renormalize down the spin stiffness from $ \rho_{bare}=\bar{\rho} $
  to $ \rho_{eff} < \bar{\rho} $, it is
  known that CSGL theory can not give precise numerical values on energy gaps even in SLQH.
  In fact, the spin stiffness  $ \rho_{s} $ which is defined as $ \frac{ \rho_{s} }{2} ( \nabla \theta_{-} )^{2} $
   in Eqn.\ref{xy} should be determined by the interlayer Coulomb interaction instead of being dependent of the band mass $ m $.

{\sl (3) Topological excitations:}

  Any topological excitations are characterized by two winding numbers
  $ \Delta \theta_{1}= 2 \pi m_{1}, \Delta \theta_{2}= 2 \pi m_{2} $, or equivalently,
  $ \Delta \theta_{+}= 2 \pi ( m_{1} + m_{2} ) = 2 \pi m_{+},
    \Delta \theta_{-}= 2 \pi ( m_{1} - m_{2} ) = 2 \pi m_{-} $.
  It is important to realize that the two fundamental angles are $ \theta_{1}, \theta_{2} $ instead of
  $ \theta_{+}, \theta_{-} $. $ m_{1}, m_{2} $ are two independent integers, while $ m_{+}, m_{-} $
  are not, because $ m_{+} - m_{-} = 2 m_{2} $ which has to be an even integer.

  There are following 4 kinds of topological excitations: $ \Delta \theta_{1} = \pm 2 \pi,
  \Delta \theta_{2} = 0 $ or $ \Delta \theta_{1} =0, \Delta \theta_{2} = \pm 2 \pi $.
  Namely $ ( m_{1}, m_{2} ) = ( \pm 1, 0 ) $ or  $ ( m_{1}, m_{2} ) = ( 0, \pm 1 ) $.
  They correspond to inserting one flux quantum in layer 1 or 2,
  in the same or opposite direction as the external magnetic field.
  Let's classify all the topological excitations in terms of $ ( q, m_{-} ) $
  where charge $ q $ is the fractional charge of the topological excitations
  in the following table.

\vspace{0.25cm}
\begin{tabular}{ |c|c|c|c|c| }
  $ ( m_1, m_2 ) $   &  $ (1,0 ) $   & $ (-1,0) $  & $ (0,1) $      &   $ ( 0, -1)  $         \\  \hline
  $ m_{-} $  &  $ 1 $        &   $ -1 $    &  $ -1 $        &  $  1 $     \\   \hline
  $ m_{+} $  &  $ 1 $        &   $ -1 $    &  $ 1 $        &  $  -1 $     \\   \hline
   $ q    $  &  $ 1/2 $        &   $ -1/2 $    &  $ 1/2 $        &  $  -1/2 $
\end{tabular}
\par
\vspace{0.25cm}
{\footnotesize  {Table 1: The fractional charge in the balanced case } }
\vspace{0.25cm}

  The fractional charges in Table 1 were determined from the constraint
  $ \nabla \times \vec{a} = 2 \pi \delta \rho $ and the finiteness
  of the energy in the charge sector:
\begin{equation}
  q= \frac{1}{ 2 \pi} \oint  \vec{a} \cdot d \vec{l}
  = \frac{1}{ 2 \pi} \times \frac{1}{2} \oint \nabla \theta^{+} \cdot d \vec{l} =
    \frac{1}{2} m_{+}
\label{chargeb}
\end{equation}

   $ m_{-} = \pm 1 $ gives the vorticities which lead to
   logarithmic interaction between the merons, while $ q= \pm 1/2 $ lead to $ 1/ r $ interaction.
   Therefore merons are confined into the following two possible pairs at low temperature.
   (1)  Charge neutral pairs: $ (\pm 1/2, \pm 1 ) $ or $ (\pm 1/2, \mp 1 ) $.
        The NGM will turn into charge neutral pairs at large wavevectors ( or short distance ).
        The pair behaves as a boson.
   (2)  Charge $ 1 $ pair $ (1/2, \pm 1 ) $ or charge $ -1 $ pair $ (- 1/2, \pm 1 ) $.
        The pair behaves as a fermion.
        They are the lowest charged excitations in BLQH and the main dissipation sources
        for the charge transports. A duality transformation ( See Eqn.\ref{dual} ) can be easily performed
   to express low energy physics in terms of the dynamics of these topological excitations.
   There is a possible
   Kosterlitz-Thouless (KT) transition above which the meron pairs are liberated into
   free meron.

  The MCF picture in the last section points to two interesting possibilities
  (1) There maybe Charge neutral bosonic excitations with $  (0, \pm 1 ) $: Note that $ m_{+} =0 $
   implies $ m_{1} = - m_{2} =m $ and $ m_{-} = 2m $. For $ m=1 $, it corresponds to
   inserting one flux quantum in layer 1 in one direction and one flux quantum
   in layer 2 in the opposite direction. So only $ (0, \pm 2 ) $ exist, while $  (0, \pm 1 ) $
   do not exist.
  (2) There may be deconfined ( or free ) $ 1/2 $ charged excitations.
  Because any excitations with non-vanishing $ m_{-} $ will be confined,
  so any deconfined excitations
  must have $ m_{-} =0 $ which implies $ m_{1} = m_{2} =m $ and $ m_{+} = 2m $.
  We find the charge $ q=  \frac{1}{2} m_{+}=m $ must be an integer. This proof rigorously
  rules out the possibility of the existence of deconfined fractional charges. We conclude
  that {\em any deconfined charge must be an integral charge}. $ m_{1}=m_{2}=1 $ corresponds to
  inserting one flux quantum through both layers which is conventional charge 1 excitation.
  Splitting the two fluxes will turn into a meron pair with the same charge.

{\sl (4) Comparison with the LLL+HF approach }

 It is constructive to compare the spin ( $ - $ ) sector  of Eqn.\ref{balance} with the EPQHF Hamiltonian
  achieved by the microscopic LLL+HF approach in \cite{moon}:
\begin{eqnarray}
   {\cal L} & = & i \frac{ \rho}{2} \vec{A}( \vec{m} ) \cdot \partial_{\tau} \vec{m}  + \beta_{m} ( m_{z} )^{2}
        - C q m_{z}(-\vec{q}) m_{z} ( \vec{q} )
                      \nonumber  \\
      & +  & \frac{ \rho_{A} }{2} ( \nabla m_{z} )^{2} + \frac{ \rho_{E} }{2}
         [ ( \nabla m_{x} )^{2} + ( \nabla m_{y} )^{2} ]
\label{fun}
\end{eqnarray}
   where $ \nabla_{\vec{m} } \times \vec{A} = \vec{m} $, $ \beta_{m} \sim d^{2} $, $ C = \frac{ e^{2} d^{2} }{ 16 \pi \epsilon} $,
   $ \rho_{A} = \frac{ e^{2} }{ 16 \pi \epsilon l } \int^{\infty}_{0} dx x^{2}
    \exp( - x^{2}/2  ) =  \frac{e^{2} }{ 16 \sqrt{2 \pi} \epsilon l} $ is determined by
    the intralayer interaction and
    $  \rho_{E}=  \frac{ e^{2} }{ 16 \pi \epsilon l } \int^{\infty}_{0} dx x^{2} \exp( - x^{2}/2- d x/l ) $
    is determined by the interlayer interaction. Note that all these numbers are achieved by assuming
    that the ground state wavefunction  is Halperin's 111 wavefunction even at finite $ d $.
    However as shown in \cite{wave,longhua}, the wavefunction at any finite $ d $ is {\em qualitatively}
    different from Halperin's (111) wavefunction which is good only at $ d=0 $. So the numbers
    calculated by the LLL+HF based on (111) wavefunctions may not even have qualitatively correct distance dependence.

    In the above equation, the first term is the Berry phase term, the second term
    is the mass ( or capacitance ) term, this term leads to the easy-plane anisotropy which
    suppresses the relative density fluctuations between the two layers,
    the third term is nonanalytic in the wave vector due to the long range nature of the Coulomb interaction,
    the fourth term is the spin stiffness term for $ m_{z} $
    and the fifth term is the spin stiffness term for easy-plane.
    At $ d=0 $ , $ \beta_{m} = C =0, \rho_{A}= \rho_{E} = \frac{e^{2} }{ 16 \sqrt{2 \pi} \epsilon l} $,
    then $ {\cal L} $ in Eqn.\ref{fun} reduces to the $ SU(2) $ symmetric QH ferromagnet as it should be.
  Note that the value of $ \rho_{A}= \rho_{E} $  at $ d=0 $ is exact, because
  the ground state wavefunction is exactly the Halperin $ (111) $ wavefunction, while at any finite $ d $,
  the estimates of $ \rho_{A} \neq \rho_{E} $ by HF approximation may be crude,
  because we still do not know what is the exact
  groundstate wavefunction which may
  be qualitatively different from the $ (111) $ wavefunction \cite{moon,wave}.
  In the presence of the easy-plane ( $ \beta_{m} $ ) term,  $ C $ and $ \rho_{A} $ terms are subleading, therefore,
  can be dropped in the long wavelength limit. However, they are
  still very important if there is a instability happening at finite
  wavevector as shown in section III-B.

 If taking the symmetry breaking direction to be along the $ \hat{x} $ direction, we can write
 $ m_{x}= \sqrt{ 1-m^{2}_{z} } \cos \theta, m_{y}= \sqrt{ 1-m^{2}_{z} } \sin \theta, m_{z} $ with
 $ m_{z} \sim 0, \theta \sim 0 $. Substituting the parameterizations into the Berry phase term in Eqn.\ref{fun},
 we find the Berry phase term to be $  i \frac{ \bar{\rho} }{4} m_{z} \partial_{\tau} \theta $, if identifying
 $ m_{z} = 2 \delta \rho_{-} $, it is the same as the linear derivative ( $ \frac{i}{2} \delta \rho_{-}  \partial_{\tau} \theta^{-} $ ) term
 in the spin sector of Eqn.\ref{balance}. However,
 we can see the leading term in Eqn.\ref{expan} which leads to the capacitive term is $ \sim d $,  but
 in Eqn.\ref{fun}, it is $ \sim d^{2} $, while the Monte-Carlo simulation in spherical geometry in \cite{wave}
 indicates it is $ \sim d $.
 The subleading term in Eqn.\ref{expan}
 is non-analytic $ \sim q $ instead of $ \sim q^{2} $, this is due to the long-range behavior of
 the Coulomb interaction. This non-analytic term is the same as that in Eqn.\ref{fun} if identifying
 $ m_{z} = 2 \delta \rho_{-} $. The third term in Eqn.\ref{expan} leads to a $  ( \nabla m_{z} )^{2} $
 term with a coefficient $ \sim d^{3} $,  while the coefficient of $  ( \nabla m_{z} )^{2} $ term
 in Eqn.\ref{fun} is $ \rho_{A} $ which approaches the constant $ \frac{e^{2} }{ 16 \sqrt{2 \pi} \epsilon l} $
 as $ d \rightarrow 0 $ as dictated by $ SU(2) $ symmetry at $ d=0 $.
 The difficulty to recover $ SU(2) $ symmetric limit at $ d=0 $ from the CB approach is due to
 that the decomposition Eqn.\ref{decom} in
 our CB approach takes advantage of the easy-plane anisotropy from the very beginning.
  This is similar to Abelian bosonization versus Non-Abelian bosonization in one dimensional
  Luttinger liquid or multi-channel Kondo model \cite{non}. The last
  ( $ \rho_{E} $ ) term in Eqn.\ref{fun} becomes $ \frac{ \rho_{E} }{2} ( \nabla \theta )^{2} $.

 By this detailed comparison between the CB approach and the microscopic LLL+HF approach, we find that
 the two approaches lead to exactly the same functional form in the spin sector,
 some coefficients such as the  {\em Berry phase} term and the {\em non-analytic term } are the same, while some
 other coefficients such as the easy-plane term and spin stiffness term are not.  Unfortunately, it is very
 difficult to incorporate the LLL projection in the CB approach.  As suggested in \cite{moon}, we should simply
 take some coefficients in CB approach as  phenomenological values to be fitted into microscopic LLL+HF
 calculations or numerical calculations or eventually experimental data. The advantage of CB approach is that it
 also keeps the charge $ + $ sector explicitly, therefore treat the QH effects in the charge sector and the inter-layer
 phase coherence in the spin sector at the same footing. While the charge sector in the LLL+HF approach is
 completely integrated out.

   Most of the results achieved in this subsection were achieved before
  \cite{yang,moon,rev} in microscopic calculations where the charge fluctuations
  were integrated out, the LLL projections were explicitly performed
  and HF approximations were made. Here, we reproduce
  these old results in a very simple way which
  keeps both spin and charge in the same footing and bring out the spin-charge connection
  in a very transparent way. We also classify all the possible excitations in this effective CB approach.

  In the next section, we will look at the effects of imbalance.

\subsection{ Im-balanced case $ \nu_{1} \neq \nu_{2} $ }

{\sl (1) Off-diagonal algebraic order and Spin-wave excitation: }

   In the im-balanced case,
   the second term in Eqn.\ref{main} includes the coupling between spin sector
   and charge sector even when neglecting vortex excitations. Expanding this term, we
   find the effective action of the coupled $ \theta_{+} $ and $ \theta_{-} $ modes:
\begin{eqnarray}
    {\cal L}_{c} & = & \frac{1}{8} \theta_{+} (- \vec{q}, - \omega )
       [ \frac{ \omega^{2} + \omega^{2}_{ \vec{q} } }{ V_{+}( q) + \frac{ 4 \pi^{2} \bar{\rho} }{m}
     \frac{1}{ q^{2} } } ] \theta_{+} ( \vec{q},  \omega )
                       \nonumber    \\
        & + & \frac{1}{8} \theta_{-} (- \vec{q}, - \omega )
       [ \frac{ \omega^{2} } { V_{-}( \vec{q} ) } + \frac{ \bar{\rho} }{m} q^{2}  ] \theta_{-} ( \vec{q},  \omega )
                         \nonumber  \\
        & + &  \frac{ \bar{\rho} }{4m} ( \nu_{1}-\nu_{2} ) q^{2}
       \theta_{-} (- \vec{q}, - \omega ) \theta_{+} ( \vec{q},  \omega )
\label{inc}
\end{eqnarray}
     where  we safely dropped a linear derivative term in $ \theta_{-} $ in Eqn. \ref{main} in the
     Interlayer Coherent QH state. However, the linear derivative term will be shown to
     play important role in the PSDW to be discussed in section IV.B.
     From Eqn.\ref{inc}, we can identify the three propagators:
\begin{eqnarray}
      < \theta_{+} \theta_{+} >  & = &
      \frac{ ( \frac{ 4 m }{ \bar{\rho} q^{2} } ) ( \omega^{2} + v^{2} q^{2} ) \omega^{2}_{q} }{
       \omega^{4} + \omega^{2}( \omega^{2}_{q} + v^{2} q^{2} ) + f  v^{2} q^{2} \omega^{2}_{q} }
                                  \nonumber  \\
      < \theta_{-} \theta_{-} >  & = &
      \frac{ 4  V_{-}( q) ( \omega^{2} + \omega^{2}_{q} ) }{
       \omega^{4} + \omega^{2}( \omega^{2}_{q} + v^{2} q^{2} ) + f  v^{2} q^{2} \omega^{2}_{q} }
                                  \nonumber  \\
      < \theta_{+} \theta_{-} > & = &
      \frac{ - 4 ( \nu_{1} - \nu_{2} ) V_{-}( q) \omega^{2}_{q} } {
       \omega^{4} + \omega^{2}( \omega^{2}_{q} + v^{2} q^{2} ) + f  v^{2} q^{2} \omega^{2}_{q} }
\label{threepro}
\end{eqnarray}
     where $ f=4 \nu_1 \nu_2 \leq 1, \omega^{2}_{ \vec{q} } = \omega^{2}_{c} + \frac{ \bar{\rho} }{m} q^{2} V_{+}( q) $,
     the cyclotron frequency  $ \omega_{c} = \frac{ 2 \pi \bar{\rho} }{ m } $ and
     the spin wave velocity in the balanced case  $ v^{2}= v^{2}(q)= \frac{ \bar{\rho} }{m}  V_{-}( \vec{q} ) $
     were defined in Eqn.\ref{disper}.

    Performing the frequency integral of the first equation in Eqn.\ref{threepro} carefully, we find
    the  {\em leading term} of the equal time correlator of $ \theta_{+} $ stays the same
    as the balanced case Eqn.\ref{eq}:
\begin{equation}
     < \theta_{+}( - \vec{q} ) \theta_{+} ( \vec{q} ) >  =
     2 \times \frac{2 \pi}{ q^{2} }  + O( \frac{1}{ q } )
\label{eqi}
\end{equation}
    which leads to the same algebraic order exponent $ 2 $ as in the balanced case Eqn.\ref{alg}.

     We conclude that the algebraic order in the charge sector is independent of the imbalance.
     This maybe expected, because the total filling factor $ \nu_{T}=1 $ stays the same.

   In the $ q, \omega \rightarrow 0 $ limit, we can extract the leading terms of
   the $ \theta_{-} \theta_{-} $ propagator:
\begin{equation}
      < \theta_{-}  \theta_{-} >  =
      \frac{ 4 V_{-}(q) }{ \omega^{2} + f  v^{2}_{0} q^{2} }
\label{imblead}
\end{equation}
    where $  v^{2}_{0}=v^{2}(q=0) $ and  we can identify the spin wave velocity in the im-balanced case:
\begin{equation}
     v^{2}_{im} = f v^{2}_{0} = 4 \nu_{1} \nu_{2} v^{2}_{0} = 4 \nu_{1}( 1-\nu_{1} )
     v^{2}_{0}
\label{vbi}
\end{equation}
     which shows that the spin-wave velocity attains its maximum at the balanced case and
     decreases parabolically as the im-balance increases. The corresponding KT transition temperature
   $ T_{KT} $ also decreases parabolically as the im-balance
   increases.

   Similarly, in the $ q, \omega \rightarrow 0 $ limit, we can extract the leading terms in
   the $ \theta_{+} \theta_{-} $ propagator:
\begin{equation}
      < \theta_{+}  \theta_{-} >  =
      \frac{ - 4 V_{-}(q) ( \nu_{1} - \nu_{2} ) }{ \omega^{2} + f  v^{2} q^{2} } = -( \nu_{1}-\nu_{2} )
      < \theta_{-}  \theta_{-} >
\end{equation}
    which shows that  the behavior of $ < \theta_{+}  \theta_{-} > $ is dictated by that of
     $  < \theta_{-}  \theta_{-} >  $ instead of  $  < \theta_{+}  \theta_{+} >  $.

    When the vortex excitations  to be discussed in the following are included,
    the spin wave velocity will be renormalized down.
    As analyzed in detail in the last subsection,
    it is hard to incorporate the Lowest Landau Level (LLL) projection in the CB approach,
    so the spin wave velocity can only taken as a phenomenological parameter to be fitted to
    the microscopic LLL calculations or experiments, its precise
    dependence on imbalance can only be determined by experiments.

{\sl (2) Topological excitations}

  There are following 4 kinds of topological excitations: $ \Delta \theta_{1} = \pm 2 \pi,
  \Delta \theta_{2} = 0 $ or $ \Delta \theta_{1} =0, \Delta \theta_{2} = \pm 2 \pi $.
  Namely $ ( m_{1}, m_{2} ) = ( \pm 1, 0 ) $ or  $ ( m_{1}, m_{2} ) = ( 0, \pm 1 ) $.
  We can classify all the possible topological excitations in terms of $ ( q, m_{-} ) $ in the
  following table.

\vspace{0.25cm}

\begin{tabular}{ |c|c|c|c|c| }
  $ ( m_1, m_2 ) $    &  $ (1,0 ) $   & $ (-1,0) $  & $ (0,1) $      &   $ ( 0, -1)  $         \\  \hline
  $ m_{-} $  &  $ 1 $        &   $ -1 $    &  $ -1 $        &  $  1 $     \\   \hline
  $ m_{+} $  &  $ 1 $        &   $ -1 $    &  $ 1 $        &  $  -1 $     \\   \hline
   $ q    $  &  $ \nu_{1} $        &   $ -\nu_{1} $    &  $ \nu_{2} $        &  $  -\nu_{2} $
\end{tabular}
\par
\vspace{0.25cm}
{\footnotesize  {Table 2: The fractional charge in im-balanced case } }
\vspace{0.25cm}

  The fractional charges in table 2 were determined from the constraint
  $ \nabla \times \vec{a} = 2 \pi \delta \rho $ and the finiteness
  of the energy in the charge sector:
\begin{eqnarray}
  q &= & \frac{1}{ 2 \pi} \oint  \vec{a} \cdot d \vec{l}
  = \frac{1}{ 2 \pi} \times \frac{1}{2} \oint [ \nabla \theta^{+} + ( \nu_{1} -\nu_{2} ) \nabla \theta^{-} ]
   \cdot d \vec{l}
    \nonumber  \\
  & = &  \frac{1}{2} [ m_{+} + ( \nu_{1}- \nu_{2} ) m_{-} ]
\label{chargeimb}
\end{eqnarray}

    In contrast to the balanced case where $ q $ only depends on $ m_{+} $, $ q $ depends on
   $ m_{+}, m_{-} $ and the filling factors $ \nu_{1}, \nu_{2} $.
  Just like in balance case, any deconfined excitations
  with $ ( m_{-} =0 , m_{+} = 2m ) $  have charges $ q=  \frac{1}{2} m_{+}=m $ which
  must be integers. While the charge of  $ ( m_{+}=0, m_{-}=2m  ) $ is $ q= ( \nu_{1}- \nu_{2} ) m $
  which is charge neutral only at the balanced case.

    The merons listed in table 2 are confined into the following two possible pairs at low temperature.
   (1)  Charge neutral pairs: $ (\pm \nu_{1}, \pm 1 ) $ or $ (\pm \nu_{2}, \mp 1 ) $. They behave as bosons.
        The NGM will turn into charge neutral pairs at large wavevectors.
   (2)  Charge $ 1 $ pair $ (\nu_{1}, 1 ) + (\nu_{2}, -1 ) $ or charge $ -1 $ pair
        $ (- \nu_{1}, -1 ) + (- \nu_{2}, 1 ) $. They behave as fermions.
        They are the lowest charged excitations in BLQH and the main dissipation sources
        for the charge transports.

        In appendix A, we will calculate the meron fractional
        charges from its trial wavefunction and find they are the
        same as those listed in table 2.

 {\sl (3) Dual action }

   A duality transformation can be easily performed
   to express low energy physics in terms of the dynamics of these topological excitations.
     Performing the duality transformation on Eqn.\ref{main} leads to the dual action in terms of
   the vortex degree of freedoms $ J^{v \pm}_{\mu} = \frac{1}{ 2 \pi} \epsilon_{\mu \nu \lambda}
    \partial_{\mu} \partial_{\nu} \theta_{\pm} = J^{v1}_{\mu} \pm J^{v2}_{\mu} $ and the corresponding
   dual gauge fields $ b^{\pm}_{\mu} $:
\begin{eqnarray}
   {\cal L}_{d} & = & - i \pi b^{+}_{\mu} \epsilon_{\mu \nu \lambda} \partial_{\nu} b^{+}_{\lambda}
   - i A^{+}_{ s \mu} \epsilon_{\mu \nu \lambda} \partial_{\nu} b^{+}_{\lambda}
    + i \pi b^{+}_{\mu} J^{v+}_{\mu}
                      \nonumber   \\
   & + & \frac{ m }{ 2 \bar{\rho} f } ( \partial_{\alpha} b^{+}_{0}
      - \partial_{0} b^{+}_{\alpha} )^{2} + \frac{1}{2} ( \nabla \times \vec{b}^{+} ) V_{+} ( \vec{q})
      ( \nabla \times \vec{b}^{+} )
                            \nonumber   \\
   & - &  i A^{-}_{ s \mu} \epsilon_{\mu \nu \lambda} \partial_{\nu} b^{-}_{\lambda}
    + i \pi b^{-}_{\mu} J^{v-}_{\mu}
     - h_{z} ( \nabla \times \vec{b}^{-} )
                      \nonumber   \\
   & + &  \frac{ m }{ 2 \bar{\rho} f } ( \partial_{\alpha} b^{-}_{0}
      - \partial_{0} b^{-}_{\alpha} )^{2} + \frac{1}{2} ( \nabla \times \vec{b}^{-} ) V_{-} ( \vec{q})
      ( \nabla \times \vec{b}^{-} )
                      \nonumber   \\
   & - & \frac{m}{ \bar{\rho} f } ( \nu_{1}-\nu_{2} ) ( \partial_{\beta} b^{-}_{0} - \partial_{0} b^{-}_{\beta} )
      ( \partial_{\beta} b^{+}_{0} - \partial_{0} b^{+}_{\beta} )
\label{dual}
\end{eqnarray}
     where $ A^{\pm}_{s \mu}= A^{1}_{ s \mu} \pm  A^{2}_{ s \mu} $ are the two source fields.
     The $ + $ sector stands for the charge sector which is
     essentially the same as SLQH. While the spin ( or $ - $ ) sector of this dual action takes similar form
     as a 3D superconductor in an external magnetic field $ h_{z} $ shown in Eqn.\ref{is}.
     The last term  which stands for the coupling between the charge and the spin sector
     can be shown to be irrelevant in the ILCQH state. The spin wave velocity in Eqn.\ref{vbi} can also be easily
     extracted from this dual action. This dual action was used to
     derive the ground state, quasi-hole and a pair of quasi-hole
     wavefunctions in \cite{longhua}.

    It is constructive to compare this dual action derived from the CB approach with the action derived
    from MCF approach Eqn.\ref{ngm}. We find the following three differences:
    (1) The  topological vortex degree of freedoms $ J^{v \pm}_{\mu} $
    are missing in Eqn.\ref{ngm}. These vortex degree of freedoms are needed to make the variable $ \phi $
     in Eqn.\ref{p} to be an angle variable $ 0 < \phi < 2 \pi $. (2) The
    $ \chi_{+}, \chi_{-} $ terms in Eqn.\ref{ngm} are extra spurious terms.
    These extra spurious terms break $ SU(2) $ symmetry even in the $ d \rightarrow 0 $ limit.
    (3)  The linear term $ - h_{z} ( \nabla \times \vec{b}^{-} ) $ is missing in Eqn.\ref{ngm}. This linear term
        is not important in the interlayer coherent excitonic superfluid state, but
        it is very important in the in-coherent excitonic solid
        state to be discussed in the following subsection. Indeed, if
    we drop all the $ \chi_{+}, \chi_{-} $ terms, use bare mass and also add
    the topological vortex currents and the linear term by hands in Eqn.\ref{ngm}, then Eqn.\ref{ngm} will be
     identical to Eqn. \ref{dual}.
     We conclude that Eqn. \ref{main} and \ref{dual} from CB approach
    are the correct and complete effective actions.

{\sl (4) Energy of a meron pair }

     We can also look at the static energy of a charge $ 1 $
     meron pair consisting a meron with charge $ \nu_{1} $
     and charge $ \nu_{2} $ separated by a distance $ R $:
\begin{equation}
     E_{mp} = E_{1c} + E_{2c} + \frac{ \nu_{1} \nu_{2} e^{2} }{ R } + 2 \pi f \rho_{s0}
     \ln \frac{ R} { R_{c} }
\label{mp}
\end{equation}
     where $ E_{1c}, E_{2c} $ are the core energies of meron 1 ( charge $ \nu_{1} $ ) and
    meron 2  ( charge $ \nu_{2} $ ) respectively, $ R_{c} $ is the core size of an isolated meron,
    $ \rho_{s0} = \frac{ \bar{\rho} }{ 8 m } $ is the spin stiffness at the balanced case.

     Minimizing $ E_{mp} $ with respect to $ R $ leads to an optimal separation:
     $ R_{o}= \frac{ e^{2} }{ 8 \pi \rho_{s0} } $ which is independent of the imbalance. Namely, the optimal separation
     of a meron pair remains the same as one tunes the imbalance.
     Plugging $ R_{o} $ into Eqn.\ref{mp} leads to the optimal energy of a meron pair:
\begin{equation}
     E_{o} = E_{1c} + E_{2c} + \nu_{1} ( 1 - \nu_{1} ) ( \frac{ e^{2} }{ R_{o} } + 8 \pi \rho_{s0}
     \ln \frac{ R_{0} } { R_{c} }  )
\label{o}
\end{equation}

   Because of logarithmic dependence on $ R_{c} $, we can neglect the $ \nu_{1} $ dependence
   in $ R_{c} $, then the second term in Eqn..\ref{o} decreases parabolically  as im-balance increases.
   Unfortunately, it is hard to know the $ \nu_{1} $ dependence of $ E_{1c}, E_{2c} $ and $ R_{c} $
   from an effective theory, so the dependence of the energy gap $ E_{o} $ on $ \nu_{1} $ is still unknown.

\subsection{ Interlayer tunneling }

    Note that  in contrast to MCF,
    the same phase factor appears in the singular gauge transformation Eqn.\ref{singb}
    for the two layers $ a=1,2 $ which get canceled exactly in $ H_{t} $:
\begin{eqnarray}
  H_{t} & =  & t \phi^{\dagger}_{1}( \vec{x} ) \phi_{2} ( \vec{x} ) + h. c.
    = 2 t  \sqrt{ \bar{\rho}_{1} \bar{\rho}_{2} } \cos( \theta_{1} - \theta_{2} )
                 \nonumber  \\
     & = &  2 t \bar{\rho} \sqrt{ \nu_{1} \nu_{2} }  \cos \theta_{-}
\end{eqnarray}
    where the exciton operator is $ c^{\dagger}_{1} c_{2} = \phi^{\dagger}_{1} \phi_{2} = \bar{\rho}
    \sqrt{\nu_{1} \nu_{2} } e^{i \theta_{-} } \sim \psi_{-} $.

    In the presence of in-plane magnetic field $ B_{||}=( B_{x}, B_{y} ) $, the effective Lagrangian is:
\begin{eqnarray}
    {\cal L}_{s}  & =  & \frac{1}{ 2 V_{-}( \vec{q} ) } (
     \frac{1}{2} \partial_{\tau} \theta^{-} + i h_{z} )^{2} +
     \frac{ \bar{\rho} }{2m}  \nu_{1} \nu_{2} ( \nabla \theta_{-} )^{2}
             \nonumber   \\
     & + & 2 t \bar{\rho}  \sqrt{ \nu_{1} \nu_{2} }  \cos ( \theta_{-} - Q_{\alpha} x_{\alpha} )
\label{field}
\end{eqnarray}
     where $ x_{\alpha}=( x, y ), Q_{\alpha} = ( - \frac{ 2 \pi d B_{y} }{ \phi_{0} },
     \frac{ 2 \pi d B_{x} }{ \phi_{0} } )  $.

   In balanced case, it was found that when the applied in-plane magnetic field
   is larger than a critical field $ B > B^{*}_{||}  \sim ( t_{0}/ \rho_{s0} )^{1/2}  $,
   there is a phase transition from a commensurate state to an incommensurate state
   with broken translational symmetry. When $ B > B^{*}_{||} $, there is a finite temperature KT transition
   which restores the translation symmetry by means of dislocations
   in the domain wall structure in the incommensurate phase.

   As can be seen from Eqn.\ref{field}, $  \rho_{s} \sim   \nu_{1} \nu_{2} $, while $ t \sim \sqrt{\nu_{1} \nu_{2}} $,
   so the critical field $ B^{*}_{||} \sim ( t/ \rho_{s} )^{1/2} \sim  ( \nu_{1} \nu_{2} )^{-1/4} $
   increases as one tunes the im-balance.

\subsection{ Summary }

    In summary of this section, the effective CB theory can be used to lead to
  correct algebraic off-diagonal long range order, low energy functional
  forms, classification of both spin wave and topological
  excitations. But it may not be used to get correct values of many
  physical measurable quantities such as spin wave velocity, its
  precise dependence on the imbalance and the energy of a meron pair.
  The EPQFM approach is a microscopic one which takes care of LLL projection
   from the very beginning. However, the charge sector was explicitly projected out,
   the connection and coupling between the charge sector
   which displays Fractional Quantum Hall effect and the spin sector
   which displays interlayer phase coherence was not obvious in this approach.
   While in the CB theory, it is hard to incorporate the LLL projection
   ( see however the attempt made in \cite{shankar} ),
   some parameters can only be taken as phenomenological parameters to be fitted into
   the microscopic LLL+HF calculations or experimental data as discussed in section III.
    The two approaches are complimentary to each other.
  The biggest advantage of the CB approach is that it can be
  extended to capture competing orders even at microscopic length
  scales, so can also be used to
  describe novel phases and phase transitions as the layer distance
  changes to be discussed in the next section.

\section{Composite Boson approach: Pseudo-spin density wave State }

   The discussions in the last section are in the homogeneous exciton superfluid state,
   now we will study the instability of this ESF state as the distance is increased.
   As stated in the introduction,
   when the distance is smaller than $ d_{c1} $, the system is in the ESF state, while it is
   sufficiently large, the system becomes two weakly coupled $ \nu =1/2 $ Composite Fermion Fermi Liquid (FL) state.
   There could be a direct first order transition between the two states.
   However, the experimental observations
   that both zero voltage tunneling peak \cite{gold} and the Hall drag resistivity \cite{hall}
   develop very gradually when $ d \sim d_{c1} $
   suggest the transition at $ d= d_{c1} $ is a 2nd order phase transition.
   Although there are very little dissipations
   in both the ESF and FL, the experiment \cite{drag} discovered strong enhancement of drag and dissipations
   in an intermediate distance regime. These experimental observations
   suggest that there must be intermediate phases between the two phases.
   In this section, we will study the nature of the intermediate phase.

  In this section, by combining the composite boson method developed
  in the last section with the Ginzburg-Landau theory developed in \cite{qgl},
  we explore the similarity between the $ ^{4}He $ system and the
  BLQH step by step.
  In the dual {\em density} representation of Eqn.\ref{boson2},
  we propose a Pseudo-spin density wave ( PSDW ) state which breaks translational symmetry as the candidate
  of the intermediate phase.
  We also construct a novel effective Ginzburg-Landau theory to
  study the novel transition from the ESF to the PSDW.
  Inspired by recent possible discovery of supersolid in  $ ^{4}He $ \cite{chan,science,qgl},
  we also discuss important physical consequences of quantum
  fluctuation generated vacancies in the PSDW state.
  Just like in the last section, we study the balanced case first, then the effects of imbalance.

\subsection{ Balanced case }

{\sl (1) Collapsing of magneto-roton minimum }

    As shown in III-A-4,  the spin stiffness $ \frac{ \bar{\rho}}{2m} $  and the $ V_{-}(q) $ in
    Eqn.\ref{xy} should be replaced by the effective ones calculated by the LLL+HF
    approximation: $ \rho_{E} $ and $ V_{E}(q) $.
\begin{equation}
    {\cal L}_{s} = \frac{1}{ 2 V_{E}( \vec{q} ) } ( \frac{1}{2} \partial_{\tau} \theta^{-} )^{2} +
     \rho_{E} ( \frac{1}{2} \nabla \theta_{-} )^{2}
\label{xye}
\end{equation}
    where $ V_{E}(q)  =  a-b q+ c q^{2}, a , b, c > 0 $,
    $ a \sim d^{2}, b \sim d^{2} $, but $ c  $ remains
    a constant at small distances \cite{sfhe}. As explained in the last subsection,
    the non-analytic term is due to the long-range Coulomb interaction
    which survives the LLL projection.

   In order to study the instability of the ESF to the formation of
   PSDW, one need to write the dispersion relation Eqn.\ref{disper}
   to include higher orders of momentum:
\begin{equation}
     \omega^{2} = [2 \rho_{E}  V_{E}( \vec{q} ) ] q^{2} = q^{2}( a-b q+ c q^{2} )
\label{disperhigh}
\end{equation}

    As shown in \cite{qgl}, the advantage to extend the dispersion relation beyond the leading
    order is that the QGL action can even capture possible phase
    transitions between competing orders due to competing interactions on microscopic length scales.
    As explained in the last section, in the LLL-HF approach in
    \cite{moon}, these coefficients were found to be $ a \sim d^{2}, b \sim d^{2} $, but $ c  $ remains
    a constant at small distances. By looking at the two conditions $ \omega^{2}|_{q=q_{0}}=0 $
    and $ \frac{ d \omega^{2} }{ d q }|_{q=q_{0}}=0 $, it can be shown that the dispersion curve
    Eqn.\ref{disperhigh} indeed has the shape shown in Fig.5a.  When
    $ b \sim d^{2} < b_{c} = 2 \sqrt{ac} \sim d $, the magneto-roton minimum has a gap at
    $ q= q_{0}=\sqrt{a/c} \sim d $,
    the system in the ESF state, this is always the case when the distance $ d $ is
    sufficiently small ( Fig.4). However, when $ b = b_{c} $, the magneto-roton minimum  collapses at $ q= q_{0}
    $ which signifies the instability of the ESF to some density wave formation as shown in \cite{qgl}.
    When $ b \sim d^{2} > b_{c} = 2 \sqrt{ac} \sim d $, the minimum
    drops to negative and is replaced by the stable density wave formation, the system  gets to the
    PSDW state, this is always the case when the distance $ d $ is
    sufficiently large ( Fig.4).

\vspace{0.25cm}

\begin{figure}
\includegraphics[width=8cm]{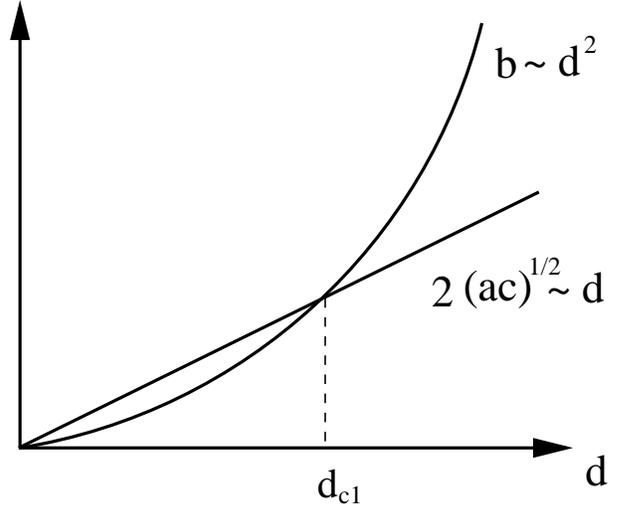}
\caption{ The critical distance $ d_{c1} $.    }
\label{fig4}
\end{figure}

\vspace{0.25cm}

    The phenomenon of the collapsing of the magento-roton minimum
    as the distance increases was also seen in the numerical
    calculations in the LLL+HF approaches \cite{jog}. It was
    estimated that $ q_{0} l \sim 1 $, so the lattice constant of
    the PSDW is of the same order of magnetic length $ l $ which is $ \sim 100 \AA $. The critical
    distance $ d_{c1} $ is also of the same order of the magnetic length.

\vspace{0.25cm}

\begin{figure}
\includegraphics[width=8cm]{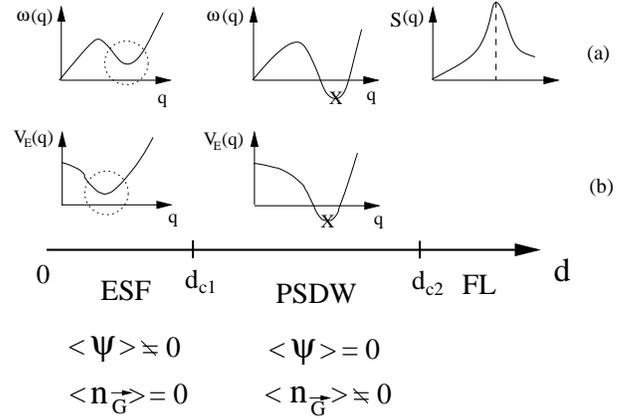}
\caption{ The zero temperature phase diagram in the balanced case as
the distance between the two layers increases. ESF where $ < \psi>
\neq 0, < n_{\vec{G} } >=0 $ stands for excitonic superfluid, PSDW
where $ < \psi> = 0, < n_{\vec{G} } > \neq 0 $ stands for
pseudo-spin density wave phase, FL stands for Fermi Liquid. (a)
Energy dispersion relation $ \omega(q) $ in these phases. (b) $
V_{E}(q) $ in these phases. The cross in the PSDW means the negative
minimum value of $ V_{E}( q ) $ is replaced by the PSDW. The order
parameters are also shown. In fact, the instability happens before
the minimum touches zero. }
\label{fig5}
\end{figure}

\vspace{0.25cm}

{\sl (2) Instability in the density channel }

    From Eqns.\ref{balance},\ref{xy},\ref{disper}, we can see that it is the original
    instability in $ V_{E}( q )= a-bq+ cq^{2} $ which leads to the magneto-roton
    minimum in the Fig.5. By looking at the two conditions $ V_{E}(\vec{q})|_{q=q_{0}}=0 $
    and $ \frac{ d V_{E}(\vec{q})}{ d q }|_{q=q_{0}}=0 $,
    it is easy to see that $ V_{E}(q) $ indeed has the shape shown in
    Fig.5b. When $ b \sim d^{2} < b_{c} = 2 \sqrt{ac} \sim d $,
    the minimum of $ V_{E}(q) $ at $ q= q_{0}=\sqrt{a/c} \sim d $ has a gap,
    the system  is in the ESF state, this is always the case when the distance $ d $ is
    sufficiently small. However, when $ b = b_{c} $, the minimum collapses  and  $ S(q) $ diverges at $ q= q_{0}
    $ ( Fig.5a), which signifies the instability of the ESF to an exciton normal solid (PSDW) formation.
    When $ b \sim d^{2} > b_{c} = 2 \sqrt{ac} \sim d $, the minimum
    drops to negative, the system gets to the PSDW state, this is always the case when the distance $ d $ is
    sufficiently large. These conditions are essentially the same as
    those achieved by looking at the dispersion relation $ \omega $ in
    Fig.5a.

{\sl (3) Effective action in the dual density representation}

    In Eqn.\ref{xy}, $ \delta \rho_{-} $ was integrated out in favor
    of the phase field $ \theta_{-} $. Because the original instability comes from the density-density interaction,
    it is important to  do the opposite: integrating out the phase field in favor of the
    density operator in the original action Eqn.\ref{balance}.
    Neglecting the vortex excitations in $ \theta_{-} $ and integrating out the $ \theta_{-} $ in
    Eqn.\ref{xye} leads to:
\begin{equation}
 {\cal L}[ \delta \rho_{-} ]=
    \frac{1}{2} \delta \rho^{-}(-\vec{q},-\omega_{n} ) [ \frac{ \omega^{2}_{n} }{ 2  \rho_{E} q^{2}}
    + V_{E} (\vec{q} ) ] \delta \rho^{-}(\vec{q},\omega_{n} )
\label{density}
\end{equation}
    where we can identify the dynamic pseudo-spin density-density correlation
    function:
\begin{equation}
    S_{-}(\vec{q},\omega_{n} )
    = < \delta \rho^{-}(-\vec{q},-\omega_{n} ) \delta
    \rho^{-}(\vec{q},\omega_{n} ) > =  \frac{ 2 \rho_{E} q^{2} }{ \omega^{2}_{n} +
   v^{2}(q) q^{2} }
\label{dd}
\end{equation}
    where $ v^{2}(q)= 2 \rho_{E} V_{E} (q) $ is the spin wave velocity defined in Eqn.\ref{disper}.

    From the pole of the dynamic density-density correlation
    function, we can identify the speed of sound
    wave which is exactly the same as the spin wave velocity.
    This should not be too surprising. As shown in liquid $ ^{4}He $,
    the speed of sound is exactly the same as the phonon
    velocity. Here, in the context of excitonic superfluid, we
    explicitly prove that the sound speed is indeed the same as the spin
    wave velocity.

    From the analytical continuation
    $ i \omega_{n} \rightarrow \omega + i \delta $ in Eqn.\ref{dd}, we can identify the dynamic
    structure factor: $ S_{-}( \vec{q},\omega ) = S_{-}(q) \delta ( \omega
    -v(q)q ) $ where $ S_{-}(q)= \rho_{E} q \pi/v(q) $ is the equal time
    pseudo-spin correlation function shown in Fig.5a. As $ q
    \rightarrow 0, S_{-}(q) \rightarrow q $.
    The {\sl Feymann relation }  in BLQH which relates the dispersion relation to the equal
    time structure factor is
\begin{equation}
     \omega (q) =  \frac{ \rho_{E} \pi  q^{2} }{  S_{-}(q) }
\end{equation}
    which takes exactly the same form as
    the Feymann relation in superfluid $ ^{4} He $.
    Obviously, the $ V_{E}(q) $ in the Fig.5b leads to the magneto-roton dispersion $ \omega^{2}= q^{2} V_{E}(q)
    $ in the Fig.5a.

    Because the instability happens near $ q=q_{0} $ instead of $ q=0 $, so the
    transition in Fig5 is {\em not} driven by vortex unbinding
    transitions like in 3D XY model, so the vortices remain tightly
    bound near the transition \cite{3dxy}. So integrating out
    the vortex excitations in $ \theta_{-} $ will only generate
    {\em weak} interactions among the pseudo-spin density $ \delta \rho_{-} $:
\begin{eqnarray}
 {\cal L}[ \delta \rho_{-} ] & = &
    \frac{1}{2} \delta \rho^{-}(-\vec{q},-\omega ) [ \frac{  \omega^{2}_{n} }{ 2 \rho_{E} q^{2}}
    + V_{E} (\vec{q} ) ] \delta \rho^{-}(\vec{q},\omega )  \nonumber  \\
    & + &  u ( \delta \rho_{-} )^{4} + w ( \delta \rho_{-} )^{6} + \cdots
\label{densityint}
\end{eqnarray}
    where the momentum and frequency conservation in the quartic and
    sixth order  term is assumed.

    In sharp contrast to the conventional classical normal liquid (NL) to
    normal solid (NS) transition \cite{tom}, the possible cubic interaction term $ ( \delta \rho_{-})^{3} $ is
    forbidden by the $ Z_{2} $ exchange symmetry between the two layers
     $ \delta \rho_{-} \rightarrow  -\delta \rho_{-} $. Note that the $  ( \omega_{n}/q )^{2} $ term in the first
    term stands for the quantum fluctuations of $ \delta \rho_{-} $ which is absent in the classical
    NL to NS transition. The density representation Eqn.\ref{densityint} is dual to the phase representation
    Eqn.\ref{xy}.
    However, the phase representation Eqn.\ref{xy} contains explicitly the superfluid order
    parameter $ \psi_{-} \sim e^{i \theta_{-} } $ which can be used
    to characterize the superfluid order in the ESF phase. All the
    results in the ESF state are achieved from phase representation
    in the last section.
    While in Eqn.\ref{densityint}, the only signature of the
    superfluid phonon mode is encoded in the density sound mode,
    because the order parameter $ \psi_{-} $ is integrated out, the superfluid
    order is hidden, so it is not as powerful as the phase representation in describing
    the ESF state.  However, as shown in the following, when describing the transition
    from the ESF to the PSDW, the density representation Eqn.\ref{densitymin} has a big advantage
    over the phase representation.

    Expanding $ V_{E}(q) $ near the minimum $ q_{0} $ in the Fig.
    3 leads Eqn.\ref{densityint} to the quantunm Ginsburg-Landau action to describe the ESF to the PSDW transition:
\begin{eqnarray}
 {\cal L}[ \delta \rho_{-} ] & = &
    \frac{1}{2} \delta \rho^{-} [ A_{\rho} \omega^{2}_{n}
    + r+ c( q^{2}-q^{2}_{0} )^{2} ] \delta \rho^{-}
        \nonumber   \\
    & + &  u ( \delta \rho_{-} )^{4} + w ( \delta \rho_{-} )^{6} + \cdots
\label{densitymin}
\end{eqnarray}
    where $  A_{\rho} \sim \frac{ 1 }{ 2 \rho_{E} q_{0}^{2}} $ which
    is non-critical across the transition.

    Because near $ q_{0} $, the coefficient $ A_{\rho} $ of $
    \omega^{2}_{n} $ in Eqn.\ref{densitymin} is a constant in the density representation,
    while the corresponding quantity $ A_{\theta} \sim S_{-}(q) $
    in the phase representation Eqn.\ref{xye} is divergent, so
    Eqn.\ref{xye} breaks down as $ q \rightarrow q^{-}_{0} $ and may
    not be used to describe the ESF to the PSDW transition.

    It is constructive to compare  the QGL Eqn.\ref{densitymin} to describe
    the ESF to the PSDW in $ 2+1 $ dimensional BLQH with that in $
    3+1 $ dimensional $ ^{4}He $ to describe the superfluid to
    normal solid transition \cite{qgl}:
\begin{eqnarray}
 {\cal L}[ \delta n ] &  = &
    \frac{1}{2} \delta n [ A_{n} \omega^{2}_{n}
    + r+ c( q^{2}-q^{2}_{0} )^{2} ] \delta n   \nonumber  \\
    & -  & w ( \delta n )^{3} + u ( \delta n )^{4} + \cdots
\label{sfdensity}
\end{eqnarray}
    where $ r \sim p_{c1}-p $ and  $  A_{\rho} \sim \frac{ 1 }{ \rho_{s} q_{0}^{2}} $ which
    is non-critical across the transition.

    The most crucial difference between Eqn.\ref{densitymin} and Eqn.\ref{sfdensity} is
    the absence of the cubic term in Eqn.\ref{densitymin} and the
    presence of the cubic term in Eqn.\ref {sfdensity}. As shown in
    the following, this crucial difference make the lowest energy
    lattice structure different in the two systems.
    Drawing the analogy of the familiar He4 phase diagram at $ T=0 $ discussed in \cite{qgl}, we
    may assume the divergence of $ S^{-}(q) $ at $ q=q_{0} $ leads to the PSDW state in the
    Fig.5. In Eqn.\ref{densitymin}, $ r $ is the gap of $ V_{E}(q) $ at the
    minimum which tunes the transition
    from the ESF to the PSDW. This is a Brazovskii type transition \cite{bra} described by a $ n=1 $
    component $ ( d+1, d) $  quantum Lifshitz action (  with $ d=2 $ in BLQH  )
    \cite{tom}. Due to the absence of the cubic term, the
    transition is 2nd order at mean field level. However, as shown by renormalization group analysis in \cite{qgl},
    the fluctuations will drive the transition into a 1st order transition.

{\sl  (4) Lattice structures of the PSDW phase }

  In the ESF, $ r > 0 $ and $ < \delta \rho_{-}> = 0  $ is uniform. In the PSDW, $ r < 0 $ and
  $ < \delta \rho_{-} > = \sum_{\vec{G}} n (\vec{G} ) e^{ i \vec{G} \cdot \vec{x} }, \ n(0)=0 $
  takes a lattice structure. In two dimension, the most common
  lattices are 1d stripe embedded in two dimension, square and triangular lattices.
  The shortest reciprocal lattice vectors of the three lattices
  were shown in Fig.5 (a) and (b) and (c) respectively.

\vspace{0.25cm}

\begin{figure}
\includegraphics[width=8cm]{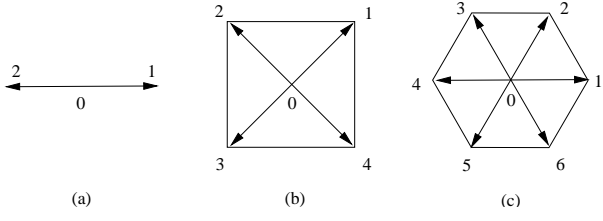}
\caption{2,4,6 shortest reciprocal lattice
  vectors in a (a) stripe (b) square lattice (c) triangular lattice  }
  \label{fig6}
\end{figure}

\vspace{0.25cm}

  (a) P=2: $ \vec{Q}_{1} = -\vec{Q}_{2}= \vec{Q} $ are a pair of anti-nodal points ( Fig.6a ).
   They are the two shortest reciprocal lattice vectors generating a 1
   dimensional lattice embedded in a 2 dimensional system.
   The 1d stripe will lead to transport anisotropy which was not
   seen in experiments, so will not be considered any more in the
   following.

  (b) P=4: $ \vec{Q}_{3} = - \vec{Q}_{1}, \vec{Q}_{4}= - \vec{Q}_{2}, \vec{Q}_{1} \cdot \vec{Q}_{2}=0 $,
   $ \vec{Q}_{i}, i=1,2,3,4 $ form the 4 corners of a square ( Fig.6b ). They are the four shortest
   reciprocal lattice vectors generating a 2 dimensional square lattice.

  (c) P=6. $ \vec{Q}_{i}, i=1,2,3,4,5,6 $ form the 6 corners of a
    hexagon ( Fig.6c ). They consist of the 6 shortest
    reciprocal lattice vectors generating a 2 dimensional triangular lattice.

  If thinking Eqn.\ref{sfdensity} as at $ 2+1 $ dimension, it can be shown that
  the triangular lattice ( Fig.6c ) is
  the favorite lattice for the normal solid. As shown in the appendix B, due to the cubic term, the
  triangular lattice is also the favorite lattice for the Wegner crystal (WC) in
  the QH to WC transition in the SLQH.
  However, due to the absence of the cubic
  term in Eqn.\ref{densitymin}, in the following, we will show that the favorite lattice is the
  square lattice. At mean field level, we can ignore the $ \omega $ dependence in Eqn.\ref{densitymin}.
 Substituting $ < \delta \rho_{-} > = \sum_{\vec{G}} n (\vec{G} ) e^{ i \vec{G} \cdot \vec{x} } $ into
 Eqn.\ref{densitymin} leads to:
\begin{eqnarray}
  f_{n} & = &  \sum_{\vec{G}} \frac{1}{2}  r_{\vec{G}} | n_{\vec{G}} |^{2}
    + u  \sum_{\vec{G}} n_{\vec{G}_1}
    n_{\vec{G}_2} n_{\vec{G}_3} n_{\vec{G}_4} \delta_{ \vec{G}_1 + \vec{G}_2 +
    \vec{G}_3 + \vec{G}_4,0 }  \nonumber   \\
    & + &  v \sum_{\vec{G}} n_{\vec{G}_1}
    n_{\vec{G}_2} n_{\vec{G}_3} n_{ \vec{G}_4 } n_{ \vec{G}_5 } n_{ \vec{G}_6 } \delta_{ \vec{G}_1 + \vec{G}_2 +
    \vec{G}_3+ \vec{G}_4 + \vec{G}_5 + \vec{G}_6, 0} + \cdots
\label{compare}
\end{eqnarray}
   where $ u $ could be positive or negative, $ v > 0 $ is always
   positive to keep the system stable.

   From Eqn.\ref{compare}, we can compare the ground state energy of the two most
   commonly seen lattices: square lattice ( Fig.6b ) and triangular lattice ( Fig.6c ).
   Square ( triangular ) lattice has $ m=4 $ ( $ m=6 $ ) shortest
   reciprocal lattice vectors $ G= 2\pi/a $ where $ a $ is the lattice constant in a given layer ( Fig.6).
   For the quadratic and quartic terms,
   all the contributions come from the paired reciprocal lattice
   vectors ( Fig.6b and 6c ). For the sixth order term, in square lattice, all the contributions are still from the
   paired reciprocal lattice vectors ( Fig.6b ), however, in triangular
   lattice, there is an {\em additional} contribution from a
   triangle where $ \vec{G}_1 + \vec{G}_2 + \vec{G}_3 =0 $ ( Fig.6c ). It is this
   {\em extra} contribution which make a crucial difference between the
   square lattice and the triangular lattice. Following \cite{tom},
   after scaling  $ n_{\vec{G}} \rightarrow m^{-1/2} n_{\vec{G}} $
   to make the quadratic term the same for both lattices, then
   Eqn.\ref{compare} is simplified to:
\begin{equation}
  f_{\alpha} = \frac{1}{2} r_{\vec{G}} | n_{\vec{G}} |^{2}+ u_{\alpha} | n_{\vec{G}}|^{4}+ v_{\alpha} | n_{\vec{G}} |^{6}
\label{simple}
\end{equation}
  where  $ u_{\Box}= u_{\triangle} =3u $ for both square and
  triangular lattice
  and $ v_{\Box}=18v, v_{\triangle}=18v+20/3 v $ for square and triangular
  lattice respectively. Obviously .
  The mean field phase diagram of Eqn.\ref{simple} is well known:
  If $ u > 0 $ ( $ u < 0 $ ), there is 2nd ( 1st ) order
  transition, there is a tri-critical point at $ u=0 $. Minimizing $
  f $ with respect to $ n_{\vec{G}} $ leads to $  n_{\vec{G}}= n_{\alpha} =
  (-2u_{\alpha} + \sqrt{ 4 u^{2}_{\alpha}-6 v_{\alpha} t } )/6v_{\alpha} $ which holds for
  both $ u_{\alpha} > 0 $ and $ u_{\alpha} < 0 $. Obviously, $
  n_{\triangle} \neq n_{\Box} $. From Eqn.\ref{simple}, we can see
  that because $ v_{\triangle} > v_{\Box} $, for any given $ n $, $ f_{\Box}(n) <  f_{\triangle}(n)
  $, then $ f_{\Box}(n_{\Box} ) <  f_{\Box}( n_{\triangle} ) < f_{\triangle}(
  n_{\triangle}) $, namely, {\sl the square lattice is the favorite
  lattice }. So it has three elastic constants instead of two.
  Neglecting zero-point quantum fluctuations, $ < \delta \rho_{-}
  > = \sum_{i} \delta( \vec{x} -\vec{R_{i}} )- \sum_{i} \delta( \vec{x} -\vec{R_{i}}-\vec{l}
  ) $ where the $ \vec{l}=\frac{1}{2}( \vec{a}_{1}+ \vec{a}_{2}  ) $
  is the shift of the square lattice in the bottom layer relative to that in the top layer ( Fig.2).

\vspace{0.25cm}

\begin{figure}
\includegraphics[width=8cm]{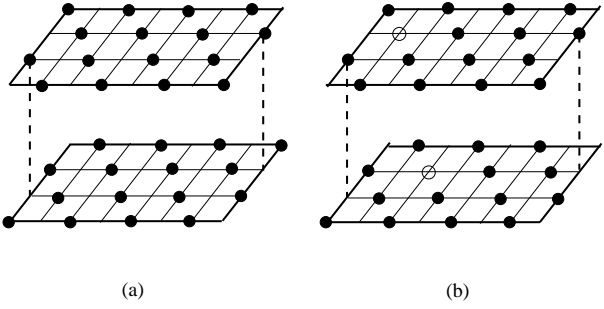}
\caption{(a) The charge distribution of
    the  PSDW in a square lattice. The "up " pseudo-spins take
    sublattice A, while the "down" pseudo-spins take sublattice B.
    (b) In general, there are quantum and thermal fluctuation generated vacancies in both layers
    denoted by a  $ \bigcirc $. }
\label{fig7}
\end{figure}

\vspace{0.25cm}

  This PSDW state not only breaks the translational
  symmetry, but also the $ Z_{2} $ exchange symmetry.
  It is very rare to get a 2d square lattice, because it is not a
  close packed lattice. Due to the special $ Z_{2} $ symmetry of
  BLQH, we show it indeed  can be realized in BLQH.
  The system is compressible with gapless phonon excitations
  determined by the 3 elastic constants.

{\sl (5)  Light scattering from the PSDW state}

    It is very interesting to see if soft X-ray or light
   scattering experiments  \cite{light} can directly test the existence of the
   PSDW when $ d_{c1} < d < d_{c2} $.
   The {\em in-plane} light scattering intensity from the PSDW is $ I( \vec{K} )
   \sim | S( \vec{K} ) |^{2} $ where  $ S( \vec{K}) $ is the
   structure factor and $ \vec{K}=\frac{2 \pi}{a}(\vec{i} + \vec{j} )  $ is the 2d in-plane
   reciprocal lattice vector of the square PSDW. For $ \vec{l}=\frac{a}{2}(\vec{i} + \vec{j}
   ) $, $ S( \vec{K} ) = 1-e^{i  \vec{K} \cdot \vec{l} } = 1-(-1)^{m+n}
   $ which is $ 0 $ and $ 2 $ for even and odd $ m+n $ respectively.
   The {\em in-plane} light scattering experiments \cite{light}
   probe the intralayer shift $ \vec{l} $ in the Fig.7, in principle, it can
   be performed in the BLQH. Note that the large zero-point quantum fluctuations may diminish the
   light scattering intensity by a Debye-Waller factor \cite{qgl}.

   The {\em out of-plane} light scattering intensity from the PSDW
   follows the Bragg  diffraction law $ 2d \sin \theta= n \lambda $ where $ \theta $ is
   the angle between the light ray and the 2d plane and the $
   \lambda $ is the wavelength of the light. It doubles or vanishes
   when $ n $ is an integer or a half integer. The {\em out of-plane} light
   scattering experiments probe the interlayer distance $ d $ in the
   Fig.7, but it is quite difficult to perform due to several $
   GaAs/AlGaAs $ layers above the 2d electron gas.

{\sl (6)  Vacancies, disorders and Coulomb Drag in the PSDW state }

  In principle, the $ \delta \rho_{+} $
  mode in Eqn.\ref{main} should also be included.
  It stands for the translational (or sliding ) motion of the PSDW lattice.
  However, any weak disorders will pin this PSDW state to make it an
  insulating state (INS). The insulating state also has a charge gap which
  has a completely different origin than that in the QH state.
  Because there are charge gaps on both sides of  the ESF/QH to
  PSDW/INS transition, it is valid to integrate out the charge $ \delta \rho_{+} $
  sector and focus on the spin sector in studying the two phases and
  the transition in Fig.5. Disorders will smear the 1st order
  transition from the ESF to the PSDW into a 2nd order transition.
  It was argued in \cite{gold} that disorders in real samples are so strong that they
  may even have destroyed the ESF state,
  so they may also transfer the long
  range lattice orders of the PSDW into short range ones. This fact makes the
  observation of the lattice structure by light scattering difficult.
  Being an insulating state, the PSDW state will not show
  any quantized Hall plateau and any zero-bias interlayer tunneling
  peak. The two square lattices in the
  top and bottom layer are locked together, so it will show huge Coulomb drag.
  However, vacancies generated by the large zero-point quantum
  fluctuations may play important roles in the drag ( Fig.2b ).
  As the distance increases to the critical
  distance $ d_{c1} $ in Fig.1, the ESF turns into the PSDW whose lattice constant $ a=\sqrt{4 \pi} l $ is
  completely fixed by the filling factor $ \nu_{1}=1/2 $, so it may not
  completely match the instability point $ 1/q_{0} $. Due to this slight mismatch, the
  resulting PSDW is likely to be an {\em in-commensurate} solid where
  the total number of sites $ N_{s} $ may not be the same as the total number of sites $
  N $ even at $ T=0 $.
  As the distance increases further $ d_{c1} < d < d_{c2} $ in Fig.1, the PSDW
  lattice constant is still {\em locked} at $ a =\sqrt{4 \pi} l $.
  Assuming zero-point quantum fluctuations favor vacancies
  over interstitials, so there are vacancies  $ n_{0} $ even at $ T=0 $ in each layer.
  At finite temperature,
  there are also thermally generated vacancies $ n_{a}(T) \sim e^{-
  \Delta_{v}/T } $ where $ \Delta_{v} $ is the vacancy excitation
  energy. So the total number of vacancies at any $ T $ is $ n_{v}(T) =n_{0}+ n_{a}(T) $.
  Obviously, the vacancies in top layer are strongly correlated with
  those in the bottom layer.
  The drag in the weakly coupled FL regime $ d>d_{c2} $ was shown to
  be very small and goes like $ T^{4/3} $ at low $ T $ \cite{momentum}. The drag in the
  ESF regime $ d < d_{c1} $ is also small and goes
  like $ e^{-\Delta_{QH}/T} $ at low $ T $. The drags in both states
  are due to momentum transfer between electrons in the two layers.
  However, in the PSDW, the holes are hopping on the square lattice,
  as shown in \cite{pump}, the microscopic origin of the drag is
  due to the correlated character of hopping transports  of the vacancies between the active and
  passive layers, it leads to a very large drag. We can estimate the
  resistance in the active layer as $ R \sim 1/n_{v}(T) $.
  As shown in \cite{pump}, the drag resistance in the passive layer is $ R_{D} \sim \alpha_{D} R $
  where $ \alpha_{D} $ should not be too small. This temperature
  dependence is indeed consistent with that found in the experiment
  \cite{drag}:  starting from $ 200 mK $, $ R_{D} $ increases exponentially until
  to  $ 50 mK $ and then saturates. This behavior is marked
  different than that at both  small and large  distance where the system is in the ESF and FL regimes respectively.
  We suggest that in the presence of disorders,
  all the properties of the PSDW are consistent with  the experimental
  observations in  \cite{gold,hall,drag} on the intermediate phase.
  The effects of very small imbalance on the PSDW
  phase will be investigated in IV-B-3 and was
  found to naturally explain the small imbalance experiment in \cite{imbexp}.
  All these facts suggest that the intermediate phase can be identified
  as the PSDW.

{\sl (7) Finite temperature phase diagram }

    Due to the spontaneous translational
  symmetry breaking in the PSDW, there are also Goldstone modes in the spin
  sector which are the two lattice phonons $
   \vec{u}(\vec{r}) $ in 2 dimension. The translational order is characterized by $ n_{\vec{Q}_{m}}(\vec{r})=
   e^{i \vec{Q}_{m} \cdot \vec{u}} $, the orientational order is characterized by
   $ \psi_{or}= e^{i 4 \phi} $ for the square lattice where $ \phi=
   \frac{1}{2} \hat{z} \cdot \nabla \times \vec{u} $
   ( in contrast, $ \psi_{or}= e^{i 6 \phi} $ for the hexagonal lattice ).
   It is known that at any finite temperature at $ d=2 $, there is no true long
   range order except possible long-range orientational order.
   The square lattice has 3 elastic constants ( due to higher symmetry, the hexagonal
   lattice only has 2 ).
   We assume the Kosterlitz-Thouless-Halperin-Nelson-Young ( KTHNY ) theory \cite{hex}
   designed for hexagonal lattice also work for the PSDW square lattice in Fig.7.
   When interlayer distance $ d=0 $, the
   symmetry is $ SU(2) $, the broken symmetry state at $ T= 0 $  in Fig. 3 is
   immediately destroyed at any
   finite $ T $ in Fig.8. The system becomes the exciton liquid (EL). When $ 0< d < d_{c1} $, the true long-range
   ordered PSDW in Fig.5 has only algebraic ODLRO in Fig.8,
   there is a finite temperature KT transition above which the system is in the EL.
   When $ d_{c1} < d < d_{c2} $, the true long-range ordered  translational order in Fig.5
   has only algebraic one, but true orientational order.
   There could also be a dislocation driven
   transition at $ T=T_{v} $ in the universality class of 2d vector Coulomb gas with the correlation length
   exponent $ \nu_{v} = 0.37 $. The system is in the Exciton Hexatic ( EH )phase with
   short range translational order and algebraic orientational
   order. There could also be a disclination driven
   transition at $ T=T_{s} $ in the universality class of  KT transition ( scalar Coulomb gas ) with the correlation length
   exponent $ \nu_{s} = 1/2 $. The system becomes the EL phase.
   The finite temperature phase diagram is shown in Fig.8.
\vspace{0.25cm}

\begin{figure}
\includegraphics[width=8cm]{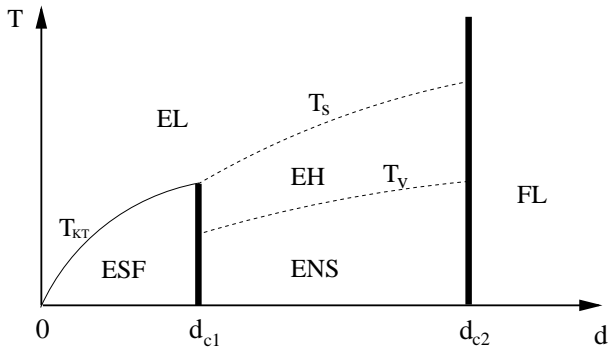}
\caption{ The Finite temperature phase diagram in the balanced case
as the distance between the two layers increases. ESF stands for
excitonic superfluid with only algebraic ODLRO. PSDW stands for
Pseudo-spin density wave with only  algebraic translational order
and true orientational order. The EH stands for the exciton hexatic
phase with only algebraic orientational order. FL stands for Fermi
Liquid. $ T_{KT} $ is the KT transition temperature driven by vortex
unbinding in the phase $ \theta_{-} $. $ T_v $ is the vector Coulomb
gas transition temperature driven by dislocation unbinding. $ T_s $
is the scalar Coulomb gas transition temperature driven by
disclination unbinding. } \label{fig8}
\end{figure}

\vspace{0.25cm}

{\sl (8) The PSDW to the FL transition }

  In Fig.5, as the distance increases to $ d_{c2} $, the PSDW will
   melt into the two weakly coupled FL.
   The flux attachment transformation Eqn.\ref{singb} which treats the
  electrons in the two layers
  on the same footing breaks down and a completely different
  flux attachment transformation within a single layer to transform an electron into a Composite Fermion
  Eqn.\ref{dec} must be used \cite{hlr}. At $ d < d_{c2} $, it is the composite boson to feel
  zero field, they either condense into the phase ordered ESF or the density ordered PSDW states, while at $ d > d_{c2} $, it is the composite fermion to feel
  zero field, they form a Fermi surface.
  Unfortunately, because completely different actions are needed in the two
  sides of the transition at $ d=d_{c2} $, this quantum phase transition can not be addressed in this
  paper and remains an interesting open question.

{\sl (9) General arguments on the existence of the PSDW state }

 When the distance between the two layers is very large, intralayer correlations dominate,
 a crystalline state is not favorable. When the distance is  sufficiently small, due to the
 strong interlayer correlation, electrons in the top layer pair with
 the holes in the bottom layer to form excitons, it is essentially a bosonic system
 in the pseudo-spin channel.  As shown in all the experiments \cite{gold,hall,drag}, the strong interlayer
 correlations go well beyond the regime of the ESF phase.
 At $ T=0 $, any bosonic system has to
 be ordered in some way, namely, breaks some sort of symmetry. If it is a homogeneous state, then it
 is phases ordered which breaks some internal symmetry such as global $ U(1) $
 symmetry,  this is the ESF phase where the kinetic energy of the
 excitons dominate. It happens at very small distance $ d < d_{c1} $.
  If it is an inhomogeneous state, then it is density
 ordered which break translational symmetry, this is the PSDW phase where the potential energy of the
 excitons dominate. This happens at some intermediate distances $ d_{c1} < d < d_{c2} $.
 There is no other possibilities.
  Some exotic state such as exciton liquid which
 does not break any symmetry is very unlikely \cite{3dxy}

   In the following, we will discuss the effects of the imbalance on
   Fig.5

\subsection{  Imbalanced case }

{\sl (1) The shape of the magnetoroton minimum }

    In order to see how the imbalance affects the shape of the
    dispersion curve shown in the Fig.5a, we need to push
    Eqn.\ref{imblead} to the order $ q^{4} $:
\begin{equation}
     \omega^{2}_{imb}= f \omega^{2} -f(1-f) \frac{ v^{4}_{0} q^{4} }{
     \omega^{2}_{c} }
\label{q4}
\end{equation}
     where $ f=4 \nu_{1} \nu_{2} $ and $ \omega^{2} $ is defined in Eqn.\ref{disperhigh}.

     Scaling out the common factor $ f $, one can see $ a $ and $ b $ stay the
     same, while $ c $ is reduced by the imbalance. Therefore $
     q_{0} $ increases, namely, the lattice constant $ a $ of the PSDW
     may decrease ( See Fig.10 ). While $ b_{c} $ decreases, namely, the critical distance $ d_{c1} $
     also decreases. Unfortunately, as warned in Sec.III.D, the CB
     approach may not describe the magnetoroton minimum shape dependence on
     imbalance correctly. In the subsection (3), we will approach
     the phase boundary from the PSDW side.

{\sl (2) Dual density representation in the ESF state }

  Just like in the balanced case, in order to focus on the {\em original} instability,
  we need to keep $ \delta \rho_{-} $. The form of Eqn.\ref{inc}
  {\em before} integrating out $ \delta \rho_{-} $ is
\begin{eqnarray}
    {\cal L}_{c} & = & \frac{1}{8} \theta_{+} (- \vec{q}, - \omega )
       [ \frac{ \omega^{2} + \omega^{2}_{ \vec{q} } }{ V_{+}( q) + \frac{ 4 \pi^{2} \bar{\rho} }{m}
     \frac{1}{ q^{2} } } ] \theta_{+} ( \vec{q},  \omega )
                          \nonumber  \\
        & + &  \frac{ \bar{\rho} }{4m} ( \nu_{1}-\nu_{2} ) q^{2}
       \theta_{-} (- \vec{q}, - \omega ) \theta_{+} ( \vec{q},  \omega )
                       \nonumber    \\
     & + & \frac{i}{2} \delta \rho^{-}  \partial_{\tau} \theta^{-} +
          \frac{ \bar{\rho} }{2m} ( \frac{1}{2} \nabla \theta_{-} )^{2}
          + \frac{1}{2} \delta \rho^{-} V_{-} (\vec{q} )  \delta \rho^{-}
          - h_{z}  \delta \rho^{-}
\label{incrhoi}
\end{eqnarray}

    Obviously, integrating out $ \delta \rho_{-} $ and dropping out
    the linear derivative term in $ \theta^{-} $, we recover
    Eqn.\ref{inc}.

    By defining $ \delta \rho_{-} = \frac{ h_{z} }{ a } + \delta
    \rho^{\prime}_{-} $, we can absorb the last linear term in
    Eqn.\ref{incrhoi} into the quadratic term. Integrating out both $
    \theta_{+} $ and $ \theta_{-} $ leads to:
\begin{equation}
 {\cal L}[ \delta \rho^{\prime}_{-} ]=
    \frac{1}{2} \delta \rho^{\prime}_{-}(-\vec{q},-\omega ) [ \frac{ m \omega^{2}}{\bar{\rho} q^{2}}
    ( \frac{ \omega^{2} + \omega^{2}_{q} }{ \omega^{2} +  f \omega^{2}_{q} } ) +
    V_{-} (\vec{q} ) ] \delta \rho^{\prime}_{-} (\vec{q},\omega )
\label{densityi}
\end{equation}
    where we can identify the dynamic pseudo-spin density-density correlation
    function $ S^{\prime} (\vec{q},\omega )= < \delta \rho^{\prime}_{-} \delta \rho^{\prime}_{-} > $ :
\begin{equation}
    < \delta \rho^{\prime}_{-} \delta \rho^{\prime}_{-} >
    = \frac{  \frac{\bar{\rho}}{m} q^{2} ( \omega^{2} +  f \omega^{2}_{q} ) }{
       \omega^{4} + \omega^{2}( \omega^{2}_{q} + v^{2} q^{2} ) + f  v^{2} q^{2} \omega^{2}_{q} }
\label{densityci}
\end{equation}
    where $ v^{2}= v^{2}(q) $ was defined in Eqn.\ref{disper}.
    Obviously $ S^{\prime} (\vec{q},\omega=0 )= \frac{1}{ V_{-}(q) }
    $.

    Again, From the pole of the dynamic density-density correlation
    function Eqn.\ref{densityci}, we can identify the speed of sound
    wave in the imbalanced case which is exactly the same as the spin wave velocity in the imbalanced case
    Eqn.\ref{q4} achieved from the phase representation.
    From both the phase representation in Sec.III and the density
    representation in this subsection, we conclude that imbalance is
    irrelevant in the ESF side, but it may play important role in
    the PSDW side to be discussed in the following.

{\sl (3) The quantum phase transitions driven by small imbalances in
         the PSDW state }

    As shown in section III, the imbalance is irrelevant in the ESF
    side, but we expect it is important in the PSDW side.
    Starting
    from the ESF side, it would be useful to calculate how the mageto-roton minimum
    depends on the imbalance, this was attempted in (1).
    As said in (1), the precise dependence of $ d_{c1} $ on the imbalance is hard to
    achieve from the CB effective theory.  Here we take a different strategy: starting
    from the PSDW side and studying how a small imbalance affects the PSDW.
    If the imbalance is sufficiently small,
    we expect the C-PSDW at the balanced case in Fig.7a is a very good reference state.
    In the following, we ignore the small number of vacancies  in the PSDW shown
    in Fig.7b. Because it is a square lattice with the "up" pseudospins
    taking sublattice $ A $ and the "down" pseudospins taking
    sublattice $ B $, it is reasonable to start from a lattice model from the PSDW
    side. If we think the PSDW as a charge density wave (CDW )  of bosons at half filling
    on a square lattice, then we can view the ESF to the PSDW as a
    superfluid to CDW transition in a boson Hubbard model of {\em hard core} bosons  {\em near } half filling
    hopping on square lattice of bosons:
\begin{eqnarray}
   H  & = & -t \sum_{ < ij > } ( b^{\dagger}_{i} b_{j} + h.c. )
          - \mu \sum_{i} n_{i}    \nonumber   \\
    & + &  V_{1} \sum_{ <ij> } ( n_{i}-1/2 )( n_{j}-1/2)  \nonumber  \\
     &  +  &   V_{2} \sum_{ <<ik>> } (n_{i}-1/2) ( n_{k}-1/2 ) + \cdots
\label{bosonla}
\end{eqnarray}
    where  $ S^{+} = b^{\dagger}=c^{\dagger}_{1} c_{2}, S^{-}= b = c^{\dagger}_{2} c_{1},
     S^{z}= \frac{1}{2}( c^{\dagger}_{1} c_{1} - c^{\dagger}_{2} c_{2} )=  b^{\dagger} b-1/2 $
     are the pseudo-spin density and boson operators.
    At the total filling factor $ \nu_{T}=1 $, we can impose the local constraint
    $ c^{\dagger}_{1} c_{1} + c^{\dagger}_{2} c_{2} =1 $.
    $ n_{i} = b^{\dagger}_{i} b_{i} $ is the boson density, $ t $ is the nearest neighbor hopping
    amplitude, $ V_{1}, V_{2} $ are the nearest and next nearest neighbor repulsive
    interactions between the bosons. The $ \cdots $ may include further neighbor interactions.
    Because of the long-rang Coulomb interaction in Eqn.\ref{main},
    it is important to keep all the long-range interactions in
    the lattice model Eqn.\ref{boson}.
    If the chemical potential $ \mu=0 $, the bosons are at the half filling $ < n_{i} >=1/2
    $ which corresponds to the balanced case $ \nu=1/2 $. The
    particle-hole symmetry of Eqn.\ref{boson} corresponds to the $ Z_{2} $ exchange
    symmetry of the BLQH.
    If the chemical potential $ \mu \neq 0 $, the bosons are slightly away from the half filling
    which corresponds to the slightly imbalanced case.

    The boson Hubbard model Eqn. \ref{bosonla} in square lattice
    at generic commensurate filling factors $ f=p/q $ ( $ p, q $ are relative prime numbers ) were
    systematically studied in \cite{pq1}
    by performing the charge-vortex duality transformation.
    Recently, we applied the dual approach to study
    extended boson Hubbard model
    Eqn.\ref{bosonla} in bipartite lattices such as square and honeycomb lattice
    which maybe directly relevant to the Helium 4 and
    Hydrogen adsorbed on various structures of substrates \cite{he,honey}.
    The key result achieved in this paper is that
    there are two consecutive transitions at zero temperature {\em driven by the coverage }: a
    Commensurate-Charge Density Wave (CDW) at half filling to a narrow window of supersolid, then to
    an Incommensurate-CDW. In the Ising limit, the supersolid is a CDW supersolid;
    whereas in the easy-plane limit, it is a  valence bond supersolid.
    The first transition is a second order one in the same universality as the Mott insulator to the superfluid
    transition, therefore has the exact critical exponents $
    z=2, \nu=1/2, \eta=0 $ ( with possible logarithmic corrections
    ). The second is a first order transition. In the following,
    we simply straightforwardly apply the results near half filling in
    the square lattice achieved in \cite{nature} to the present
    problem on square lattice near half filling. At $ q=2 $, there
    are two dual vortex fields $ \psi_{a} $ and $ \psi_{b} $.
   Moving {\em slightly} away from half filling $ f=1/2 $ corresponds to adding
   a small {\em mean} dual magnetic field $ H \sim  \delta f= f-1/2 $ in the dual action.
   The most general action invariant under all the MSG transformation laws upto quartic
   terms is \cite{pq1,nature}:
\begin{eqnarray}
    {\cal L} & = & \sum_{\alpha=a/b} | (  \partial_{\mu} - i A_{\mu} ) \psi_{\alpha} |^{2} + r | \psi_{\alpha} |^{2}
    +  \frac{1}{4} ( \epsilon_{\mu \nu \lambda} \partial_{\nu} A_{\lambda}
    - 2 \pi \delta f \delta_{\mu \tau})^{2}   \nonumber  \\
       & +  & \gamma_{0} ( | \psi_{a} |^{2} + |\psi_{b} |^{2} )^{2} -
                       \gamma_{1} ( | \psi_{a} |^{2} - |\psi_{b} |^{2} )^{2} + \cdots
\label{away}
\end{eqnarray}
     where $ A_{\mu} $ is a non-compact  $ U(1) $ gauge field. Upto
     the quartic level, Eqn.\ref{away} is the same in
     square lattice and in honeycomb lattice.
     If $ r > 0 $, the system is in the superfluid state $  < \psi_{l} > =0 $ for every $ l=a/b $.
     If $ r < 0 $, the system is in the insulating state $ < \psi_{l} > \neq 0 $ for
     at least one $ l $.  We assume $ r < 0 $ in Eqn.\ref{away}, so the system is in the insulating state.
     $ \gamma_{1} > 0 $ ( $ \gamma_{1} < 0 $ ) corresponds to the
     Ising ( or Easy-plane ) limit. The insulating state takes the CDW  state ( or valence
     bond solid (VBS) state ).

     In the balanced case $ \delta f=0 $, the SF to the VBS transition
     in the easy plane limit was argued to be 2nd order through a novel deconfined quantum
     critical point \cite{senthil}. However, the boson Hubbard model
     Eqn.\ref{boson} on the PSDW side
     corresponds to the Ising limit in the dual model Eqn.\ref{away}, therefore $ \gamma_{1} > 0 $.
     The SF to the CDW transition in the Ising limit
     is first order.
     This is consistent
     with the first order ESF to PSDW transition driven by the collapsing of magnetoroton minimum studied in
     Sec.IV-A.

    In the CDW order side, the mean field solution is $ \psi_{a}=1, \psi_{b}=0 $ or vice versa.
    In the slightly imbalance case $ \delta f \neq 0 $,
    setting $ \psi_{b}=0 $ in Eqn.\ref{away} leads to:
\begin{eqnarray}
  {\cal L} & = & | (  \partial_{\mu} - i A_{\mu} ) \psi_{a} |^{2} + r | \psi_{a}
  |^{2} + \gamma_{0} | \psi_{a} |^{4} + \cdots     \nonumber  \\
     & + &  \frac{1}{4} ( \nabla \times \vec{A} - 2 \pi \delta f )^{2}
\label{is}
\end{eqnarray}
   where $ \rho_{A} =  \psi^{\dagger}_{a} \psi_{a} $ should be interpreted as the
   vacancy number, while the vortices in its phase winding  are interpreted as  boson number.
   Of course, a negative imbalance can simply achieved by a particle
   hole transformation $ \psi_{a} \rightarrow \psi^{\dagger}_{a},
   \delta f \rightarrow - \delta f $ in Eqn.\ref{is}.

   Eqn.\ref{is} has the structure identical to the conventional $ q=1 $ component
   Ginzburg-Landau model for a  type II " superconductor "  in a "magnetic"
   field. It was well known that as the magnetic field increases, there are
   two first order phase transitions: $ H < H_{c1} $, the system is in the
   Messiner phase, $ H_{c1} < H < H_{c2} $, it is in the vortex
   lattice phase, $ H > H_{c2} $ it is in the normal phase.
   In the present boson problem with the nearest neighbor interaction $
   V_{1} > 0 $ in Eqn.\ref{bosonla} which stabilizes
   the CDW state at $ f=1/2$, this corresponds to C-CDW to IC-CDW to
   superfluid transition shown in Fig.9a. Transferring back to the
   original BLQH problem, the small imbalance will first drive the C-PSDW to
   the In-commensurate normal solid (IC-PSDW), then drive a 1st order transition from
   the IC-PSDW to the ESF shown in Fig.9b.

\vspace{0.25cm}

\begin{figure}
\includegraphics[width=8cm]{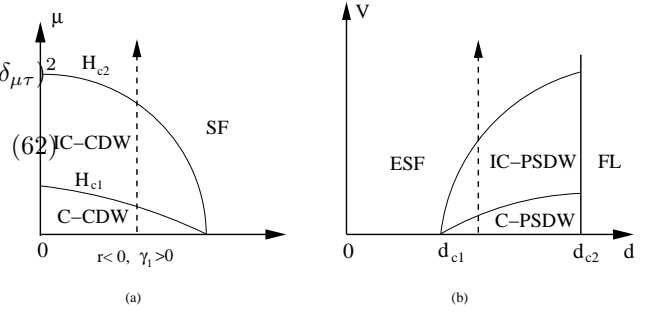}
\caption{(a) The phase diagram of the boson
      Hubbard model Eqn.\ref{boson} slightly away from the half filling.
      $ \mu $ is the chemical potential. (b) The bias voltage $  V $ versus
      distance $ d $ phase diagram at zero temperature. IC-PSDW
      stands for the incommensurate PSDW. The dashed line
      is the experimental path investigated in \cite{imbexp}. All the transitions are first order
      transitions.  }
\label{fig9}
\end{figure}

\vspace{0.25cm}

    As shown in Fig.9b, the bias voltage increases, the imbalance will first
    introduce interstitials in the top layer and vacancies in the
    bottom layer, namely, turn the C-PSDW into the IC-PSDW whose charge distributions are shown in
    Fig.7 and 10 respectively, then the whole IC-PSDW melts into the ESF through a 1st
    order transition.

\vspace{0.25cm}

\begin{figure}
\includegraphics[width=8cm]{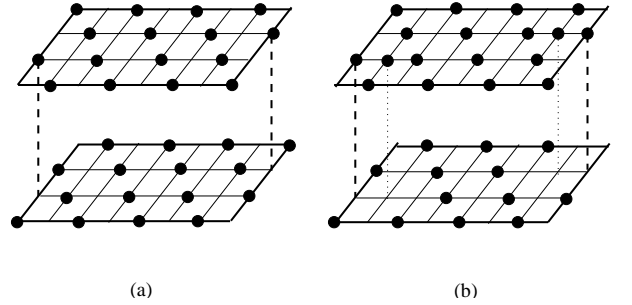}
\caption{(a) The charge distribution of the PSDW in a square
lattice. The dashed line stand for "up" pseudo-spins which take
sublattice A, while "down" pseudo-spins take sublattice B. (b) The
charge distribution of the IC-PSDW in a square lattice. The dotted
line stand for "inverted" pseudo-spins which play the role of
interstitials in the top layer or vacancies in the bottom layer. The
number of interstitials in the top layer is equal to that of
vacancies in the bottom layer. Compare with Fig.7 }
 \label{fig10}
\end{figure}

\vspace{0.25cm}

   The dashed line in Fig.9b was investigated in a recent experiment
   \cite{imbexp}. But the first phase transition in Fig.9b was not paid attention in the
   experiment where the phase diagram was drawn against fixed charge imbalance $
   h_{z} $ instead of fixed bias voltage $ V $.
   So the C-PSDW phase was crushed into the horizontal axis.
   A simple mean field argument leads to the linear scaling of the second
   transition line $ V \sim d-d_{c1} $.
   A parabolic behavior $ h^{2}_{z} \sim d-d_{c1} $ was found for the
   shape of the second transition at very small imbalances.
  Disorders will smear all the 1st order
  transitions in Fig.9b into 2nd order transitions.
  The disorders may also transfer the long
  range lattice orders of the C-PSDW and IC-PSDW into short range ones. The fact make the
  observation of the commensurate and incommensurate lattice structures by light scattering experiment \cite{light}
  difficult. We expect the disorders may transform the linear behavior $ V \sim d-d_{c1} $ in clean case to the
  parabolic one $ h^{2}_{z} \sim d-d_{c1} $ in the dirty case.
  In the presence of disorders, all the properties of the C-PSDW and IC-PSDW are consistent with all the experimental
  observations in \cite{imbexp} on the intermediate phase at small imbalance.

  When the distance of the two layers is further increased to larger than a second critical
  distance $ d_{c2} $, then all the signature of the interlayer
  coherent state are lost, the two layers are decoupled into two separate $ \nu=1/2 $ CF Fermi liquid state ( Fig.1b ).
  We expect that there is a level crossing and associated first order transition at $ d_{c2} $.
  When $ d > d_{c2 } $, increasing the bias voltage may not transform the two decoupled
  FL state back into the ESF.

{\sl (4) Large imbalance }

    At large imbalance, all the states in Fig.5 at the balanced case
    are not good reference states anymore.
    Fig. 7 is valid only for small imbalance
    such that $ \nu_{1}, \nu_{2} $ do not fall onto any fractional Quantum Hall
    plateaus for separate layers. For large imbalance such as two layers with $ \nu_{1}=1/3, \nu_{2}=2/3 $
    still with $ \nu_{T}=1 $,
    then the two weakly coupled layers show $ \nu_{1}=1/3 $ and $ \nu_{2}=2/3 $  fractional Quantum Hall states
    separately when $ d > d_{c2} $. Because there are intralayer gaps, they are more robust against the interlayer
    correlations, we expect $ d_{c2} $ for $ \nu_{1}=1/3, \nu_{2}=2/3 $
    to be {\em smaller} than $ \nu_{1}=\nu_{2} =1/2 $.   If we bring them closer, there could be a direct 1st order transition
    at $ d_{c1}= d_{c2} $ from the weakly coupled pair to the ESF, or through a narrow regime of IC-PSDW.
    The extreme imbalanced case $ \nu_{1} \ll 1, \nu_{2}=1-\nu_{1}
    \sim 1 $ where coupled electron and hole Wigner crystal (EHWC) was discussed in \cite{kun2}.
    As shown in the appendix B, we believe this EHWC should take a
    triangular lattice in contrast to the square lattice at the
    balanced case.

\subsection{ Comparison with earlier work }

     Although the ESF phase and FL phase at the two extreme distances are well established,
    the picture of how the ESF phase evolves into the two weakly-coupled FL states is still not clear, namely,
    the nature of the intermediate phase at $  d_{c1} < d < d_{c2} $ is still under debate.
    There are previous work also based on the instability due to magnetoroton minimum collapsing \cite{wigner}.
    These work proposed different kinds of translational symmetry
    breaking ground states as candidates of the intermediate state .
    All these previous work used either HF approximation or trial
    wavefunctions approximation. Especially, Cote, Brey and Macdonald
    in \cite{wigner} found that the lowest energy lattice structure of the PSDW is square
    lattice.
    The microscopic calculations are powerful and accurate in studing quantum Hall states with a bulk gap in SLQH
    or BLQH systems, but its accuracy and power drop considerably
    in the  $ \nu=1/2 $ FL system in SLQH and  the $ \nu=1 $ BLQH  system with a broken symmetry state and its associated Goldstone
    mode. The Quantum Ginsburg-Landau theory presented in this paper is complementary to and goes well beyond the
    previous microscopic calculations.
    As shown in the text, our QGL theory Eqn.\ref{densitymin} very quickly and firmly leads to
    the conclusion that the square lattice is the most favorite lattice. We also explicitly point out
    the hopping mechanism for the large Coulomb drag in the PSDW state. The two approaches are
    complementary to each other and reach similar conclusions.
    Furthermore, the QGL theory  circumvents the difficulty
    associated with the unknown wavefunction at any finite $ d $ \cite{wave} and treat both the interlayer
    and the intralayer correlations on the same footing, so
    can be used to capture competing
    orders on microscopic length scales and naturally leads to the PSDW as
    the intermediate state which breaks translational symmetry. The theory puts the ESF state and
    the PSDW state on the same footing, characterize the symmetry breaking patterns in the two
    states by corresponding order parameters and describe the universality
    class of the quantum phase transition between the two states.
    It can also be used to determine the nature, properties
    and lattice structure of the PSDW state.
    The properties of the PSDW proposed in this paper are also consistent
    with earlier experimental observations in \cite{wignerexp}.

    By the microscopic LLL+ HF approach, the authors in \cite{han,jog,exp} claimed that
    claimed that the transition at
    $ h_{z}=0, d=d_{c1} $ is an instability through a 1st order transition
    to a pseudospin density wave state driven by the gap closing of magneto-roton minimum
    at a finite wave-vector. Starting from the ESF side, their numerical results indicated that
    the imbalance increases the spin stiffness and also the critical
    distance $ d_{c1} $.
    All the calculations in LLL+HF approach assumes
    that the ground state wavefunction at any finite $ d $ is still the  (111) wavefunction.
    However as shown in \cite{wave,longhua}, the wavefunction at any finite $ d $ is {\em qualitatively}
    different from  the (111) wavefunction which is good only at $ d=0 $. So the excitation spectra
    in the ESF side calculated by the LLL+HF based on the (111) wavefunction may not have quantitatively correct
    distance dependence. The LLL+HF theory in \cite{jog} completely breaks down in the PSDW side.
    Furthermore, in this LLL+HF approach, the charge fluctuations
    are completely integrated out, therefore can not
    address the interplay between the QH effects in the charge sector and the interlayer phase
    coherence in the spin sector. The CB field theory approach in this paper circumvents this difficulty
    associated with the unknown wavefunction at any finite $ d $ (see
    Sec.V) and was used to show how the pseudo-spin density wave state is formed and its properties.
    We look at the effects of small imbalance from the PSDW side and
    map its effect as a chemical potential of a hard core bosons
    hopping on a square lattice near half filling, namely, mapping Fig.9a to Fig.9b. Our upper phase boundary
    in Fig.9b is consistent with that in \cite{jog}. We also worked
    out  the lattice structures of the C-PSDW and IC-PSDW
    and the whole physical picture  along the dashed line in
    Fig.9b. However, as pointed out in Sec.III,
    it is hard to incorporate the LLL projection in the CB. Some parameters can only be estimated in the LLL+HF approaches.
    So the CB approach in this paper and the microscopic LLL+HF approach in \cite{jog} are
    complementary to each other.

    There are other proposals which are not based on the collapsing of magnetoroton minimum.
    Recently, a trial wavefunction involving the
    coexistence of composite bosons and composite fermions at finite distance was proposed in \cite{mil}.
    However, it seems to the author that
    there is no phase transitions in this CF-CB coexistence trial wavefunction.
    This is in-consistent with the experimental observations. As
    further elucidated in Sec.V wavefunction approach is unlikely
    to be efficient in describing transitions.
    The authors in \cite{sep} argued that the state $ d_{c1} < d < d_{c2}
    $ is a phase separated state between the ESF and the FL. Based
    on this phase separation picture, Wang investigated the effect of the interlayer tunneling \cite{wang2}.

\section{Conclusions}
 {\sl (1) Comments on different approaches in SLQH}

     Three common approaches to SLQH systems are the wavefunction ( or first
   quantization ) \cite{laugh,jain}, Composite Fermion Field Theory (CFFT) \cite{frad,hlr,shankar} and
   Composite Boson Field Theory (CBFT) approaches \cite{off,cb}.
   The Wavefunction  approach has been
   very successfully applied to study SLQH at Jain's series
   at $ \nu= \frac{ p}{ 2sp \pm 1 }  $.
   CFFT has been used to study the CF Fermi liquid at $ \nu=1/2 $ which is at the end point of
   Jain's series in the limit $ p \rightarrow \infty  $. However, there are many problems
   with CFFT approach whose equivalence to the wavefunction approach is still not obvious even in SLQH.
   These problems have been vigorously addressed in \cite{shankar}.
   CBFT approach has so far been only limited to Laughlin's series $ \nu= \frac{1}{ 2s+1 } $ which
   is only a $ p=1 $ subset
   of Jain's series. It can not be used in any simple way to study $ p > 1 $ Jain series and $ \nu =1/2 $ CF Fermi
   liquid. Despite having played very important roles historically,
   except describing topological properties elegantly, both CFFT and CBFT have not
   been able to give much new information beyond those
   achieved in the wavefunction approach in describing bulk Quantum Hall States.
   One of main reasons for the success of
   the wavefunction approach is that  there is a gap in the bulk,
   suitable wavefunctions \cite{laugh,jain} can describe both the groundstate and low
   energy excitations quite accurately. Its accuracy can be checked
   easily by exact diagonalization in a finite size system whose size
   is beyond only a few magnetic length. Spherical geometry can be used to
   get rid of edge state effects quite efficiently .
   However, there are at least three
   exceptions even in SLQH which can not be described in the framework of wavefunction
   approach (1) The CF Fermi liquid at $ \nu=1/2 $. Because it is a
   gapless system, the wavefunction may not be very useful.
   unfortunately, a simple field theory description is still
   lacking. (2) The edge state: because the gap
   vanishes on the edge, field theory is very convenient to study the gapless edge excitations \cite{wen2}
   (3) The quantum phase transition from QH to QH or QH to insulating state.
   It is known that wavefunction approach is not convenient to describe any phase transition.
   A Field Theory is the most powerful method to study phase transitions. Unfortunately, so far, no field theory has been able to
   describe the transitions except in some artificial lattice models \cite{ye1,ye2}.
   In appendix B, we will construct a quantum Ginsburg Landau theory to describe QH to WC transition.

  {\sl (2) Summary of our results on BLQH }

    As shown in \cite{wave,longhua}, because of the gapless nature, at any finite $ d $,
    the wavefunction is qualitatively different from the $ (111) $ wavefunction,
    so the wavefunction approach to BLQH is far less powerful in BLQH than in SLQH.
    So Field theory approaches are
    much more powerful in BLQH than in SLQH, especially in the
    pseudo-spin sector which contains the new phenomena not
    displayed in SLQH.
      In this paper, we used both CFFT and CBFT approach to study the balanced and im-balanced BLQH systems.
      In Section II, we used a MCF approach to study  balanced and im-balanced
     BLQH systems. We achieved some limited success, but also run into many troublesome problems.
     We explicitly identified these problems and motivated the alternative CB approach.
     Extension of Murthy-Shankar's formalism \cite{shankar}
     in SLQH to BLQH is not expected to fix these problems.
     In Section III, we developed a simple and effective CB theory to study
     the BLQH.  The CB theory naturally fixed all the problems suffered
     in the MCF theory presented in Sec. II.
     By using this CB theory in its phase representation, we first studied the excitonic superfluid state and re-derived
     many previous results in a simple and transparent way. The
     theory puts spin and charge degree freedoms in the same
     footing, explicitly brought out
     the spin-charge connection in a straightforward way and classified all the possible
     excitations in a systematic way.
     We made detailed comparisons between the spin sector
     of the CB theory with EPQFM derived from microscopic LLL approach. Although
     some parameters in the spin sector can only be taken as phenomenological parameters, the functional form
     is identical to the EPQFM.  In Sec. III, using this CB theory in its dual density representation,
     we then analyzed the instability in the pseudo-spin sector and
     found the magneto-roton minimum collapsing leads to a new
     ground state which breaks the translational symmetry:
     Pseudo-spin density wave ( PSDW ) state. We proposed that
     for balanced BLQH, there are two critical distances $ d_{c1} < d_{c2} $.
     When $ 0 < d < d_{c1} $, the system is in the ESF state,
     when $ d_{c1} < d < d_{c2} $, the system is in the PSDW state,
     there is a first order transition at $ d_{c1} $ driven by the collapsing of magneto-roton minimum in
     the pseudo-spin channel. When $ d_{c2} < d < \infty $, the PSDW melts into two weakly coupled
     $ \nu =1/2 $ Composite Fermion Fermi Liquid (CFFL) state.
     There is also a first order transition at $ d= d_{c2} $.
     However, disorders could smear the two first order transitions into two second order transitions.
     The transition from the ESF to the PSDW is unusual because the
     two states break two completely different symmetries: the global internal $
     U(1) $ symmetry and  the translational symmetry respectively.
     We construct an effective theory in the dual density representation to describe
     this novel quantum phase transition. The effective action is a $ n=1 $ component $ (d+1,d) $ with $ d=2
     $ quantum Lifshitz action. The most favorable lattice is a
     square lattice instead of a hexagonal lattice. Because a 2d square lattice is not a close packed lattice,
     so it is hard to be realized experimentally. This is
     probably the only experimental realization of a 2d square lattice in any 2 dimensional system.
     The ESF to PSDW transition is similar to the
     superfluid to normal solid transition in $ ^{4} He $ system
     with the distance playing the role of the pressure.
     The correlated hopping of vacancies in the active and passive layers in the PSDW state leads
     to very large and temperature dependent drag consistent with the experimental data.
     The PSDW could be
     the true ground state responsible for all the experimental
     observations in intermediate distances.
     We also applied the  CB theory to study the effects of imbalance on
     both  the ESF and the PSDW side in corresponding sect. III and IV.
     On the ESF side, as we tune the im-balance, the system supports
     continuously changing fractional charges,
     We also  derived the dual action of the CB theory
     in terms of topological currents and dual gauge fields. By comparing this dual action with the action derived
     from MCF theory, we can explicitly identify the missing and the arti-facts of the MCF approach in section II.
     On the PSDW side, we map the square lattice PSDW at the balanced
     case into a hard core bosons hopping on a square lattice at half
     filling, then adding a small imbalance in the BLQH corresponds to
     adding a small chemical potential in the boson model. Through this mapping, we find
     the imbalance drives two quantum phase transitions:
     the first one is from the commensurate PSDW state to
     an incommensurate PSDW (IC-PSDW), the second one is from IC-PSDW state to
     the ESF state. Both transitions are first order transitions.
     We discuss the effects of disorders
     and finite temperature. We compare our results with the previous results achieved from the LLL+HF
     approach, the other proposals on the possible intermediate phases
     and available experimental data.
     We concluded that CBFT approach is superior than CFFT approach in the
     BLQH having ground states with different kinds of broken symmetries.
     It would be interesting to
     see if the PSDW can also be achieved in Tri-layer quantum Hall systems studied in \cite{tri}.
    Only the interlayer coherent
    $ (111) $ state was discussed in this paper, it can be easily generalized to other interlayer coherent
    $ (m,m,m) $ ( with $ m $ odd ) states at total filling factors $ \nu_{T}= 1/m $.

{\sl (3) General remarks on bosonic and fermionic approaches }

    It is general true that when there is an ordered state with broken symmetry
    and an associated order parameter, bosonic approach is superior than fermionic
    approach. It is well known that the bosonic approach can be easily
    applied to capture competing orders at long wavelength
    scales. As explicitly and systematically shown in \cite{qgl} and this paper,
    it can also be used to capture competing orders at microscopic length scales.
    For example, in Quantum Antiferromagnet, fermionic approach can only
    address the disordered phase \cite{aff}, while bosonic approach \cite{rs} can address not only
    the disordered phase, but also the ordered phase with broken symmetry and the
    quantum phase transition between the disordered phase and ordered phase.
    Similarly, in quantum spin glass, bosonic approach can study spin liquid, spin glass and
    the quantum phase transition between the two, while the fermionic representation can only study
    the spin liquid phase \cite{sg}.
    In SLQH system, there is only algebraic long range order, but
    no broken symmetry and no
    true ordered state, so CF approach could be very successful.
    Fermionic theory can be even used to describe QH to insulating
    transitions. For example, the QH to insulating transitions in
    a periodic potential was described in terms of $ 2+1 $ dimensional Dirac
    fermion \cite{ye1,ye2}. For general BLQH $ (m, m^{\prime}, n ) $ states
    with $ m, m^{\prime} $ are odd and
    $ m m^{\prime}-n^{2} \neq 0 $ , because there are no broken symmetry in the ground states and no
    associated gapless Goldstone modes, we expect the Composite Fermion approach works better.
    For example, $ (3, 3, 1 ) $ state at $ \nu_{T}=1/2 $ can be described in terms of the Entangled Composite
    Fermion (ECF) discussed at section II.
    While $ (m, m, m) $ BLQH system has a
    true $ U(1) $ broken symmetry ground state and an associated order parameter and a Goldstone mode,
    the CB approach becomes
    more effective as demonstrated in this paper. Most importantly,
    we use the CB approach to explore the PSDW state which breaks the translational symmetry
    instead of the $ U(1) $ symmetry.

{\sl (4) The synthesis of roton collapsing in $ d=1,2,3 $ }

    As shown in \cite{kun2}, due to the long-range Coulomb interactions between electrons, in the effective low energy theory
   describing the edge reconstruction in the FQHE,
   there are also two low energy sectors at $ k=0 $ and $ k=k_{r} $.
   It is the magneto-roton minimum collapsing  at $ k=k_{r} $ is  responsible for
   edge reconstruction in the edge state of FQHE. In one dimensional
   edge, the roton manifold at $ k=k_{r} $ becomes two disconnected
   points. Of course, 1 dimension is always special. Higher dimensions could be completely different.
   This paper showed that the magneto-roton minimum collapsing
   at $ d=2 $ leads to the PSDW state. At 2d, the roton manifold at $ k=k_{r} $
    is a circle. In \cite{qgl}, it was shown
   that the roton minimum collapsing driven by pressure could lead
   to a normal solid or a supersolid state in $ ^{4}He $. At 3d, the roton manifold at $ k=k_{r} $
   is a sphere. Combining all these results, we find that the roton
   minimum collapsing could lead to novel physics in all possible
   experimental accessible dimensions.

{\bf  Acknowledgement }

    I thank M. Chan, H. Fertig, E. Fradkin, S. M. Girvin, B. Halperin,
    J. K. Jain, Longhua Jiang, A. H. Macdonald, G. Murthy  and Jun Zhu for helpful discussions.
    I also thank Prof. Haiqing Lin
    for hospitality during my visit at Chinese University of Hong Kong in the summer of 2004.

\appendix

\section{ Meron fractional charges in imbalanced case }

   In this appendix, we will evaluate the fractional charges of
   merons in imbalanced case from their wavefunctions and find  they
   are indeed the same as those listed in Table 2. We will first
   discuss the ground state wavefunction, then the meron wavefunction.
   The balanced case was discussed in \cite{wave}. Here we extend the analysis to the imbalanced case.

\subsection{ Ground state wavefunction in first and second quantization: }

    In first quantization, the ground state trial wavefunction of BLQH in $ d \rightarrow 0 $ limit
    is Halperin's $ (111) $ state \cite{bert}:
\begin{equation}
  \Psi_{111} (z,w)=  \prod^{N_{1}}_{i=1} \prod^{N_{2}}_{j=1} (z_{i}-w_{j} )
   \prod^{N_{1}}_{ i<j } ( z_{i}-z_{j} )   \prod^{N_{2}}_{ i<j } ( w_{i}-w_{j} )
\label{111}
\end{equation}
    where $ z $ and $ w $ are the coordinates in layer 1 and layer 2 respectively.
    As explicitly written, there are $ N_{1} $ electrons in the top
    layer and $ N_{2} $ electrons in the upper layer.

    In the EPQFM picture \cite{rev,fer}, the ground state EPQFM
     wavefunction was written in second quantization form:
\begin{equation}
  | G; \theta, \phi > =  \prod^{ M-1 }_{ m=0 }
  (  \cos \frac{\theta}{2} C^{\dagger}_{m \uparrow} + \sin \frac{\theta}{2}
  e^{i \phi} C^{\dagger}_{m \downarrow} ) |0>
\label{phase}
\end{equation}
  where $ M=N $ is the angular momentum quantum number corresponding to the edge of the system.

  In the following, we will show that the EPQFM wavefunction in the
  second quantization form of Eqn.\ref{phase} is equivalent to $ (111) $ wavefunction in Eqn.\ref{111}.

    The wavefunction Eqn.\ref{phase} can also be interpreted as the
    pairing of an electron in one layer and a hole in another which
    leads to exciton condensation. It is easy to show that the charges
    on the top layer are $ N_{1}= N \cos^{2} \theta/2= N \nu_{1} $,
    while charges on the bottom layer are $ N_{2} = N \sin^{2} \theta/2 = N
    \nu_{2} $. Obviously $ N_{1}+N_{2}=N $.

     We can expand Eqn.\ref{phase} into:
\begin{eqnarray}
 & & | G; \theta, \phi  > =  ( \cos \frac{\theta}{2} )^{N} \sum^{N}_{ N_{1}=0 } \frac{ \sqrt{N !}}{ N_{1} ! N_{2} ! }
  ( tan \frac{\theta}{2} )^{N_{2}} e^{i N_{2} \phi}    \nonumber  \\
 & & \sum_{ k_{1},\cdots, k_{N} }
  (-1)^{ P( k_{1},\cdots, k_{N}  ) } C^{\dagger}_{ k_{1}, \uparrow} \cdots C^{\dagger}_{ k_{ N_1}, \uparrow}
            C^{\dagger}_{k_{N_1+1}, \downarrow} \cdots  C^{\dagger}_{k_{N_1+ N_2},\downarrow}
            |0>
\end{eqnarray}
   where $ N= N_1+ N_2 $ and $ P $ is the permutation of $ N $ variables.

   Projecting onto the state with $ N_{1} $ electrons in layer 1 and $ N_{2} $ electrons in layer 2
\begin{equation}
    |G; \theta, N_{1},  N_{2} >= \int^{2 \pi}_{0} \frac{ d \phi}{2
     \pi} e^{-i N_{2} \phi} |G;  \theta, \phi >
\label{number}
\end{equation}
   and then transforming into the first quantization form leads to:
\begin{eqnarray}
& &  | G; \theta, N_{1}, N_{2} > =  ( \cos \frac{\theta}{2} )^{N} (
tan \frac{\theta}{2} )^{N_{2}}
  \frac{ \sqrt{N !} }{ N_{1} ! N_{2} ! }
  \sum_{ k_{1},\cdots, k_{N} } (-1)^{ P( k_{1},\cdots, k_{N}  ) }
                           \nonumber  \\
& &  \frac{1}{ \sqrt{ N ! }} {\cal A}  [ \phi_{k_1}(z_{1}) \cdots
\phi_{ k_{ N_1}}( z_{N_1})
   \phi_{k_{ N_1+1}} (w_{1}) \cdots \phi_{ k_N}( w_{N_2} )
                         \nonumber  \\
& &  (1, \uparrow ) \cdots (N_1, \uparrow ) ( N_1+1, \downarrow )
\cdots ( N, \downarrow) ]
\end{eqnarray}
   where $ {\cal A} $ stands for anti-symmetrization.

   Moving the summation over the orbital states into $ {\cal A} $, the sum
   is essentially the $ (111) $ state:
\begin{eqnarray}
& & \sum_{ k_{1},\cdots, k_{N} } (-1)^{ P( k_{1},\cdots, k_{N}  ) }
   \phi_{k_1}(z_{1}) \cdots \phi_{ k_{ N_1}} ( z_{N_1} )
   \phi_{k_{ N_1+1}} (w_{1}) \cdots \phi_{ k_N}( w_{N_2} )   \nonumber  \\
&& = \prod^{N_{1}}_{i=1} \prod^{N_{2}}_{j=1} (z_{i}-w_{j} )
   \prod^{N_{1}}_{ i< j } ( z_{i}-z_{j} )   \prod^{N_{2}}_{ i< j } ( w_{i}-w_{j} ) =\psi_{111}(z,w)
\end{eqnarray}
    Finally
\begin{eqnarray}
& &  | G; \theta, N_{1}, N_{2} > =  ( \cos \frac{\theta}{2}  )^{N} (
tan \frac{\theta}{2} )^{N_{2}}
    \frac{1}{ N_{1} ! N_{2} ! }   \nonumber  \\
& &  {\cal A}  [ \psi_{111}( z,w)
   (1, \uparrow ) \cdots (N_1, \uparrow ) ( N_1+1, \downarrow ) \cdots ( N, \downarrow) ]
\end{eqnarray}

   This concludes that the orbital part of the projection of the EPQFM wavefunction Eqn.\ref{phase}
   onto $ ( N_{1}, N_{2} ) $ sector is exactly the same as $ (111) $
   state. If putting $ \theta= \pi/2 $, we recover the balanced case in \cite{wave}.
   It is clear that the $ (111) $ wavefunctions with different $
   ( N_{1}, N_{2} ) $ provide a complete basis, the ground state
   wavefunction Eqn.\ref{phase} can be expanded in terms of this
   basis, the angle $ \theta $ controls the expansion coefficients.
   $ \theta =\pi/2 ( \neq \pi/2 ) $ corresponds to the balanced (imbalanced ) case.

\subsection{ The fractional charges of merons}

   In the subscetion, we will calculate the  fractional charges  of the excitations above the the $ (111) $ state.

   The charge density on each of the layers in the ground state Eqn.\ref{phase} is:
\begin{equation}
 \rho_{G1}(z)= \nu_{1}\sum_{m=0}^{M-1}|\phi_{i}(z)|^{2},~~~
 \rho_{G2}(z)= \nu_{2} \sum_{m=0}^{M-1}|\phi_{i}(z)|^{2}
\label{charge0}
\end{equation}

    The wavefunction of a meron in the bottom layer is:
\begin{equation}
  |M, \theta, \phi, 2> = \prod_{m=0}^{M-1} (\cos{\theta/2} C^{\dagger}_{m \uparrow}+
\sin{\theta/2}e^{i\phi}C^{\dagger}_{m+1\downarrow} )|0>
\end{equation}

    The charge density on each layer is:
\begin{equation}
 \rho_{m1}(z)=\nu_{1}\sum_{m=0}^{M-1}|\phi_{i}(z)|^{2},~~~~~
 \rho_{m2}(z)=\nu_{2} \sum_{m=1}^{M}|\phi_{i}(z)|^{2}
\label{charge1}
\end{equation}

     Subtracting Eqn.\ref{charge0} from Eqn.\ref{charge1} leads to the fractional charge of the meron:
\begin{equation}
 \delta\rho_{m1}(z)=0,~~~~
 \delta\rho_{m2}(z)=- \nu_{2}[|\phi_{0}(z)|^{2}-|\phi_{M}(z)|^{2}]
\label{mcharge}
\end{equation}

    The charges in top layer remains untouched,
    while $ \nu_{2} $ charge is moved from the origin to the boundary ( Fig.1 ),
    namely, there is a hole with charge $ \nu_{2} $ at the origin and extra charge $ -\nu_{2} $ at the
    boundary.

\vspace{0.5cm}

\begin{figure}
\includegraphics[width=8cm]{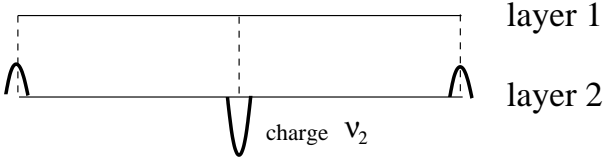}
\caption{ The charge densities of the smallest meron wavefunction }
 \label{A1}
\end{figure}

\vspace{0.25cm}

   As concluded in \cite{wave}, because the meron wavefunction
   ignores the strong interlayer correlation, the charge
   distributions will be strongly modified by the neutral gapless
   mode. However, as shown in III B (2), the total charge is a topological
   property. Although the meron wavefunction may not even be a
   qualitatively a good one, it still should give a correct answer
   on the total charge. Indeed,  Eqn. \ref{mcharge} gives the  correct total charge  $ \nu_{2} $
   listed in Table 2.

\section{Quantum Hall state to Wigner crystal transition in single layer QH systems }

    Despite the success of wavefunction approach in SLQH, the
    nature of quantum phase transitions from QH to insulating state is beyond the scope of
    the wavefunction approach.  It remains one of the outstanding
    problems in QH. Field theory is a must to study the transitions. Quantum Hall transitions in periodic potential
    in the presence of Coulomb interaction was investigated in
    \cite{ye1} in terms of Dirac fermion. The effects of both Coulomb interaction and
    disorders on the transition were discussed in \cite{ye2}.
    But these investigations may not be directly relevant to the
    real QH transitions which happen in a continuous system. In this appendix, inspired by the ESF to
    PSDW transition investigated in Sec.IV, we study the QH to Wigner
    Crystal (WC) transition in SLQH.

    It is instructive to point out the difference between
    SLQH and BLQH. In the SLQH, there is no true symmetry
    breaking, the collective mode at $ k=0 $ turns out to be a local maximum.
    There is still a magneto-roton minimum at $
    q_{0} \sim l $ (Fig.B1), the collapsing of the minimum also signifies the
    collapsing of the QH gap, the system gets to a possible Wigner
    crystal state (Fig.B1). Due to lack of true symmetry breaking, it is not known how to
    characterize the QH order by a local order parameter. The QH
    states may possess possible topological orders which need to be characterized by non-local
    order parameters. So QH and WC also have two completely
    different orders: topological order and translational order.
    In the BLQH, the pseudo-spin sector is a charge
    neutral sector, as shown in \cite{qgl}, both low energy modes at $ k=0 $
    and $ k=k_{0} $ are important. The collapsing of the magneto-roton  minimum  in the charge neutral sector
    leads to the PSDW in Fig.5.

\vspace{0.25cm}

\begin{figure}
\includegraphics[width=8cm]{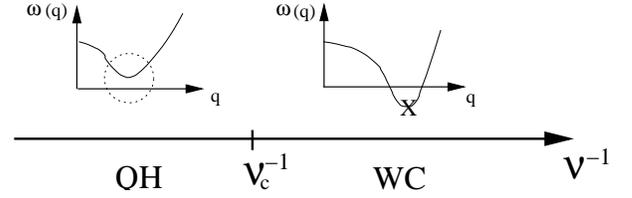}
\caption{The zero temperature phase diagram in a single layer
quantum Hall system as the filling factor $ \nu $ changes. QH stands
for Quantum Hall state, WC stands for Wigner Crystal state. The
dispersion relations of the collective modes in the two phases are
also shown. The cross in the WC means the negative magneto-roton
minimum is replaced by the WC formation.  }
\label{B1}
\end{figure}

\vspace{0.25cm}

  At fixed electron density, as the magnetic field increases, the
  filling factor $ \nu $ decreases, strong Quantum Hall states are
  discovered in $ \nu=1/3,1/5,1/7 $, but not in $ 1/9 $ where the
  system may be in the WC state. The WC state will be pinned by disorders and become an insulating
  state. The quantunm Ginsburg-Landau action to describe the QH to the WC
transition is:
\begin{eqnarray}
 {\cal L}_{wc}[ \delta \rho ] & = &
    \frac{1}{2} \delta \rho( -\vec{q},- \omega ) [ A_{\rho} \omega^{2}
    + r+ c( q^{2}-q^{2}_{0} )^{2} ]  \delta \rho( \vec{q},\omega )
        \nonumber   \\
    & - & w ( \delta  \rho )^{3} + u ( \delta \rho )^{4} + \cdots
\label{densitywc}
\end{eqnarray}
    where $ \rho= c^{\dagger} c $ is the electron density, $ \delta \rho= \rho- \rho_{0} $ is the deviation
    from the average electron density, the momentum and frequency conservations in the
    cubic and quartic terms are assumed.

 In sharp contrast to the QGL action to describe the ESF to the PSDW transition Eqn.\ref{densitymin}
 where the possible cubic interaction term is forbidden by  the $ Z_{2} $ exchange symmetry between the two
 layers, the cubic term is allowed. So Eqn.\ref{densitywc} is similar to
 that Eqn.\ref{sfdensity} describing the SF to NS transition in Helium 4, except the former is in $ 2+1 $ dimension
 and the latter is in $ 3 + 1 $ dimension.
 Unfortunately, the QH order is not transparent in Eqn.\ref{densitywc}.

  In Eqn.\ref{densitywc}, $ r $ is the magnetoroton minimum at $ q=q_{0} $
  which tunes the 1st order transition from the QH to the WC.
  In the QH fluid state, $ r > 0 $ and $ \rho $ is uniform, $ < \delta \rho >=0 $. In the WC, $ r < 0 $ and
  $ < \delta \rho > = \sum^{\prime}_{\vec{G}} n (\vec{G} ) e^{ i \vec{G} \cdot \vec{x} } $
  where $ \prime $ stands for the sum over all non-zero reciprical
  lattice vectors takes a lattice structure. Due to the cubic term, it can be shown
  that the most favorable lattice structure is the hexagonal
  lattice Fig.6c. This is in sharp contrast to the PSDW square
  lattice Fig.5 in BLQH.

  Disorders will turn the 1st order transition into a 2nd order one,
  so scaling behaviors at low temperature are expected. As shown in
  \cite{ye1,ye2}, Coulomb interactions will make the dynamic
  exponent of the 2nd order transition $ z=1 $. We believe the Coulomb interaction is automatically
  incorporated in the action Eqn.\ref{densitywc}. How to incorporate the effects of disorder into
  the action remains an outstanding problem.

\section{  Bilayer quantum Hall system in a periodic potential }

    The Bilayer quantum Hall system in a periodic potential can be described by the following relativistic
    model:
\begin{eqnarray}
  {\cal L }_{b} & =  & | ( \partial_{\mu} - i a_{\mu} ) z_{i} |^{2} + m^{2} | z |^{2}
     +  u  | z |^{4} +  \frac{ i}{ 4 \pi \theta } \epsilon_{\mu \nu \lambda} a_{\mu} \partial_{\nu} a_{\lambda}
                                       \nonumber   \\
     & +  &  U ( | z_{1} |^{2} - | z_{2} |^{2} )^{2}
\label{rela}
\end{eqnarray}
    where $ i=1,2 $ are the two components, $ |z|^{2}=| z_{1} |^{2} + | z_{2} |^{2}  $,
    $ \theta $ is the statistical angle and
    $ U >0 $ ( $ U<0 $ ) plays the role of the easy-plane  ( Ising ) anisotropy which breaks
    the symmetry in the pseudo-spin sector from $ SU(2) $ down to $ U(1) \times Z_{2} $.

    The SLQH case with just one complex field was
    investigated in \cite{wu}. It can be considered as a model
    describing the transition from a quantum Hall state to a Mott
    insulator (MI) in a periodic potential as the strength of the periodic potential is varied.
    The fermionic version of the
    QH to MI transition in a periodic potential  was investigated in
    \cite{ye1,ye2}. In both bosonic and fermionic versions, the CS
    term is exactly marginal, the transition is a second order
    transition with continuously changing exponents depending on the
    statistical angle $ \theta $.
    One can view Eqn.\ref{rela} to describe the transition
    in a BLQH system in a two 2-dimensional periodic potentials as the distance between
    the two layers is varied. We assume there is a critical distance
    $ d_{c} $ such that  $ m^{2} ( d<d_{c} )< 0, m^{2} ( d=d_{c} )=0, m^{2} ( d>d_{c} )>0
    $. Due to the exact marginality of the CS term, we expect the
    transition is still 2nd order with continuously changing exponents depending on the
    statistical angle $ \theta $.

    In the imbalanced case, we add a Zeeman-like term to Eqn.\ref{rela}
\begin{equation}
    {\cal L }_{imb}= {\cal L}_{b}- h_{z} ( | z_{1} |^{2} - | z_{2} |^{2} )
\label{imb}
\end{equation}

     The advantage of the relativistic theory Eqn.\ref{imb} over the non-relativistic  model Eqn.\ref{main}
     in the main text is that it puts the fluctuations of spin sector and charge sector at the same footing.
     Of course, its physics could be different from the real BLQH
     studied in the main text, but it is also interesting in its own
     right. It would be also interesting to compare its phases and
     phase diagram with those of the real BLQH discussed in the main
     text. In the following, we will first work out the phase diagrams of two
     closely  related models with non-compact and compact Maxwell gauge
     theories in (a) and (b), then we study the phase diagram of Eqn.\ref{imb} in (c).

{\sl (a) Two component non-compact Maxwell theory}

    If replacing the CS term in Eqn.\ref{imb} by a {\em non-compact} Maxwell
    term, we get
\begin{eqnarray}
  {\cal L }_{non} & =  & | ( \partial_{\mu} - i a_{\mu} ) z_{i} |^{2} + m^{2} | z |^{2}
     +  u  | z |^{4} +  \frac{1}{ 4 e^{2} } ( f_{\mu \nu}  )^{2}
                                       \nonumber   \\
     & +  &  U ( | z_{1} |^{2} - | z_{2} |^{2} )^{2} - h_{z} ( | z_{1} |^{2} - | z_{2} |^{2} )
\label{nonc}
\end{eqnarray}
    where $ f_{\mu \nu}=  \partial_{\mu} a_{\nu}- \partial_{\nu} a_{\mu} $.

    This model had been studied in \cite{decon} in both isotropic $
    U=0 $ and easy plane $ U > 0 $ limit. Here we also study the Ising limit $ U<0 $.
    In the easy-plane limit $ U>0 $, the model was shown to be self-dual
    and there is a 2nd order transition through a
    deconfined quantum critical point \cite{decon}.
    Under the duality transformation, the spin and charge sectors exchange their roles on the two sides of
    the QCP. There is one QPT driven by the imbalance
    on both sides of the QCP which are XY and inverted XY
    transitions respectively, so the two transitions on the two sides are indeed dual to each other
    ( Fig.C1a ) \cite{decon}. In the Ising limit $ U < 0 $,  on the $ P^{*} $ side,
    the transition is still in the universality class of the inverted
    XY transition, but in the Ising order side, there is no transition, the theory is not self-dual
    anymore ( Fig.C1b ). The transition may even be a 1st order
    transition driven by the gauge field fluctuations.

\vspace{0.25cm}

\begin{figure}
\includegraphics[width=8cm]{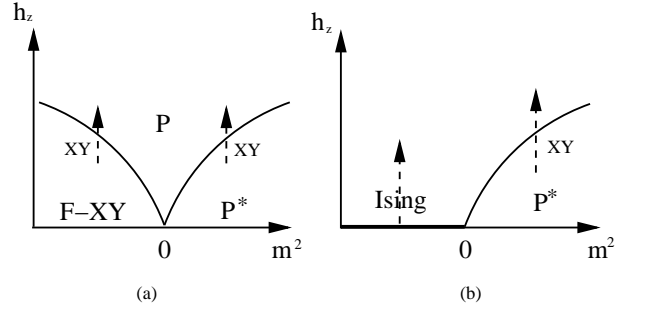}
\caption{The phase diagram of Eqn.\ref{nonc} at
      $ T=0 $ (a) Easy-plane limit $ U>0 $ is self-dual.
      The three phases: XY Ferromagnet ( $ F-XY $ ), exotic paramagnet ( $ P^{*} $ )
      and paramagnet ( $ P $ ) meet at
      the Tetra- quantum critical point which is a
      deconfined quantum critical point \cite{decon}. The  $ P^{*} $ phase has a gapless photon mode.
      (b) Ising limit $ U<0 $, the transition from the Ising ordered phase to the  $ P^{*} $
           may be first order driven by gauge field fluctuations indicated by a dot.  }
\label{C1}
\end{figure}

\vspace{0.25cm}

    The finite temperature phase diagram at the balanced case $
    h_{z}=0 $ is shown in the Fig.2. There is one QPT driven by the
    temperature on both sides of the QCP which is in the universality class of the KT
    transition ( Fig.C2a ) \cite{decon}. In the Ising limit $ U < 0 $, on the $ P^{*} $ side,
    the transition is still in the universality class of the KT
    transition, but in the Ising ordered side, the transition is in the universality class of the
    Ising transition, the theory is not self-dual anymore.( Fig.C2b ).

\vspace{0.25cm}

\begin{figure}
\includegraphics[width=8cm]{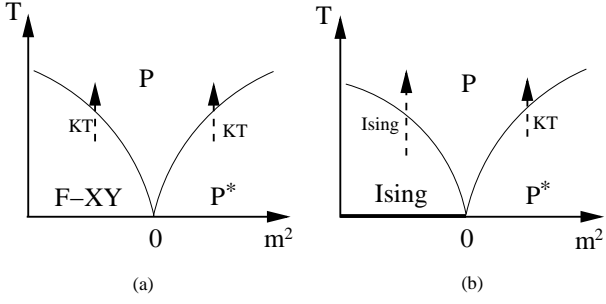}
\caption{The finite temperature phase diagram of
      of Eqn.\ref{nonc} at $ h_{z}=0 $. (a) Easy-plane limit $ U>0 $ which is self
      dual. (b)  Ising limit $ U<0 $ which is not self dual  }
\label{C2}
\end{figure}

\vspace{0.25cm}

{\sl (b) Two component Compact Maxwell theory }

    However, if the Maxwell term in Eqn.\ref{nonc} is {\em compact}, then the compact gauge field
    confine the $ z_{\alpha} $
    fields into a single unit vector $ \vec{n}= z^{*}_{\alpha}( \vec{\sigma} )_{\alpha \beta} z_{\beta} $,
    it can be shown that the compact model becomes:
\begin{equation}
  {\cal L}_{comp}= \frac{1}{2g} ( \partial_{\mu} \vec{n} )^{2} + U (
  n_{z} )^{2}-h_{z} n_{z}
\label{comp}
\end{equation}

    In the easy plane limit, the model is {\em dual } to
    a one component superconductor in an external magnetic field $ h_{z} $ as described by Eqn.\ref{is}. So there
    is one ( two ) QPT(s) for type I ( type II ) superconductor on the QD side only driven by the imbalance.
    The type II case was already shown in Fig.9. The type I case was
    shown in Fig.C3a. The zero density transition was discussed in \cite{nature}, it has
    the exact exponent $ z=2, \nu=1, \eta=0 $. In the Ising limit,
    there is no transition on both sides as shown in Fig.C3b.

\vspace{0.25cm}

\begin{figure}
\includegraphics[width=8cm]{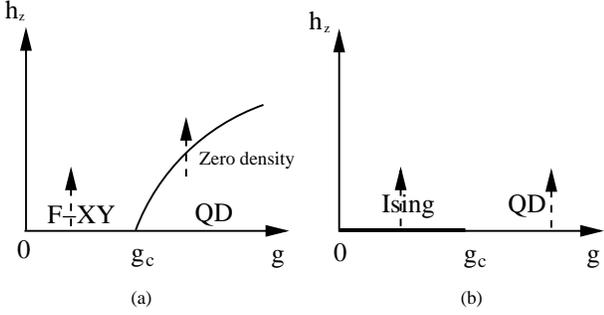}
\caption{The phase diagram of the Eqn.\ref{comp}
  at $ T=0 $. (a) Easy-plane limit $ U>0 $ (b)  Ising limit $ U<0 $.  } \label{C3}
\end{figure}

\vspace{0.25cm}

   The finite temperature phase diagram at the balanced case $
   h_{z}=0 $ is shown in the Fig.C4. On the
   ordered side, there is a KT transition in the easy-plane limit (
   Fig.C4a ) and an Ising transition  in the Ising limit ( Fig.C4b ).

\vspace{0.5cm}

\begin{figure}
\includegraphics[width=8cm]{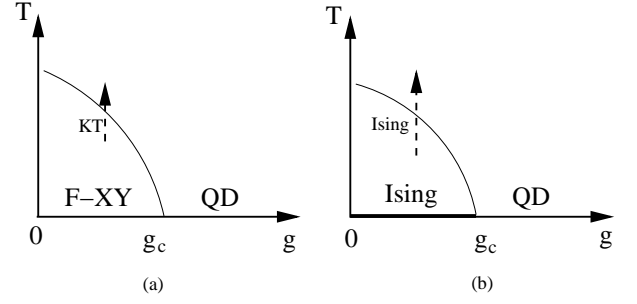}
\caption{ The finite temperature phase diagram
   of Eqn.\ref{comp} (a) Easy-plane limit $ U>0 $ (b)  Ising limit $ U<0 $.  } \label{C4}
\end{figure}

\vspace{0.25cm}

    So the phase diagram  of the two component Maxwell theory is very different depending
    on the Maxwell term is compact or non-compact.

{\sl (c)  Two components Chern-Simon theory  }

    In the present model Eqn.\ref{imb} with the CS term, it was shown in \cite{cscom} that
    in a pure compact Maxwell-Chern-Simon
    theory, monopoles are linearly confined by a string of magnetic flux due to the CS term.
    Following the procedures in \cite{rs}, it is easy to see that in the model Eqn.\ref{imb} where
    the two component relativistic bosonic matter fields $ z_{\alpha}, \alpha =1,2 $ are coupled to a compact CS term,
    the monopoles remain confined, therefore the two component bosonic fields are deconfined.
    So the CS term is compact or not
    in Eqn.\ref{imb} will not lead to qualitative changes in the phase diagram.
    In the balanced case $ h_{z}=0 $, in the easy-plane limit,
    due to the $ Z_{2} $ symmetry, $ z_{1} $ and $ z_{2} $ have to condense at the same time.
    When $ d < d_{c} $ ( $ m^{2} < 0  $ ), the system is in the ILCQH which is a
    Quantum Hall state  with $ \nu=1 $ integer Hall conductance $ \sigma_{xy} = e^{2}/h $ with a charge gap
    and a superfluid state with gapless Goldstone mode in the spin sector.  While $ d> d_{c} $ ( $ m^{2} > 0  $ ),
    the system is in a quantum disordered state with a spin gap, but charge gapless.
    In the imbalanced case $ h_{z} > 0 $,
    because $ z_{1} $ and $ z_{2} $ are {\em deconfined}, they do not have to condense at the same distance anymore.
    In the following, we discuss the Easy plane and Ising limit  respectively.
    (1) Easy-palne limit, $ U>0 $ ( Fig.C5a ):
    On the QD side, when $ h_{z} >  m^{2} $, $ z_{1} $ becomes condensed, the QPT is in the same universality
    class as the QH transition in a single layer system, namely, one component complex scalar coupled to a CS gauge field
    in $ 2+1 $ dimension. Let's call this transition: QD-SLQH.
    Starting from the SLQH at fixed $ h_{z} $, if decreasing the distance further, the
    $ z_{2} $ will also condense, the system will get into the BLQH
    state with interlayer phase coherence, let's call this transition: SLQH-BLQH, because the $ U(1) $ CS
    gauge field is already broken by the $ z_{1} $ condensation, this transition is
    in the 3d $ XY $ universality class as shown in Fig.C5a. We can look at the
    transition from a different way:
    on the ILCQH side, when $ h_{z} > - m^{2} $, $ z_{2} $ becomes
    uncondensed.
    Obviously, the two transitions are
    not self-dual, in contrast to the non-compact Maxwell case in Fig.1. The QCP at $ d=d_{c} $ in
    Fig.C5a is essentially a Tetra-quantum-critical point.
    (2) Ising limit $ U<0 $ ( Fig.C5b ). At the balanced case, on
    the ordered side, there is only a SLQH state which has no gapless
    Goldstone mode, so there is no BLQH state. At finite $ h_{z} $,
     there is  a QD-SLQH  transition in the disordered side,
     while there is no transition in the SLQH side.
      Note that the transitions in both easy plane limit (a) and Ising limit (b) are second order due to the exact
      marginality of the CS term, even the theory is not self-dual in both cases. This is in sharp contrast to the
      corresponding phase diagram  Fig.C1 in the non-compact Maxwell
      theory.

\vspace{0.25cm}

\begin{figure}
\includegraphics[width=8cm]{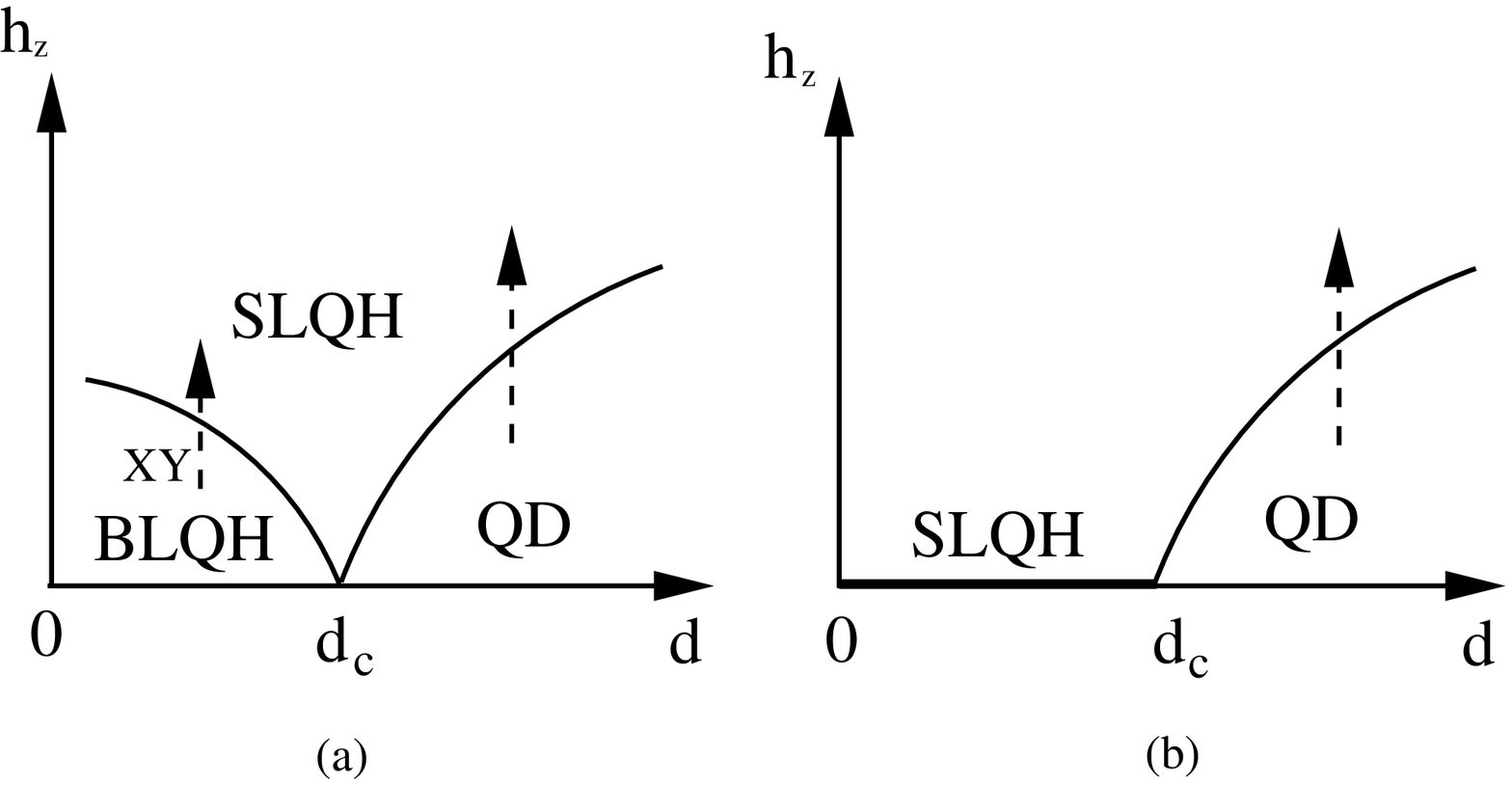}
\caption{The phase diagram of
      of Eqn.\ref{imb} at $ T=0 $. (a) Easy-plane limit $ U>0 $,
      there are three phases meeting at the Tetra- quantum critical point: (1) Bilayer Quantum Hall state
      with interlayer coherence $ <z_{1} > \neq 0, <z_{2}> \neq 0 $  ( BLQH+ILCH  )  (2) Single Quantum Hall state
      $ < z_{1} > \neq 0, < z_{2} >=0 $  (SLQH )
      (3) Quantum Disordered state $ < z_{1} > = < z_{2} >=0 $.
      (b)  Ising limit $ U<0 $, there is no BLQH state. The SLQH and the QD meets at a Bi-quantum critical point.   }
 \label{C5}
\end{figure}

\vspace{0.25cm}

   The finite temperature phase diagram at the balanced case $
   h_{z}=0 $ is shown in the Fig.C6. On the
   ordered side, there is a KT transition in the easy-plane limit (
   Fig.C6a ) and an Ising transition  in the Ising limit ( Fig.C6b ).
   Fig.C6 is similar to Fig.C4 to some extent.

\vspace{0.25cm}

\begin{figure}
\includegraphics[width=8cm]{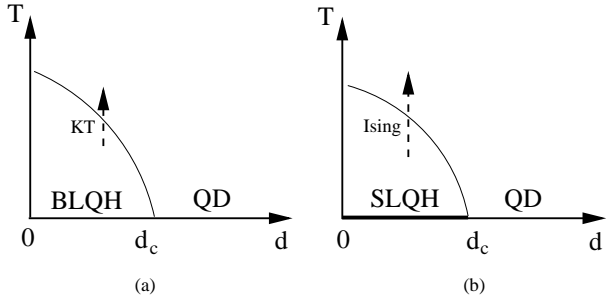}
\caption{The finite temperature phase diagram
       of Eqn.\ref{imb}. (a) Easy-plane limit $ U>0 $ (b)  Ising limit $ U<0 $   }
 \label{C6}
\end{figure}

\vspace{0.25cm}

    One can compare Fig.C5(a) ( Fif.C6(a) )  with Fig.9(b) ( Fig.8 )
    in the main text. As expected, they are different, because they are two different systems. The former is
    the BLQH system in a periodic potential, while the latter is the real
    BLQH in a continuous system.


\begin{thebibliography}{99}


\bibitem{bert} B. I. Halperin, Helv. Phys. Acta 56, 75 (1983); Surf. Sci. 305, 1 (1994)

\bibitem{rev} For reviews of bilayer quantum Hall systems, see S. M. Girvin and A. H. Macdonald,
in {\sl Perspectives in Quantum Hall Effects}, edited by S. Das Sarma and Aron Pinczuk ( Wiley, New York, 1997).

\bibitem{supp} J. P. Eisenstein, L. N. Pfeiffer and K. W. West, Phys. Rev. Lett. 69, 3804 (1992);
               Song He, P. M. Platzman and B. I. Halperin, Phys. Rev. Lett. 71, 777 (1993).

\bibitem{gold} I. B. Spielman {\sl et al}, Phys. Rev. Lett. 84, 5808 (2000). {\sl ibid}, 87, 036803 (2001).

\bibitem{hall} M. Kellogg, {\sl et al}, Phys. Rev. Lett. 88, 126804 (2002).

\bibitem{counterflow} M. Kellogg, {\sl et al}, cond-mat/0401521.

\bibitem{counterflow2}  J.P. Eisenstein and A.H. MacDonald, cond-mat/0404113.

\bibitem{fer} H. Fertig, Phys. Rev. B 40, 1087 (1989).

\bibitem{wen} X. G. Wen and A. Zee, Phys. Rev. Lett. 69, 1811 (1992).

\bibitem{jap} Z. F. Ezawa and A. Iwazaki, Phys. Rev. B. 47, 7259; 48, 15189 (1993). Phys. Rev. Lett. 70, 3119 (1993).

\bibitem{yang} Kun Yang {\sl et al}, Phys. Rev. Lett. 72, 732 (1994). Phys. Rev. B 54, 11644 (1996).

\bibitem{moon} K. Moon {\sl et al}, Phys. Rev. B 51, 5138 (1995).

\bibitem{balents} A. Stern, S. M. Girvin, A. H. Macdonald and N. Ma, {\sl ibid} 86,
1829 (2001). Y. N. Joglekar and A. H. Macdonald, Phys. Rev. Lett.
87, 196802 (2001), Ziqiang Wang, Phys. Rev. Lett. 94, 176804 (2005).


\bibitem{transfer}  Enrico Rossi, Alvaro S. N¨²nez, A.H. MacDonald,  Phys. Rev. Lett. 95, 266804
(2005).

\bibitem{phasetran} J. Schliemann, S. M. Girvin, A. H. Macdonald, Phys. Rev. Lett. 86, 1849 (2001).

\bibitem{benice}  A. L\' opez and E. Fradkin, Phys. Rev. B 51, 4347 (1995); {\sl ibid}, 63, 085306, 2001.
  See also a review article by the two authors in
  {\sl Composite Fermions:  A unified view of the Quantum Hall
  Regime}, edited by Olle Heinonen. World Scientific (Singapore, 1998).

\bibitem{kim} Yong Baek Kim {\sl et al}, cond-mat/0011459.

\bibitem{phi2} M. Y. Veillette, L. Balents and M. P. A. Fisher, Phys. Rev. B. 66, 155401 (2002).
      Due to using $ \tilde{\phi}=2 $, the authors concluded that the mutual Hall drag resistivity is $ 2
      h/e^{2} $ which is in-consistent with the experimental result in \cite{hall}.

\bibitem{jain1} V. W. Scarola and J. K. Jain, Phys. Rev. B 64, 085313 (2001).

\bibitem{drag} M. Kellogg, {\sl et al}, Phys. Rev. Lett. 90, 246801 (2003).

\bibitem{he4} P. Kapitza,  Nature 141, 74 (1938).

\bibitem{he3} D. D. Osheroff, R. C. Richardson, and D. M. Lee, Phys. Rev. Lett. 28, 885¨C888 (1972)

\bibitem{chan} E. Kim and M. H. W. Chan, Nature 427, 225 - 227 (15 Jan 2004).

\bibitem{science}  E. Kim and M. H. W. Chan, Science 24 September 2004; 305: 1941-1944.

\bibitem{andreev} A.  Andreev and I. Lifshitz, Sov. Phys. JETP {\bf 29}, 1107
(1969).

\bibitem{ches}  G. V. Chester, Phys. Rev. A {\bf 2}, 256 (1970).

\bibitem{ander} P. W. Anderson, W. F. Brinkman, David A. Huse, Science 18, 310: 1164-1166 (2005).

\bibitem{qgl} Jinwu Ye, Phys. Rev. Lett. 97, 125302 (2006);  cond-mat/0603269,

\bibitem{and} P. W. Anderson, cond-mat/9812063.

\bibitem{random} Jinwu Ye, Phys. Rev. Lett. 86, 316 (2001).


\bibitem{quantum} Jinwu Ye, Phys. Rev. Lett. 87, 227003 (2001); Phys. Rev. B. 65, 214505 (2002).

\bibitem{ex} The concept of MCF was implied previously in \cite{wen,benice,jain1} in different languages.
    Here, we explicitly write down the relation between a MCF and an original electron by Eqn.\ref{sing}
    which is more convenient to the following developments in the paper.

\bibitem{jain} J. K. Jain, Phys. Rev. Lett. 63, 199 (1989).

\bibitem{frad} A. Lopez and E. Fradkin, Phys. Rev. B. 44, 5246 (1991).

\bibitem{hlr} B. I. Halperin, P. A. Lee and N. Read, Phys. Rev. B47, 7312 (1993).





\bibitem{shankar} See the review by G. Murthy and R. Shankar, Rev. of Mod. Phys. 75, 1101, 2003.


\bibitem{off} S. M. Girvin and A. H. Macdonald, Phys. Rev. Lett. 58, 1252 (1987).

\bibitem{cb} S. Z. Zhang, T. H. Hansson and S. Kivelson, Phys. Rev. Lett. 62, 82 (1989).
 S. Z. Zhang, Int. J. Mod. Phys. B 6, 25 (1992).


\bibitem{wave} Gun Sang Jeon and Jinwu Ye, Phys. Rev. B 71, 035348 (2005).


\bibitem{non} For discussions on relations between  Abelian bosonization and Non-Abelian bosonization
     approaches to multi-channel Kondo model, see Jinwu Ye, Phys. Rev. Lett. 77, 3224 (1996).

\bibitem{ring} Feymann originally conceived the 3d roton as drifting
   vortex loop. But this point of view is very controversial. If taking this view, then the  3d roton
   condensation can be considered as
   the vortex loop condensation. In 2d BLQH, the magneto-roton condensation
   in Fig.11 may be considered as that of the charge neutral meron
   pairs listed below Eqn.\ref{chargeb} and Eqn.\ref{chargeimb}.

\bibitem{sfhe} In charge neutral superfluid $ ^{4} He $, the roton
dispersion relation is $ \omega^{2} \sim  q^{2} (  a- bq^{2} +
cq^{4} ) $. See \cite{qgl}.


\bibitem{momentum} I. Ussishkin and A. Stern, Phys. Rev. B, 56, 4013
(1997); S. Sakhi, Phys. Rev. B, 56, 4098 (1997); Y. B. Kim and A.J.
Millis, Physica E ( Amsterdam)4, 171 (1999).

\bibitem{pump} M.E. Raikh and F. von Oppen, Phys. Rev. Lett. 89, 106601 (2002).
   V. M. Apalkov and M. E. Raikh, Phys. Rev. B 71, 245109 (2005).


\bibitem{3dxy} If the transition were at $ q=0 $ in the Fig.5, the
intermediate phase would be a featureless Exciton Liquid (EL) phase
which respects all the symmetries even at $ T=0 $. This kind of
phase is extremely unlikely despite it was claimed by several
authors in different contexts. This transition at $ q=0 $ will be
discussed in the appendix C. This EL phase is the same as the
quantum disordered (QD) phase in Fig.C3a and Fig.C4a.

\bibitem{un} Jinwu Ye, unpublished.

\bibitem{hex} For a nice review on the KTHNY theory on two
dimensional lattice, see the wonderful book by  P. M. Chaikin and T.
C. Lubensky, Principles of Condensed Matter Physics, Cambridge
university press,1995. Note that the hexatic  phase in the KTHNY
theory is still not convincingly established in experiments.

\bibitem{bra} S. A. Brazovskii,  JETP 41, 85 (1975).

\bibitem{tom} For  discussions on Classical Lifshitz Point (CLP) and their applications in nematic to
smectic-A and -C transitions in liquid crystal, see the book in
\cite{hex}.


\bibitem{pq1} L. Balents, L. Bartosch, A. Burkov, S. Sachdev, K. Sengupta, Phy. Rev. B 71, 144508 (2005).

\bibitem{nature} Jinwu Ye,  cond-mat/0503113.

\bibitem{he} P. A.  Crowell and J. D. Reppy, Phys. Rev. Lett. 70, 3291¨C3294
(1993); Phys. Rev. B 53, 2701¨C2718 (1996).

\bibitem{honey} H. Wiechert and K. Kortmann, Phys. Rev. B 70, 125410 (2004).

\bibitem{senthil} T. Senthil {\sl et. al}, Science 303, 1490 (2004); Phys. Rev. B 70, 144407 (2004).

\bibitem{han} C. B. Hanna, Bull, Am. Phys. Soc. 42, 553 (1997)

\bibitem{jog} Y. N. Joglekar and A. H. Macdonald, Phys. Rev. B 65, 235319 (2002).

\bibitem{exp}  W. R. Clarke, {\sl et. al}, preprint

\bibitem{imbexp}   I.B. Spielman,  {\sl et al}, cond-mat/0406067.

\bibitem{laugh} Laughlin, in {\sl The Quantum Hall Effects}, 2nd ed. edited by R. E. Prange
   and S. M. Girvin ( Springer-Verlag, New York, 1990).

\bibitem{light} For light scattering experiments, see Aron Pinczuk,
Chap. 8 in the book in \cite{rev}.

\bibitem{bert2} I thank B. Halperin for discussions leading to this
point.









\bibitem{dual} M. Peskin, Ann. Phys. 113, 122 (1978); C. Dasgupta and B. I. Halperin, Phys. Rev. Lett. 47, 1556 (1981).


\bibitem{kun1} Kun Yang, Phys. Rev. Lett, 87, 056802 (2001).

\bibitem{tri} Jinwu Ye, Phys. Rev. B 71, 125314 (2005).

\bibitem{longhua}  Longhua Jiang and Jinwu Ye, Phys. Rev. B 74, 245311 (2006).


\bibitem{mil} S. Simon, E. H. Rezayi and M. V. Milovanovic,  Phys. Rev. Lett. 91, 046803 (2003) .

\bibitem{sep}  A. Stern and B. I. Halperin,   Phys. Rev. Lett. 88,
106801 (2002).

\bibitem{wang2} Ziqiang Wang, Phys. Rev. Lett. 92, 136803 (2004).

\bibitem{wigner} L. Brey, Phys. Rev. Lett, 65, 903 (1990);
 R. Cote, L. Brey and A. H. Macdonald, Phys. Rev. B 46, 10239 (1992);
 X. M. Chen and J. J. Quinn, Phys. Rev. Lett, 67, 895 (1991),
 Phys. Rev. B 45, 11054 (1992), Phys. Rev. B 47, 3999 (1993);
  L. Zheng and H. A. Fertig, Phys. Rev. B 52, 12282 (1995)
  S. Narasimhan and Tin-Lun Ho, Phys. Rev. B 52, 12291 (1995)

\bibitem{wignerexp} H. C. Manoharan, {\sl et al}, Phys. Rev. Lett, 77, 1813
(1996);  M. Shayegan in Chap. 9 in the book in \cite{rev}.

\bibitem{wen2} X. G. Wen, Advances in Physics, 44, 405-473 (1995).

\bibitem{aff} I. Affleck and J. B. Marston, Phys. Rev. B 37, 3774 (1988).

\bibitem{rs} N. Read and S. Sachdev, Phys. Rev. B 42, 4568 (1990).

\bibitem{sg} S. Sachdev and Jinwu Ye, Phys. Rev. Lett. 70, 3339 (1993).

\bibitem{kun2} Kun Yang, Phys. Rev. Lett, 91, 036802 (2003).



\bibitem{ye1} Jinwu Ye and S. Sachdev, Phys. Rev. Lett. 80, 5409 (1998).

\bibitem{ye2} Jinwu Ye, Phys. Rev. B 60, 8290 (1999).

\bibitem{decon}  O I. Motrunich and A. Vishwanath;  cond-mat/0311222.

\bibitem{wu} X. G. Wen and Y. S. Wu; Phys. Rev. Lett. 70, 1501 (1993).


\bibitem{cscom} P. D. Pisarski, Phys. Rev. D 34, 3851 (1986); I. Affleck, {\sl et. al}, Nucl. Phys. B 328, 575 (1989);
  M . C.  Diamantini {\sl et al}: Phys. Rev. Lett. 71, 1969 (1993).

\end{thebibliography}
\end{document}